\documentclass[a4]{aa}

\usepackage[dvipsnames]{xcolor}
\usepackage{xspace}
\usepackage{graphicx}
\usepackage{txfonts}
\usepackage[pdftex,pagebackref,hyperindex=true,colorlinks=true,bookmarks=true,bookmarksopen=true,bookmarksnumbered=true,citecolor=OliveGreen]{hyperref}
\usepackage{siunitx}
\usepackage[normalem]{ulem}
\usepackage{float}
\usepackage{caption}
\usepackage{subcaption}
\usepackage{longtable}
\usepackage{xtab}
\usepackage{pdflscape}
\usepackage{adjustbox}
\usepackage{booktabs}

\DeclareSIUnit\parsec{pc}
\DeclareSIUnit\Year{y}
\DeclareSIUnit\angstrom{\text{Å}}
\DeclareSIUnit\mag{\text{mag}}
\DeclareSIUnit\dex{\text{dex}}
\newlength\defaultparindent

\usepackage{natbib}
\bibpunct{(}{)}{;}{a}{,}{,}

\newcommand{\ie}{\emph{i.e.}\xspace}
\newcommand{\eg}{\emph{e.g.}\xspace}

\newcommand{\snr}{\ensuremath{\mathrm{S}/\mathrm{N}}\xspace}
\newcommand{\abratio}[2]{[\mathrm{#1}/\mathrm{#2}]\xspace}

\newcommand{\Gaia}{\emph{Gaia}\xspace}

\newcommand{\GES}{\Gaia-ESO\xspace}
\newcommand{\GESlg}{\Gaia-ESO\xspace}
\newcommand{\GESsh}{GES\xspace}
\newcommand{\gspspec}{GSP-Spec\xspace}

\newcommand{\revun}[1]{#1\xspace}
\newcommand{\revdeux}[1]{#1\xspace}

\begin{document}

\title{The \Gaia-ESO Survey DR5.1 and \Gaia DR3 \gspspec: a comparative analysis\thanks{Based on observations collected with the FLAMES instrument at VLT/UT2 telescope (Paranal Observatory, ESO, Chile), for the \Gaia-ESO Large Public Spectroscopic Survey (188.B-3002, 193.B-0936, 197.B-1074).}}
\titlerunning{\GESsh DR5.1 and \Gaia DR3 \gspspec}
\authorrunning{Van der Swaelmen et al.}

\author{M. Van der Swaelmen
  \inst{\ref{oaa}}
  \and
  C. Viscasillas V{\'a}zquez
  \inst{\ref{vilnius}}
  \and
  L. Magrini
  \inst{\ref{oaa}}
  \and
  A. Recio-Blanco
  \inst{\ref{oca}}
  \and
  P.~A. Palicio
  \inst{\ref{oca}}
  \and
  C. Worley
  \inst{\ref{cambridge},\ref{canterbury}}
  \and
  A. Vallenari
  \inst{\ref{oapd}}
  \and
  L. Spina
  \inst{\ref{oaa}}
  \and
  P. Fran{\c c}ois
  \inst{\ref{gepi}}
  \and
  G. Tautvai{\v s}ien{\. e}
  \inst{\ref{vilnius}}
  \and
  G.~G. Sacco
  \inst{\ref{oaa}}
  \and
  S. Randich
  \inst{\ref{oaa}}
  \and
  P. de~Laverny
  \inst{\ref{oca}}
}

\institute{
  INAF -- Osservatorio Astrofisico di Arcetri, Largo E. Fermi 5, 50125, Firenze, Italy\\
  \email{mathieu.van+der+swaelmen@inaf.it}\label{oaa}
  \and
  Institute of Theoretical Physics and Astronomy, Vilnius University, Sauletekio av. 3, 10257 Vilnius, Lithuania\label{vilnius}
  \and
  Université Côte d’Azur, Observatoire de la Côte d’Azur, CNRS, Laboratoire Lagrange, Bd de l'Observatoire, CS 34229, 06304 Nice Cedex 4, France\label{oca}
  \and
  Institute of Astronomy, University of Cambridge, Madingley Road, Cambridge CB3 0HA, United Kingdom\label{cambridge}
  \and
  School of Physical and Chemical Sciences- Te Kura Mat\={u}, University of Canterbury, Private Bag 4800, Christchurch 8140, New Zealand\label{canterbury}
  \and
  INAF -- Padova Observatory, Vicolo dell'Osservatorio 5, 35122 Padova, Italy\label{oapd}
  \and
  GEPI, Observatoire de Paris, PSL Research University, CNRS, Université Paris Diderot, Sorbonne Paris Cité, 61 avenue de l'Observatoire, 75014 Paris, France\label{gepi}
}

\date{Submitted: 12/04/2024; accepted: 26/06/2024}

\abstract
{The third data release of \Gaia, has provided stellar parameters, metallicity $\abratio{M}{H}$, $\abratio{\alpha}{Fe}$, individual abundances, \revun{broadening parameter} from its Radial Velocity Spectrograph (RVS) spectra for about \num{5.6} million objects thanks to the \gspspec \revun{module, implemented in the \Gaia} pipeline. \revun{The catalogue also publishes the radial velocity of \num{33} million sources.} In recent years, many spectroscopic surveys with ground-based telescopes have been undertaken, including the public survey \GESlg, designed to be complementary to \Gaia, \revun{in particular towards faint stars}.}
{We took advantage of the intersections between \Gaia RVS and \GESlg to compare their stellar parameters, abundances and radial and rotational velocities. We aimed at verifying the overall agreement between the two datasets, considering the various calibrations and the quality-control flag system suggested for the \Gaia \gspspec parameters.}
{For the targets in common between \Gaia RVS and \GESlg, we performed several statistical checks on the distributions of their stellar parameters, abundances and velocities of targets in common. For the \revun{\Gaia} surface gravity and metallicity we considered both the uncalibrated and calibrated values.}
{Overall, there is a good agreement between the results of the two surveys. \revun{We find an excellent agreement between the \Gaia and \GESlg radial velocities given the uncertainties affecting each dataset. Less than \num{25} out of the $\approx \num{2100}$ \GESlg spectroscopic binaries are flagged as non-single stars by \Gaia.} For the effective temperature \revdeux{and in the bright regime ($G \le 11$)}, we found a very good agreement, with an absolute residual difference of \revun{about \SI{5}{\kelvin} ($\pm \SI{90}{\kelvin}$) for the giant stars and of about \SI{17}{\kelvin} ($\pm \SI{135}{\kelvin}$) for the dwarf stars; in the faint regime ($G \ge 11$)}, we found a worse agreement, with an absolute residual difference of \revun{about \SI{107}{\kelvin} ($\pm \SI{145}{\kelvin}$) for the giant stars and of about \SI{103}{\kelvin} ($\pm \SI{258}{\kelvin}$) for the dwarf stars.} For the surface gravity, the comparison indicates that the calibrated gravity should be preferred to the uncalibrated one. For the metallicity, we observe in both the uncalibrated and calibrated cases a slight trend whereby \Gaia overestimates it at low metallicity; for $\abratio{M}{H}$ \revun{and $\abratio{\alpha}{Fe}$}, a \revun{marginally} better agreement is found using the calibrated \Gaia results; finally for the individual abundances (Mg, Si, Ca, Ti, S, Cr, Ni, Ce) our comparison suggests to avoid results with flags indicating low quality ($\mathrm{XUncer} = 2$ \revun{or higher}). \revun{These remarks are in line with the ones formulated by \gspspec.} \revun{We confirm that the \Gaia \texttt{vbroad} parameter is loosely correlated with the \GESlg $v \sin i$ for slow rotators.} \revun{Finally, we note that the quality (accuracy, precision) of the \gspspec parameters degrades quickly for objects fainter than $G \approx 11$ or $G_{\mathrm{RVS}} \approx 10$.}}
{\revdeux{We find that the somewhat imprecise \gspspec abundances due to its medium-resolution spectroscopy over a short wavelength window and the faint $G$ regime of the sample under study can be counterbalanced by working with averaged quantities.} We extended our comparison to star clusters \revdeux{using averaged abundances}, using not only the stars in common, but also the members of clusters in common between the two samples, still finding a very good agreement. Encouraged by this result, we studied some properties of the open-cluster population, using both \GESlg and \Gaia clusters: our combined sample traces very well the radial metallicity and $\abratio{Ca}{Fe}$ gradients, the age-metallicity relations in different radial regions, and allows us to place the clusters in the thin disc.}


\maketitle

\section{Introduction}
\label{Sec:introduction}

The large public spectroscopic survey \GESlg (\GESsh) observed for \num{340} nights at the Very Large Telescope (VLT) from the end of 2011 to 2018. It obtained about \num{190000} spectra, for nearly \num{115000} targets with a wide variety of scientific objectives, covering all Galactic populations. In the two survey articles \citet{gilmore22} and \citet{randich22} announcing the final data release, the survey design and structure and some of the main scientific achievements are described. The \GESlg survey is still the only one dedicated stellar spectroscopic survey using \SI{8}{\metre} class telescopes, with the explicit aim of being complementary to the data obtained by the \Gaia satellite, which were not yet available at the time the survey began. \GESlg was, indeed, among the first of the many completed and ongoing large projects\footnote{For the spectroscopic surveys listed in this sentence, we provide when possible a reference of an early publication presenting the survey and a reference linking to a recent data release.}, such as Radial Velocity Experiment \citep[RAVE;][]{2006AJ....132.1645S,2020AJ....160...82S}, Large sky Area Multi-Object fibre Spectroscopic Telescope (LAMOST; low resolution: \citealp{2012RAA....12.1197C,2022MNRAS.517.4875L}; medium resolution: \citealp{2020arXiv200507210L, 2021ApJS..256...14Z}), Apache Point Observatory Galactic Evolution Experiment \citep[APOGEE-1 and APOGEE-2;][]{2014ApJS..211...17A,2020AJ....160..120J}, Galactic Archaeology with HERMES \citep[GALAH;][]{2015MNRAS.449.2604D,2021MNRAS.506..150B}, \Gaia Radial Velocity Spectrometer \citep[\Gaia-RVS;][]{2004MNRAS.354.1223K,2018A&A...616A...5C,2023A&A...674A..29R}, and the future multi-object spectrographs and large or massive surveys such as William Herschel Telescope Enhanced Area Velocity Explorer \citep[WEAVE;]{2016ASPC..507...97D}, Multi-Object Optical and Near-infrared Spectrograph \citep[MOONS;][]{2020Msngr.180...10C}, Multi-Object Spectrograph Telescope \citep[4MOST;][]{2019Msngr.175....3D}, the Milky-Way Mapper \citep[MWM;][]{2017arXiv171103234K}, and the Subaru Prime Focus Spectrograph \citep[PFS;][]{2014PASJ...66R...1T}. All of these surveys aim at spectroscopically sampling the Galactic stellar populations and at characterising them from a chemo-dynamical point of view.

We recall that these completed and ongoing surveys differ in terms of: \textbf{a) instrumental resolution}, from the low resolution with LAMOST ($R \approx 1800$), medium-low resolution with \Gaia-RVS ($R \approx 11500$) or RAVE ($R \approx 7500$), medium resolution with \GESsh GIRAFFE and APOGEE ($R \approx 20000$), medium-high resolution with GALAH ($R \approx 28000$) and high resolution with the \GESsh UVES ($R \approx 47000$); \textbf{b) spectral coverage} (bands and range lengths), \eg optical \revdeux{+ near-infrared \revdeux{($I$-band)}} with \GESsh and GALAH, near-infrared \revdeux{only} with \Gaia-RVS and RAVE \revdeux{($I$-band)}, and APOGEE \revdeux{($H$-band)}; \textbf{c) sampled stellar types} (hot/warm/cool main-sequence stars, red-giant-branch stars); \textbf{d) sky coverage}, \eg Northern sky for APOGEE-1, Southern sky for \GESsh, GALAH and APOGEE-2, and all-sky for \Gaia; \textbf{e) magnitude range} of the science targets; \textbf{f) selection functions}; \textbf{g) analysis methods}, \eg single main pipeline for APOGEE and GALAH along with possible re-analyses using third-parties pipelines or multiple independent pipelines merged into a single set of results after homogenisation for \GESsh. \revun{It is also worth noting that unlike all other spectroscopic surveys, \Gaia-RVS is space-based spectroscopy, and therefore, the spectra are not affected by the Earth atmosphere absorption and emission.} The main high-value-added data-products of these spectroscopic surveys comprise the radial velocities of the targets, the three atmospheric parameters $\left\{T_{\mathrm{eff}}, \log g, \abratio{Fe}{H}\right\}$ and a series of individual abundances for various ions.

Far from being duplicated works, these surveys are complementary to each others since they map different stars belonging to different stellar populations and located in different parts of the Galaxy. Since it is possible to find non-empty intersections between them, they are crucial to answer numerous questions animating the stellar community such as validating the methods, building the largest unbiased sample of chemo-dynamically characterised stars, or rejecting or confirming the findings about the build-up history of the Milky Way. Because of the diversity of resolution, wavelength coverage and analysis methods, the high-value-added data-products released by the aforementioned surveys may come with different accuracy and precision. An important task is therefore to take advantage of the non-empty intersections between two given surveys to discover and correct possible biases in order to control the overall cross-survey agreement and safely combine the results from different works according to their biases and uncertainties. Such an effort is already ongoing and has led to a number of publications. We quote among others: the comparison APOGEE DR14\,+\,LAMOST DR3 for radial velocities and atmospheric parameters in \citet{2018A&A...620A..76A}, the comparison APOGEE DR17\,+\,GALAH DR3\,+\,GES DR5 for atmospheric parameters in \citet{2023A&A...670A.107H}, the comparison \Gaia DR3\,+\,APOGEE DR17\,+\,GALAH DR3\,+\,GES DR3\,+\,LAMOST DR7\,+\,RAVE DR6 for radial velocities in \citet{2023A&A...674A...5K}, the comparison \Gaia DR3\,+\,APOGEE DR17\,+\,GALAH DR3\,+\,RAVE DR6 for atmospheric parameters and abundances in \citet{2023A&A...674A..29R}, the combination of radial velocities \Gaia DR2\,+\,APOGEE DR16\,+\,GALAH DR2\,+\,GES DR3\,+\,LAMOST DR5\,+\,RAVE DR6 in \citet{2022A&A...659A..95T}, the combination of surveys by label-transfer APOGEE\,+\,GALAH in \citet{2022MNRAS.513..232N} or APOGEE\,+\,LAMOST in \citet{2017ApJ...836....5H}, or again the validation of \Gaia spectroscopic orbits with LAMOST and GALAH radial velocities in \citet{2022MNRAS.517.3888B}.

\GESlg \revun{science operations} have ended in July 2023 with the fifth and final public data release (DR5.1)\footnote{\url{https://www.eso.org/qi/catalog/show/411}}, corresponding to the sixth internal data release (iDR6). Now that \Gaia is in its third data release (DR3), which includes, for the first time, the results of the analysis of RVS spectra, providing spectroscopically-derived temperature, surface gravity, metallicity $\abratio{M}{H}$ and $\alpha$-content $\abratio{\alpha}{Fe}$, but also individual abundances of many elements \citep{2023A&A...674A..29R}, a comparison with the final results of \GESlg is definitely timely. As far as we know, already published articles comparing \GESsh results to other surveys have used \GESsh DR3 \revun{(\eg \citealt{2022A&A...659A..95T})}, which was made available in December 2016 and contains the spectroscopic chemo-kinematic information for only \num{26000} unique objects based on \num{30} months of observations. The work presented here is the first to compare \GESsh DR5.1 to \Gaia DR3 in both their kinematical and chemical data-products and the aim of this paper is to exploit the intersection between the two surveys in order to suggest practical criteria for selecting the best spectroscopic abundances from \Gaia.

The paper is structured as follows: in Section~\ref{Sec:data_samples} we describe the \GESlg and \Gaia RVS catalogues, and their intersections. In Section~\ref{Sec:radial_velocities}, we discuss the radial velocities and the census of multiple stars for the targets in common to the two surveys. Section~\ref{Sec:Rotational_velocities} compares the rotational velocities of stars to the broadening parameter. In Section~\ref{Sec:comparison_stars_in_common}, we compare the stellar parameters and abundances, while in Sect.~\ref{Sec:astero} we discuss the agreement of spectroscopic gravities and metallicities to the same quantities obtained with asteroseismic constraints for a smaller subset of stars. Finally, Sections~\ref{Sec:open_clusters} and \ref{Sec:science_with_ocs} show that \GESlg and \Gaia can be combined to derive some properties of open-cluster member stars and of Milky Way open clusters.

\section{Data and samples of stars in common}
\label{Sec:data_samples}

\subsection{The \GESlg DR5}

In this section, we recall the main aspects of the \GESlg; more information can be found in the two papers accompanying the final release \citep{gilmore22,randich22}. The \GESlg used the multi-object spectrograph FLAMES \citep{2002Msngr.110....1P} equipped with UVES and GIRAFFE fibres and mounted on the Nasmyth focus of VLT/UT2: while \num{130} fibres are feeding the GIRAFFE spectrograph, eight fibres simultaneously feed the UVES spectrograph. The survey worked in two resolution modes, a medium spectral resolution of about \num{20000} for GIRAFFE observations and a high spectral resolution of \num{47000} for UVES observations. In addition, more than one setup were used with a given instrument: two UVES setups and nearly ten different GIRAFFE setups. Though it adds a complexity to the data management and data analysis \citep[see][]{2023A&A...676A.129H, Worley2024}, this choice allowed the consortium to select the wavelength range of interest according to the targeted stellar type, the stellar population or the science case. The GIRAFFE setup HR15N \revun{($[\SI{6470}{\angstrom}, \SI{6790}{\angstrom}]$)} was mainly used for Milky Way star clusters and, for instance, it gives access to the lithium line at \SI{6707.8}{\angstrom}, whose abundances are crucial for stellar physics \citep[\eg][]{2022A&A...668A..49F} or cosmology \citep[\eg][]{Bonifacio18}. On the other hand, the GIRAFFE setups HR10 \revun{($[\SI{5339}{\angstrom}, \SI{5619}{\angstrom}]$)} and HR21 \revun{($[\SI{8484}{\angstrom}, \SI{9001}{\angstrom}]$)} were mainly used for the Milky Way field stars since they contain crucial lines for the determination of stellar parameters and of several abundances \citep{gilmore22}. One interest of HR21 for the validation of techniques and results is that its wavelength range overlaps that of the \Gaia spectrograph, encompassing the three lines of the near-infrared \ion{Ca}{II} triplet. More than three-quarter of the \GESsh observations were carried out with HR15N, HR10 and HR21.

\GESlg is organised in working groups (WG) composed of one or more analysis nodes responsible for deriving the atmospheric parameters and stellar abundances of the observed targets \citep{Worley2024}. Radial velocities and rotational broadening are instead provided in a centralised way by data reduction pipelines. WG10 analyses FGK field, open-cluster and globular-cluster stars observed with GIRAFFE, while WG11 analyses the same kinds of stars observed with UVES \citep{Worley2024}; WG12 analyses main and pre-main-sequence stars observed in young open clusters with UVES and GIRAFFE; WG13 analyses OBA-type stars in young clusters observed with UVES and GIRAFFE \citep[\eg][]{2022A&A...661A.120B}; WG14 identifies and characterises stellar peculiarities (multiplicity, \eg \citealp{2023arXiv231204721V}; emission lines).

In order to provide a final set of results, WG15 implements several sophisticated homogenisation procedures which maximise the regions of the parameter space in which the nodes perform best \citep{2023A&A...676A.129H}. At the end of the parameter determination phase, a first homogenisation occurs to create a unique set of atmospheric parameters $\left\{T_{\mathrm{eff}}, \log g, \abratio{Fe}{H}\right\}$ that was then injected into the next phase for the abundance determination. A second homogenisation occurs to create a unique set of individual abundances after the abundance determination phase. A third independent homogenisation occurs to combine the independent estimates of the radial velocities.

To help the homogenisation \citep{2023A&A...676A.129H}, subsets of objects play a dedicated role \citep[see][]{2017A&A...598A...5P}: the \num{29} \Gaia Radial Velocity Standards \citep{2013A&A...552A..64S} fix the zero-point of the radial velocity scale; the \num{42} \Gaia Benchmark Stars \citep[\eg][]{Heiter15} serve as absolute calibrators for the parameter scales; \num{15} open and globular clusters serve as absolute calibrators to control the internal quality of the metallicity and other chemical abundances; stars in common between two different GIRAFFE and/or UVES setups serve as inter-setup calibrators for radial velocities (all other setups being put on the GIRAFFE HR10 velocity scale), stellar parameters and abundances; a number of stars observed in Kepler K2 and CoRoT fields can be used as asteroseismic calibrators thanks to their independent asteroseismic estimate of the surface gravity and are used for a posteriori quality checks \citep[\eg see][]{Worley20}.

Finally, each star's identifier (called CNAME) may come with a list of TECH flags reporting analysis issues and comments and a list of PECULI flags indicating a peculiarity (\eg suspected multiplicity, emission lines). These flags are thought as a helper for the end-user to clean their sample or, on the contrary, to focus on peculiar objects. We refer the reader to \citet{2023A&A...676A.129H} for a description of the decision trees adopted and a detailed description of the homogenisation procedures.

The \GESsh DR5.1 publishes the results for \num{114916} unique stars, plus the Sun\footnote{However, the public catalogue hosted by ESO does not contain the data for the Sun {that can be accessed from the final internal data release, {\sc iDR6}}}: without any filtering on uncertainties and flags, we count \num{111348} stars with a radial velocity $v_{\star}$, \num{88353} stars with all three atmospheric parameters $\left\{T_{\mathrm{eff}}, \log g, \abratio{Fe}{H}\right\}$ and \num{39406} stars with a rotational velocity $v \sin i$.

\subsection{The \Gaia DR3}

The \Gaia mission operated by the European Space Agency (ESA) was launched in December 2013 from the Kourou spaceport in French Guiana, and since this date, it has indisputably become a game-changer for astronomers thanks to its unprecedented deep and all-sky coverage \citep{1997ESASP.402..743P,2016A&A...595A...1G}. The \Gaia collaboration publishes releases once every several years. The latest release, \Gaia DR3\footnote{\url{https://www.cosmos.esa.int/web/gaia/dr3}} \citep{2023A&A...674A...1G}, has become public in June 2022 and it comprises the full astrometric solution for nearly \num{1.5} billion sources from $G \approx \SI{3}{\mag}$ and up to $G \approx \SI{21}{\mag}$. Compared to the previous public release -- the intermediate early DR3 (eDR3) --, \Gaia DR3 provides a wealth of new data-products of particular interest for the stellar and spectroscopic community: \num{1.59} billion sources bear an object classification; \num{33} million stars with $G_{\mathrm{RVS}} < 14$ and $T_{\mathrm{eff}} \in [\SI{3100}{\kelvin}, \SI{14500}{\kelvin}]$ possess a mean radial velocity \citep{2023A&A...674A...5K}; \num{470} million objects have an estimate for $\left\{T_{\mathrm{eff}}, \log g, \abratio{Fe}{H}\right\}$ from the BP/RP spectra \citep[Apsis\/GSP-Phot:][]{2023A&A...674A..27A}; \num{5.6} million objects have an estimate for $\left\{T_{\mathrm{eff}}, \log g, \abratio{Fe}{H}\right\}$, global $\abratio{\alpha}{Fe}$ and individual abundances for up to twelve species from RVS spectra (Apsis/\gspspec: \citealp{2023A&A...674A..29R}; general presentation of Apsis: \citealp{2023A&A...674A..26C,2023A&A...674A..28F}); mean BP/RP spectra \citep{2023A&A...674A...2D} and mean RVS spectra are available for \num{219} million and \num{1} million sources respectively; \num{3.5} million objects with $G_{\mathrm{RVS}} < 12$ possess a broadening parameter that can be used as a proxy for the rotational velocity \citep{2023A&A...674A...8F}. In addition, the study of genuine variability in physical quantities has given access to time-dependent physics \citep[\eg][]{2023A&A...674A..13E} in \Gaia DR2 and more specifically in \Gaia DR3 \citep{2023A&A...674A..34G} thanks to radial-velocity and mainly photometric variability.

The \Gaia RVS and its use are described in \citet{2004MNRAS.354.1223K} and \citet{2018A&A...616A...5C}. We recall its main features hereafter. The RVS is an integral field spectrograph working at an instrumental resolution of $R = 11500$ and covering the wavelength range $[\SI{8450}{\angstrom}, \SI{8720}{\angstrom}]$. For each transit three different spectra are recorded by the three CCDs along the scan direction, with a total exposure-time amounting to \SI{13.3}{\second}. Since the CCDs are illuminated by the spectra of all stars crossing the \Gaia field of view during a given transit, a deblending procedure is needed to separate each single RVS spectrum. The first release to make use of the RVS is \Gaia DR2 \citep{2019A&A...622A.205K} with the publication of radial velocities for \num{7.2} million sources $G_{\mathrm{RVS}} < 12$ and $T_{\mathrm{eff}} \in [\SI{3550}{\kelvin}, \SI{6900}{\kelvin}]$; \Gaia DR3 has increased the catalogue of radial velocities by a factor of 4.5 and has extended the range of magnitudes ($G_{\mathrm{RVS}} < 14$) and of effective temperatures (up to \SI{14500}{\kelvin}) for which the mean RVS radial-velocity could be measured. The measurement of the RVS radial velocities relies on the standard technique of cross-correlation computation. The specific procedures of spectra deblending, template selection, cross-correlation computation, specific handling of hot stars and faint stars, computation of the mean velocity and its associated uncertainty, final sample cleaning are explained in \citet{2018A&A...616A...5C}, \citet{2018A&A...616A...6S},  \citet{2019A&A...622A.205K}, and \citet{2023A&A...674A...5K}. 

A first set of \Gaia effective temperatures came with \Gaia DR2 \citep{2018A&A...616A...8A} using the three \Gaia bands $G$, $G_{\mathrm{BP}}$ and $G_{\mathrm{RP}}$, thus forming a catalogue of \num{160} million $T_{\mathrm{eff}}$ with -- quoting the original article -- a "likely underestimated" precision of $\approx \SI{300}{\kelvin}$. The picture has improved with \Gaia DR3 using the General Stellar Parametriser from photometry on BP/RP spectra \citep{2023A&A...674A..27A} and the General Stellar Parametriser from spectroscopy \citep[\gspspec;][]{2023A&A...674A..29R} on RVS spectra. \gspspec is one module of the Astrophysical parameters inference system (Apsis; \citealp{2023A&A...674A..26C}), which is the pipeline run by the coordination unit 8 (CU8) "Astrophysical Parameters" and \revun{which, among other goals, aims at} exploiting both low-resolution BP/RP spectra and medium-resolution RVS spectra to derive a number of spectroscopic parameters characterising the physics and chemical composition of stellar atmospheres. Despite its short wavelength range of \SI{240}{\angstrom} and the medium resolution of $R = 11500$, \citet{2021A&A...654A.130C} showed that more than \num{30} atomic and molecular absorption features can be successfully used in a {typical} RVS spectrum to measure the abundances of up to \num{13} chemical species. \gspspec runs two different workflows to obtain the estimates of the atmospheric parameters and abundances, namely MatisseGauguin and ANN. In this paper, we use the set of results obtained with MatisseGauguin since it is the unique pipeline providing individual abundances in addition to the three atmospheric parameters and the global $\abratio{\alpha}{Fe}$. \citet{2023A&A...674A..29R} employed APOGEE DR17 \citep{2022ApJS..259...35A}, GALAH DR3 \citep{2021MNRAS.506..150B} and RAVE DR6 \citep{2020AJ....160...83S} to assess the quality of the \gspspec MatisseGauguin parameters and fitted a series of polynomial functions to calibrate the following \gspspec parameters: surface gravity $\log g$, metallicity $\abratio{M}{H}$, global $\abratio{\alpha}{Fe}$ and individual abundances $\abratio{X}{Fe}$. A second calibration of the metallicity $\abratio{M}{H}$ is specifically computed for open cluster stars. In this paper, the three sets of \gspspec MatisseGauguin parameters will be respectively referred as 'uncalibrated', 'calibrated' and 'calibratedOC' (for the specific calibration for open clusters). Onwards, the expression ``\Gaia'' and ``\gspspec'' will be used interchangeably when it relates the parameters and abundances obtained from the mean RVS spectra by \citet{2023A&A...674A..29R}. We refer the reader to \citet{2023A&A...674A..29R} for details on the estimating of these physical quantities, their associated uncertainties, their quality flags and the cross-surveys calibrations. Thus, \Gaia DR3 brings to the community the largest catalogue of homogeneously obtained atmospheric parameters and chemical abundances for \num{5.5} million stars. A striking demonstration of the use of these results to understand the Milky Way can be found in \citet{2023A&A...674A..38G}.

\subsection{The \GESsh DR5.1--\Gaia DR3 intersections}
\label{Sec:gaia_ges_intersection}

As shown in Fig.~2 of \citet{2023A&A...674A..29R}, the \GESsh and \Gaia surveys sample stars in different magnitude ranges. For this reason, the joint sample is expected to become limited in number \emph{when} we request specific physical quantities. The cross-match between \Gaia DR3 and \GESsh DR5.1 with a cone search of \SI{2}{\arcsecond} returns \num{114864} matches (parent sample $\mathcal{S}_{0}$); we find ambiguous matches for \num{52} stars and they are simply removed from the analysis. However, among the \num{114864} stars, \num{19855} of them have a \Gaia DR3 radial velocity $v_{\mathrm{rad,Gaia}}$ (subsample $\mathcal{S}_{1}$), \num{2094} of them have a \Gaia DR3 broadening parameter $v_{\mathrm{broad,Gaia}}$ (subsample $\mathcal{S}_{2}$) and \num{2079} of them have the three \Gaia \gspspec spectroscopic parameters $\left\{T_{\mathrm{eff}}, \log g, \abratio{Fe}{H}\right\}$ (subsample $\mathcal{S}_{3}$). Finally, \num{404} stars among the \num{114864} ones of the parent sample are flagged as non-single stars in \Gaia DR3 (subsample $\mathcal{S}_{4}$). In details, we find \num{251} astrometric binaries (AB), \num{112} spectroscopic binaries (SB), \num{19} eclipsing binaries (EB), \num{21} AB+SB, and one EB+SB. In the next sections, the subsamples $\mathcal{S}_{1}$ to $\mathcal{S}_{4}$ will be used as a starting selection to carry out the comparison between \Gaia DR3 and \GESsh DR5.1.

\begin{table*}
  \centering
  \caption{\label{Table:Intersections_summary} Numbers of stars with valid (set of) measurements for a given chemo-physical parameter found in \GESlg (second column) and \Gaia (fourth column) and numbers of stars valid (set of) measurements found in the intersection between \GESlg and \Gaia (fifth column). The third column gives the number of stars with valid (set of) measurements found in \GESlg when restricted to the sample $\mathcal{S}_{j}$ indicated in the title of the table's sub-block. $\abratio{\alpha}{Fe}$ is obtained by averaging individual abundances of $\alpha$ elements for \GESlg, while it is a globally-fitted estimate for \Gaia. We note that the intersections listed in the last column of the table may have a smaller number of data-points if one requests also valid estimates for the uncertainties of the studied parameters.}
  \begin{tabular}{lS[table-format=6.0]S[table-format=5.0]S[table-format=5.0]cS[table-format=5.0]}
    \toprule
    {Parameter/condition} & {\GESsh DR5.1} & {\GESsh DR5.1} & \multicolumn{2}{c}{\Gaia DR3} & {Intersection}\\
     & {(full catalogue)} & {(in $\mathcal{S}_{j}$ only)} & & & \\
    \midrule
    \multicolumn{6}{c}{Radial velocity ($\mathcal{S}_{1}$)}\\
    
    $v_{\mathrm{rad}}$                                                                         & 111303 & 19636 & 19855 & ($\mathcal{S}_{1}$) & 19636\\
    $v_{\mathrm{rad}}$ \& \texttt{no\_binary}                                                  & 109197 & 18888 & 19500 &                    & 18888\\

    $v_{\mathrm{rad}}$ \& \texttt{no\_binary} \& \texttt{no\_emission}                         &  91485 & 16951 & 19500 &                    & 16951\\
    $v_{\mathrm{rad}}$ \& \texttt{no\_binary} \& \texttt{no\_emission} \& \texttt{ges\_qf}     &  90664 & 16784 & 19500 &                    & 16784\\
    \midrule
    \multicolumn{6}{c}{Rotational/broadening velocity ($\mathcal{S}_{2}$)}\\
    
    $v \sin i$ & 39379 & 1832 & 2094 & ($\mathcal{S}_{2}$) & 1832\\
    \midrule
    \multicolumn{6}{c}{Atmospheric parameters ($\mathcal{S}_{3}$)}\\
    
    $T_{\mathrm{eff}}$ \& $\log g$ \& $\abratio{Fe}{H}$ & 88330 & 1566 & 2079 & ($\mathcal{S}_{3}$) & 1566\\
    $T_{\mathrm{eff}}$                                  & 97470 & 1939 & 2079 &                    & 1939\\
    $\log g$                                        & 89772 & 1575 & 2079 &                     & 1575\\
    $\abratio{Fe}{H}$                               & 94223 & 1904 & 2079 &                     & 1904\\
    \midrule
    \multicolumn{6}{c}{Chemical composition ($\mathcal{S}_{3}$)}\\
    
    $\abratio{Mg}{Fe}$     & 54201 & 1045 &  140 & &  116\\
    $\abratio{Si}{Fe}$     & 34536 & 1160 &  140 & &  114\\
    $\abratio{Ca}{Fe}$     & 49333 & 1195 &  816 & &  503\\
    $\abratio{Ti}{Fe}$     & 39126 & 1195 &  198 & &  147\\
    $\abratio{\alpha}{Fe}$ & 17995 & 1037 & 2079 & & 1037\\
    $\abratio{S}{Fe}$      &  8823 &  213 &   27 & &    4\\
    $\abratio{Cr}{Fe}$     & 46141 & 1102 &   22 & &   19\\
    $\abratio{Ni}{Fe}$     & 46572 & 1174 &  102 & &   90\\
    $\abratio{Ce}{Fe}$     &  2745 &  689 &   19 & &   13\\
    $\abratio{Nd}{Fe}$     & 10534 & 1008 &    1 & &    0\\
    \midrule
    \multicolumn{6}{c}{Multiplicity ($\mathcal{S}_{4}$)}\\
    
    non-single stars & 2117\tablefootmark{a} & 22 & 404 & ($\mathcal{S}_{4}$) & 22\\
    non-single stars & 1113\tablefootmark{b} &  8 & 404 & ($\mathcal{S}_{4}$) &  8\\
    \bottomrule
  \end{tabular}
  \tablefoot{
    \tablefoottext{a}{For \GESlg, the final census of spectroscopic binaries is not yet available (Van der Swaelmen et al., \emph{in prep}). This number is the preliminary number of SB1, SB2, SB3 and SB4 uncovered in the final \GESsh DR5.1.}
    \tablefoottext{b}{This number relies on published data: it takes into account the SB1 discussed in \citet{2020A&A...635A.155M} and the SB2, SB3 and SB4 discussed in \citet{2023arXiv231204721V}. However, these two publications are based on the fifth internal \GESsh data release (iDR5), so a dataset slightly smaller than the \GESsh DR5.1.}
    }
\end{table*}

\begin{figure}
  \centering
  \includegraphics[width=\columnwidth,clip]{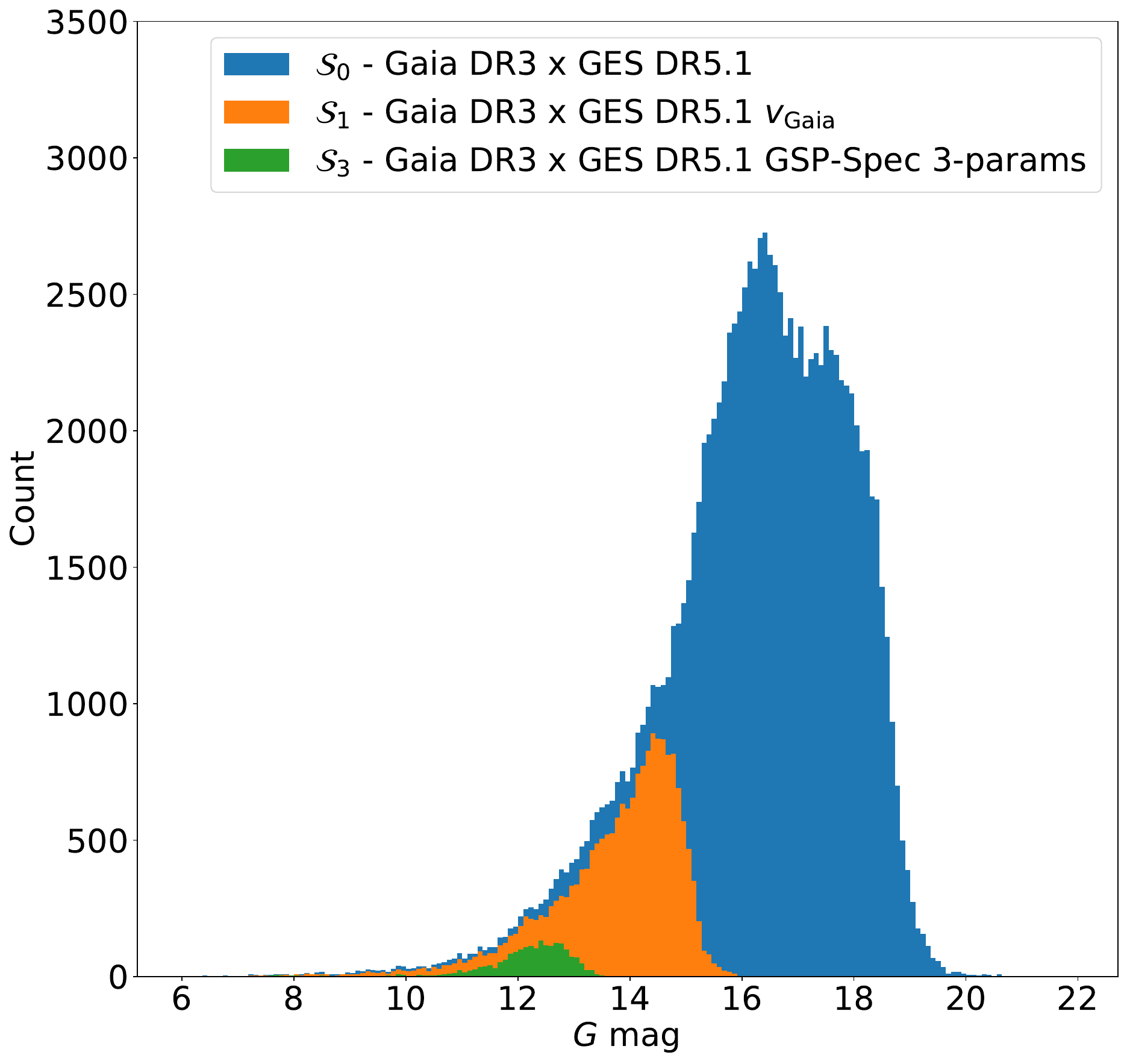}
  \caption{\label{Fig:histogram_magnitude} Distributions of the $G$ magnitudes of the \GESsh parent sample ($\mathcal{S}_{0}$; blue), of the \GESsh stars having a \Gaia DR3 radial velocity ($\mathcal{S}_{1}$; orange), and of the \GES stars having the three main \Gaia \gspspec atmospheric parameters ($\mathcal{S}_{3}$; green). The bin positions and width are identical for the three histograms; the bin width was adjusted using the Freedman-Diaconis rule.}
\end{figure}

In Figure~\ref{Fig:histogram_magnitude}, we show the distribution of $G$ magnitudes of the {whole \Gaia -- \GESlg intersection} ($\mathcal{S}_{0}$), of the \GES stars having a \Gaia DR3 radial velocity ($\mathcal{S}_{1}$), and of the \GES stars having the three main \Gaia \gspspec atmospheric parameters ($\mathcal{S}_{3}$). The mode for the parent sample $\mathcal{S}_{0}$ is located around $G = 16$; the mode of the subsample $\mathcal{S}_{1}$ is around $G = 14.5$; the mode of the subsample $\mathcal{S}_{3}$ is around $G = 12.5$. The faintest star in $\mathcal{S}_{1}$ has a $G$ magnitude of \SI{16.2}{\mag}, while the faintest star in $\mathcal{S}_{3}$ has a $G$ magnitude of \SI{13.9}{\mag}. Thus, Figure~\ref{Fig:histogram_magnitude} illustrates the fact that a vast majority of the \GESlg targets are much fainter than the range of magnitudes where \Gaia performs best: this is a feature of the \GESlg survey to complement the \Gaia spectroscopy with a good amount of objects fainter than $G \approx 15$.

\begin{figure}
  \centering
  \includegraphics[width=\columnwidth,clip]{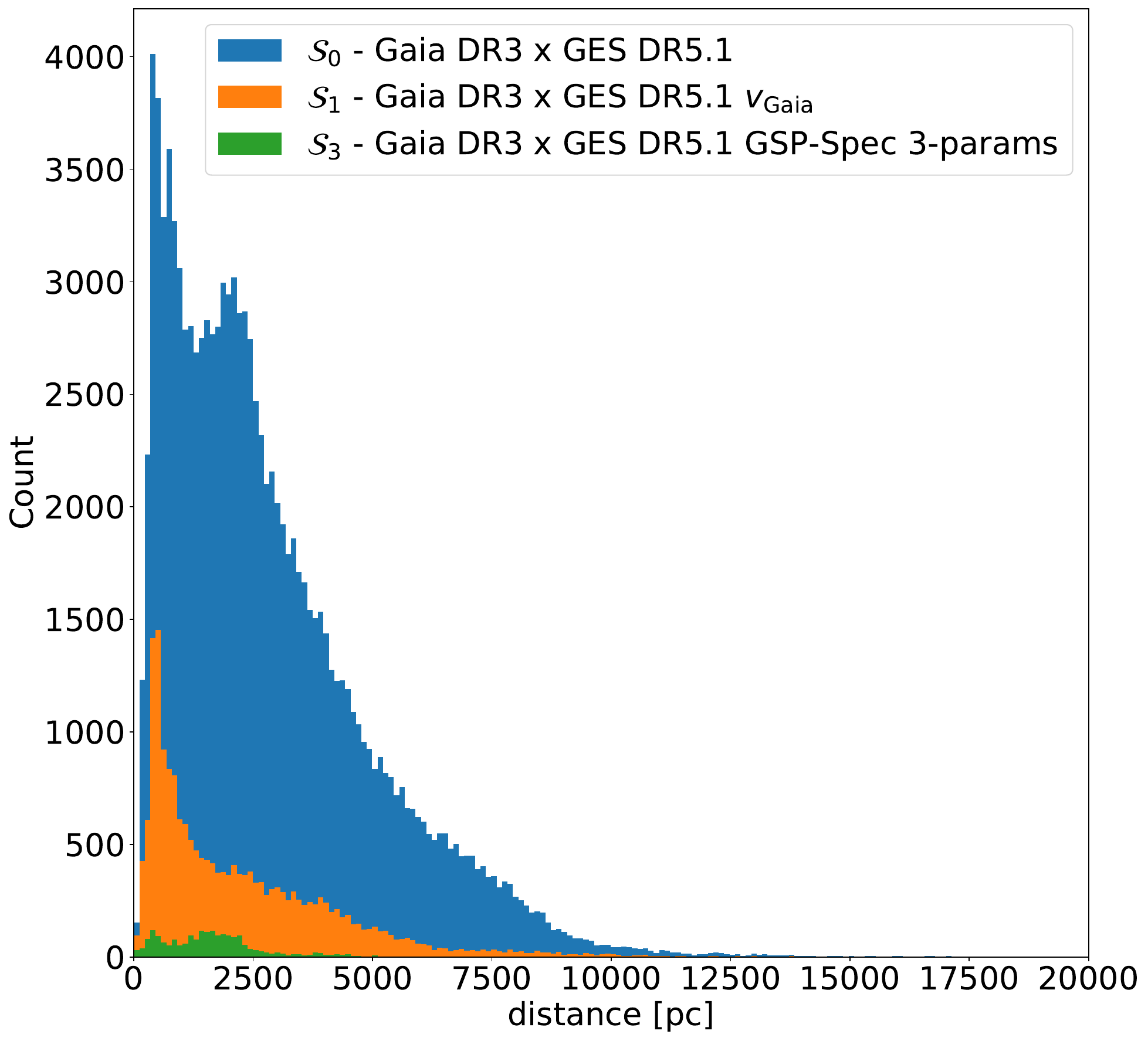}
  \caption{\label{Fig:histogram_distance} Distributions of the \Gaia distances (\revun{Bayesian distances from \citealp{2021AJ....161..147B}}) of the  of the \GESsh parent sample ($\mathcal{S}_{0}$; blue), of the stars having a \Gaia DR3 radial velocity ($\mathcal{S}_{1}$; orange), and of the stars having the three main \Gaia \gspspec atmospheric parameters ($\mathcal{S}_{3}$; green). The bin positions and width are identical for the three histograms; the bin width was adjusted using the Freedman-Diaconis rule.}
\end{figure}

\begin{figure}
  \centering
  \includegraphics[width=\columnwidth,clip]{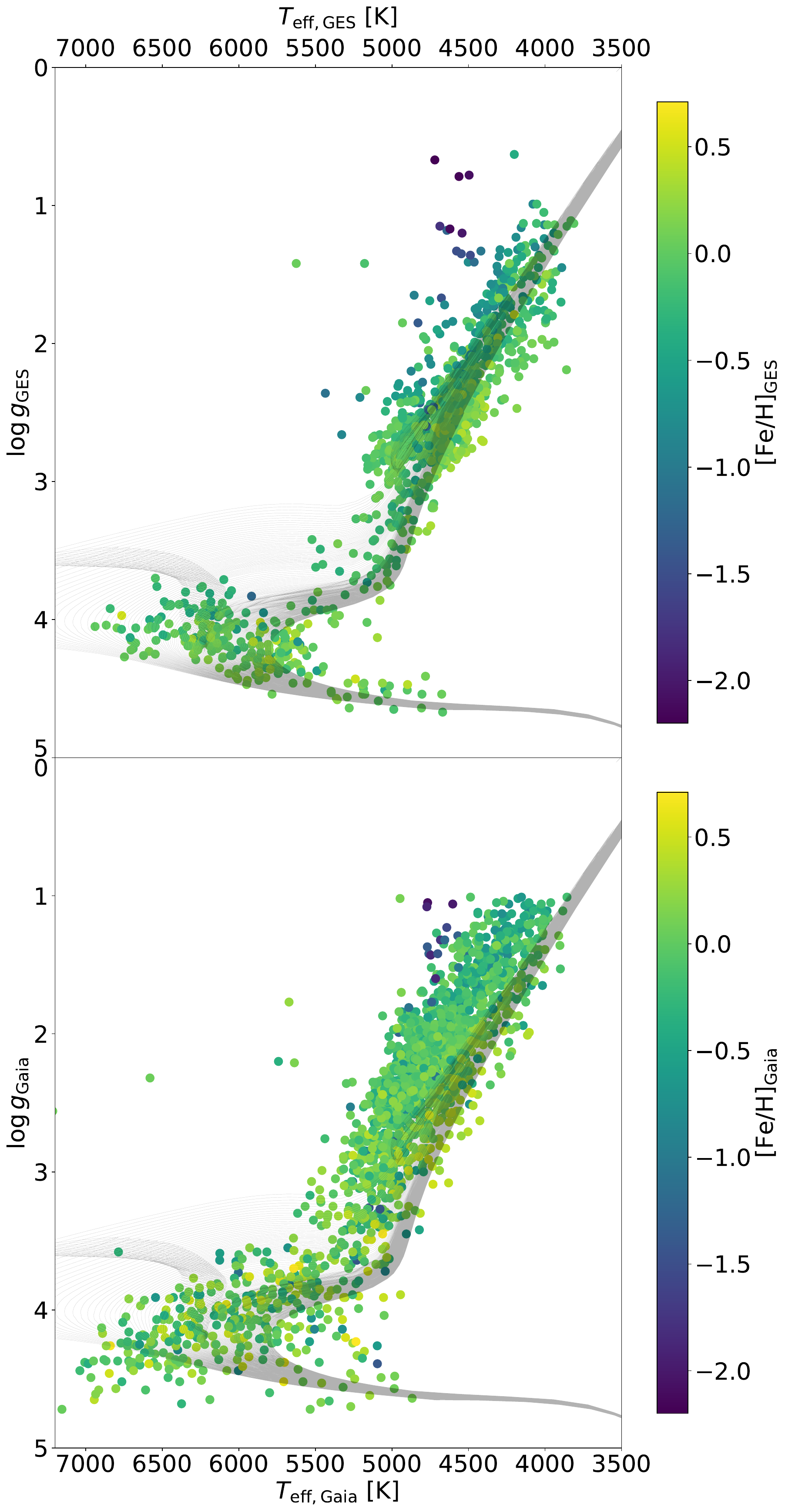}
  \caption{\label{Fig:Kiel_gspspec} Kiel diagram of the \num{2079} stars in the subsample $\mathcal{S}_{3}$, colour-coded by metallicity. A grid of Parsec isochrones \citep{bressan12} with solar metallicity and ages ranging from \num{0.1} to \SI{14}{\giga\Year} is superimposed. Top panel: based on the \GESlg atmospheric parameters; bottom panel: based on the uncalibrated \Gaia \gspspec atmospheric parameters.}
\end{figure}

\begin{figure}
  \centering
  \includegraphics[width=\columnwidth,clip]{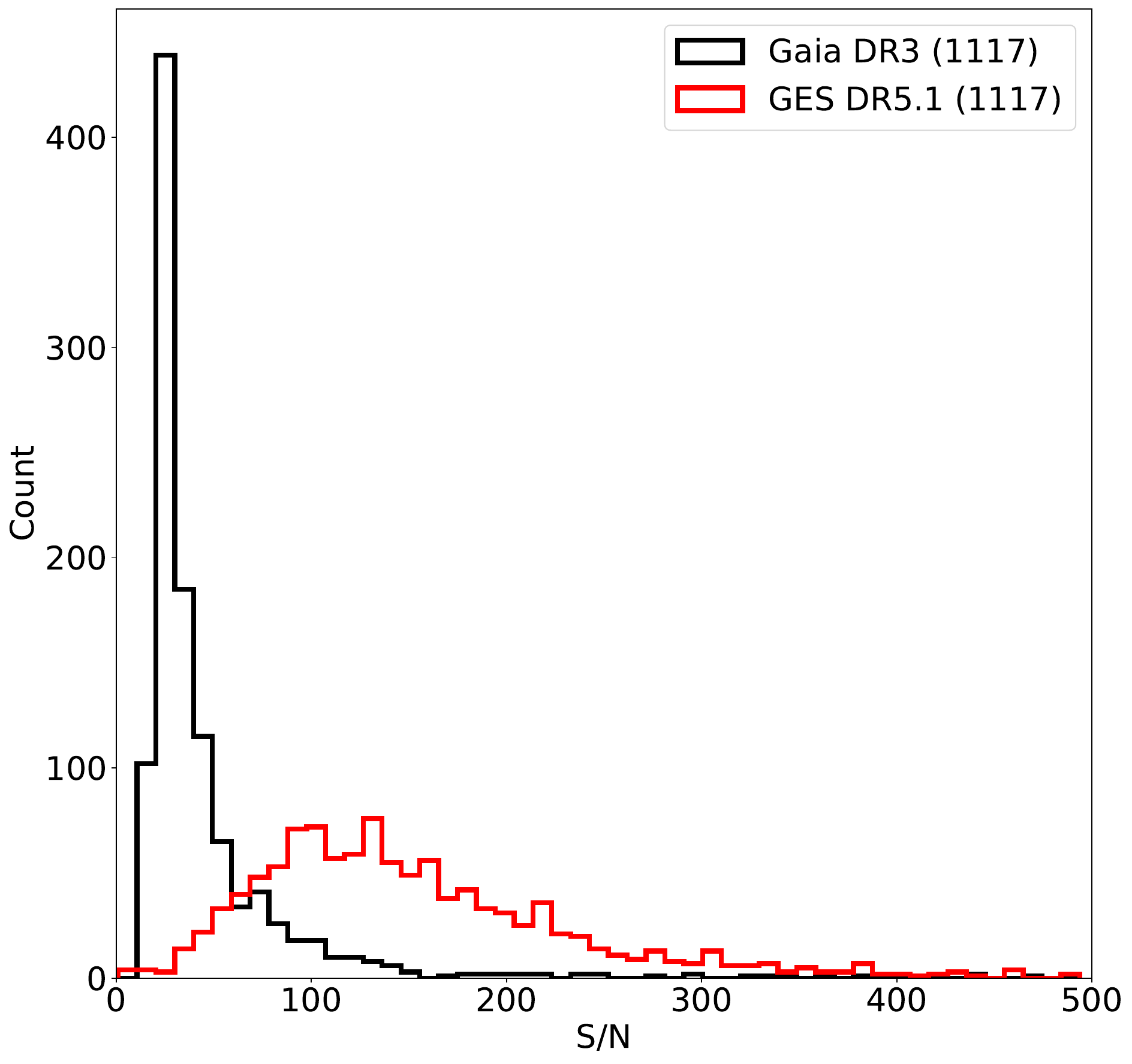}
  \caption{\label{Fig:histogram_snr} \revun{Distributions of the \GESsh (red) and \Gaia RVS (black) \snr for \num{1117} stars having both values.}}
\end{figure}

The distributions of \revun{Bayesian} distances from the Sun \citep{2021AJ....161..147B}, displayed in Fig.~\ref{Fig:histogram_distance}, show that the parent sample and subsamples probe different regions in the Galaxy: while the \GESlg parent sample reaches distances up to \revun{\SI{13}{\kilo\parsec}, \SI{75}{\percent} of $\mathcal{S}_{1}$ have a distance less than \SI{3.3}{\kilo\parsec} and \SI{75}{\percent} of $\mathcal{S}_{3}$ are located at a distance less than \SI{2.14}{\kilo\parsec}} from the Sun. Figure~\ref{Fig:Kiel_gspspec} shows the locus of the \num{2079} stars of the subsample $\mathcal{S}_{3}$ in the Kiel diagram using the \GESsh recommended atmospheric parameters (top panel) and the uncalibrated \Gaia \gspspec parameters (bottom panel): the sample stars are found from the low main-sequence (MS) to the upper red-giant-branch (RGB). The colour codes for the metallicity, ranging from about $\abratio{Fe}{H} \sim -2$ to \num{0.5}. We note already that the position of the stars in the $\left( T_{\mathrm{eff}}, \log g \right)$ plane changes with the origin of the atmospheric parameters. In particular, the main-sequence is less populated when we use the \gspspec parameters instead of the \GESsh ones. \revun{Finally, Figure~\ref{Fig:histogram_snr} shows the distribution of the \GESsh and \Gaia RVS \snr for \num{1117} stars (\texttt{rv\_expected\_sig\_to\_noise} is not systematically published in \Gaia DR3). The distribution of the \Gaia RVS \snr is skewed towards $\snr \le 50$ due to the fact that the targets under study are mainly stars fainter than $G \approx 11$. In comparison, \GESsh observations benefit from higher \snr: the mode is around \num{100}.}

\begin{figure}
  \centering
  \includegraphics[width=\columnwidth,clip]{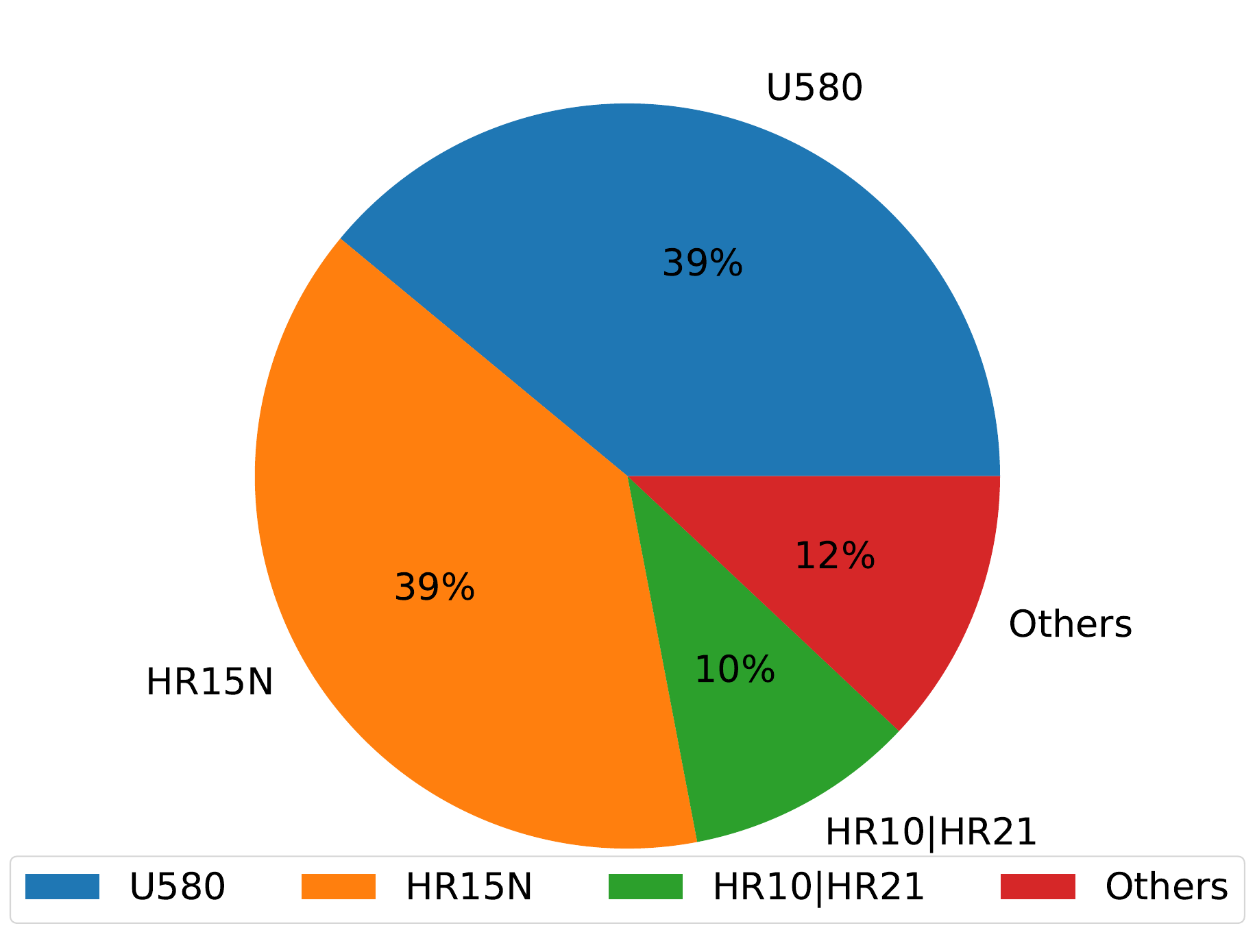}
  \caption{\label{Fig:pie_chart_ges_setup} Pie chart of the distribution of the \num{2079} stars of the subsample $\mathcal{S}_{3}$ according to the \GESsh setup used to derive the \GESsh recommended atmospheric parameters.}
\end{figure}

\begin{figure}
  \centering
  \includegraphics[width=\columnwidth,clip]{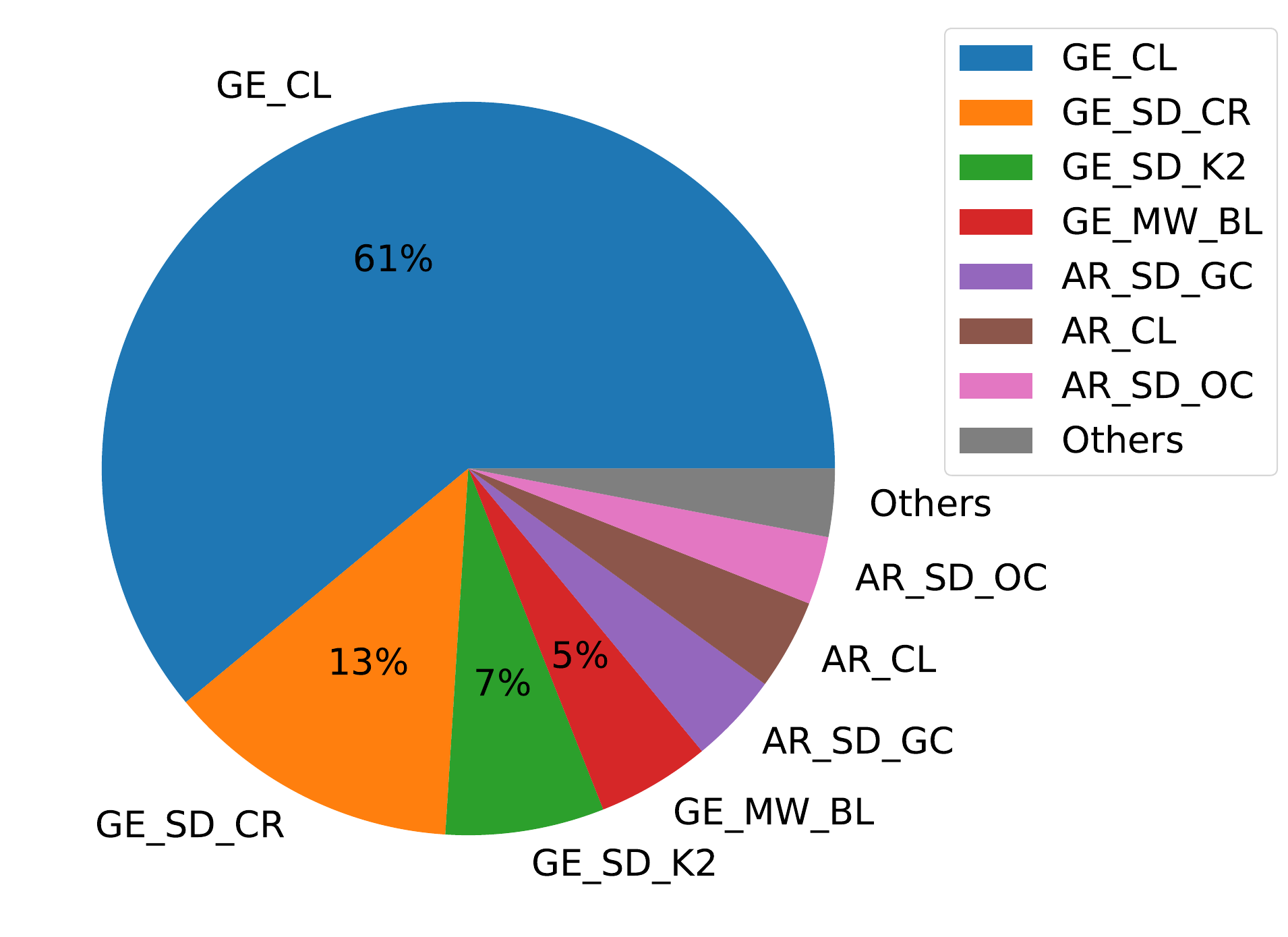}
  \caption{\label{Fig:pie_chart_ges_type} Pie chart of the distribution of the \num{2079} stars of the subsample $\mathcal{S}_{3}$ according to their \GESsh field type.}
\end{figure}

The pie chart of Fig.~\ref{Fig:pie_chart_ges_setup} shows the proportions of the different setups used by the \GESlg to derive (some of) the atmospheric parameters and abundances of the the \num{2079} stars in the subsample $\mathcal{S}_{3}$, while the pie chart of Fig.~\ref{Fig:pie_chart_ges_type} shows how are distributed these \num{2079} with respect to their \GESsh types (\texttt{GES\_TYPE}; classification system of the \GESsh targets). Most of the stars from $\mathcal{S}_{3}$ are observed with the UVES setup U580 (high resolution at $R \sim \num{47000}$), the GIRAFFE setup HR15N (medium resolution at $R \sim \num{19200}$) or the GIRAFFE setups HR10+HR21 (HR10: $R \sim \num{21500}$; HR21: $R \sim \num{18000}$). The remaining \SI{12}{\percent} are observed with less used UVES and GIRAFFE setups (U520, HR14A, HR3, HR4, HR5A, HR6, HR9B). Figure~\ref{Fig:pie_chart_ges_type} shows that about \SI{68}{\percent} of the \num{2079} stars in $\mathcal{S}_{3}$ are open cluster stars: \SI{61}{\percent} being newly observed by \GESlg (\texttt{GE\_CL}) and \SI{7}{\percent} being archival ESO data (\texttt{AR\_CL} and \texttt{AR\_SD\_OC}). One fifth of the stars in $\mathcal{S}_{3}$ are located in asteroseismic fields: \SI{13}{\percent} in CoRoT (\texttt{GE\_SD\_CR}) and \SI{7}{\percent} in K2 (\texttt{GE\_SD\_K2}). Finally, \SI{5}{\percent} of the sample are located towards the Galactic Bulge (\texttt{GE\_MW\_BL}). The remaining $\sim \SI{7}{\percent}$ of the subsample $\mathcal{S}_{3}$ comprise stars observed in the Milky Way fields, in globular clusters or they are benchmark stars.

\section{Radial velocities and detection of non-single stars}
\label{Sec:radial_velocities}

\subsection{Comparison between \Gaia and \GESsh radial velocities}

In this section, we compare the radial velocities of the stars in the sample $\mathcal{S}_{1}$. We discard the following stars: a) those flagged as SB$n \ge 1$ in \GESlg and as non-single star in \Gaia since their radial velocities are likely time-dependent; b) those having the \GESsh simplified flags \texttt{SRP} (data-reduction problems) or \texttt{SRV} (suspicious radial velocities) or \texttt{EML} (emission lines) since it may indicate a less precise or less accurate radial velocity; c) those having $\texttt{RUWE} \ge 1.4$ since it may indicate a suspicious \Gaia astrometric solution; d) those having the \Gaia \texttt{phot\_variable\_flag} set to \texttt{True}. After this cleaning, the sample $\mathcal{S}_{1}$ is downsized to \num{14692} objects.

\begin{figure}
  \centering
  \includegraphics[width=0.9\columnwidth,clip]{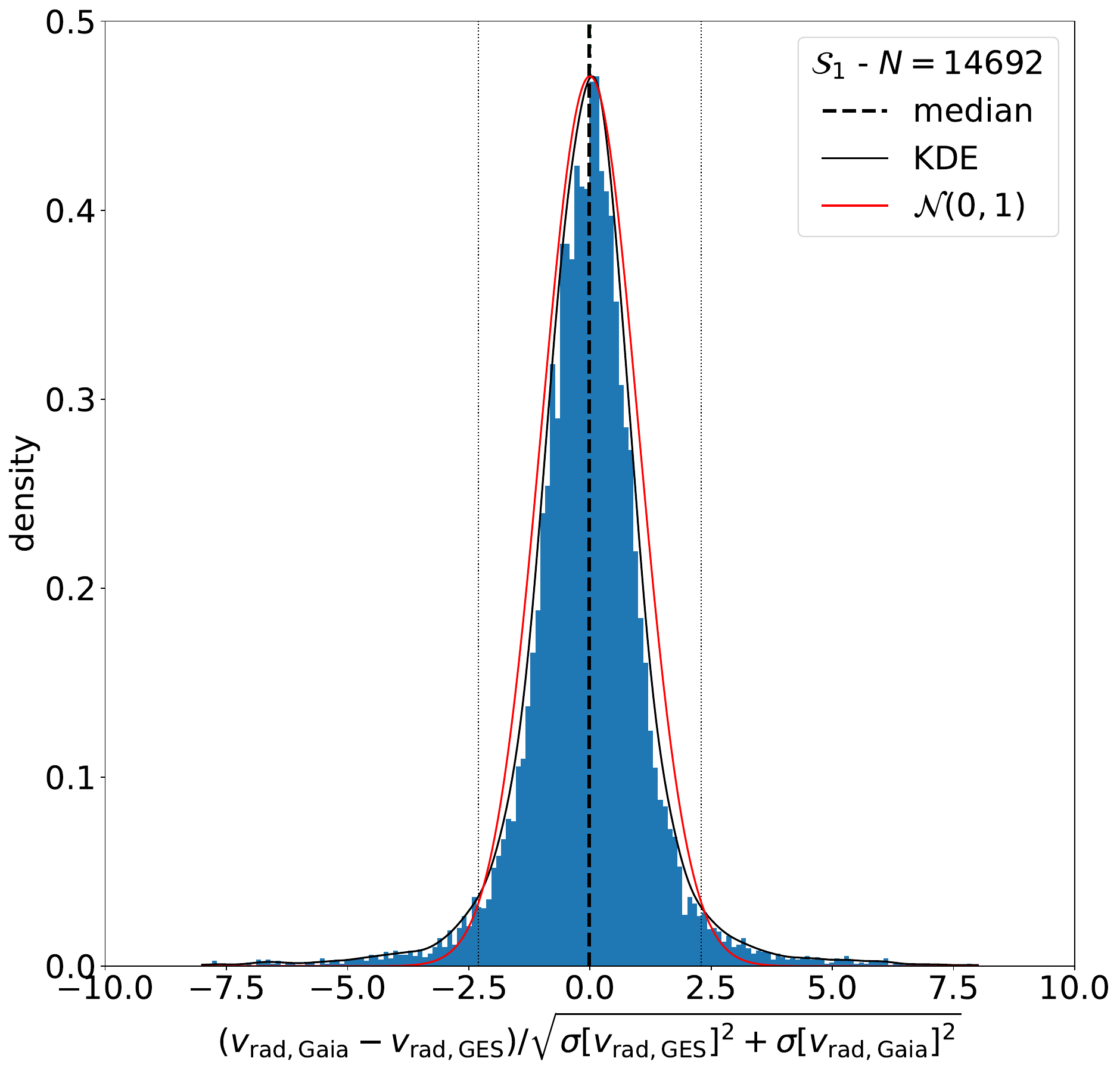}
  \caption{\label{Fig:PDF_difference_normalised_velocities} Probability distribution of the normalised velocity differences. The blue histogram (bin width = \num{0.1}) displays the distribution $\Delta_{\mathrm{norm}} v_{\mathrm{rad}}$ of the difference of the radial velocity differences $v_{\mathrm{rad,Gaia}} - v_{\mathrm{rad,GES}}$ normalised by the propagated errors $\sqrt{\sigma[v_{\mathrm{rad,Gaia}}]^2 + \sigma[v_{\mathrm{rad,GES}}]^2}$. The black line is the empirical KDE obtained from the sample distribution. The red line is the probability distribution function (PDF) of a the normal law centred in 0 and of unit variance. The dashed vertical black line indicates the mean of the distribution. The black dotted lines have equation $| \Delta_{\mathrm{norm}} v_{\mathrm{rad}} | = 2.3$ and show where the tails of the empirical distribution become heavier than those of the normal law.}
\end{figure}

Figure~\ref{Fig:PDF_difference_normalised_velocities} shows the distribution of the radial velocity differences $v_{\mathrm{rad,Gaia}} - v_{\mathrm{rad,GES}}$ normalised by the propagated errors $\sqrt{\sigma[v_{\mathrm{rad,Gaia}}]^2 + \sigma[v_{\mathrm{rad,GES}}]^2}$ (sample histogram and sample KDE) and for reference, it also displays the normal law $\mathcal{N}\left(0, 1\right)$. If we assume that for a given star, each \Gaia (\GESsh, respectively) radial velocity is a random variable distributed along a normal law $\mathcal{N}\left(v_{\mathrm{rad}}, \sigma[v_{\mathrm{rad}}]^2\right)$ where $v_{\mathrm{rad}}$ is the true radial velocity of the star and $\sigma[v_{\mathrm{rad}}]$ describes the instrumental error, then $v_{\mathrm{rad,Gaia}} - v_{\mathrm{rad,GES}}$ should follow a probability distribution $\mathcal{N}\left(0, \sigma[v_{\mathrm{rad,Gaia}}]^2 + \sigma[v_{\mathrm{rad,GES}}]^2\right)$, or equivalently, the normalised differences $\Delta_{\mathrm{norm}} v_{\mathrm{rad}} = \left( v_{\mathrm{rad,Gaia}} - v_{\mathrm{rad,GES}} \right) / \sqrt{\sigma[v_{\mathrm{rad,Gaia}}]^2 + \sigma[v_{\mathrm{rad,GES}}]^2}$ should follow a probability distribution $\mathcal{N}\left(0, 1\right)$.

The mean of $\Delta_{\mathrm{norm}} v_{\mathrm{rad}}$ is \num{-0.03} and its standard deviation is \num{1.68}. Fig~\ref{Fig:PDF_difference_normalised_velocities} shows that the sample probability distribution deviates marginally from the normal law: while the core of the sample distribution follows closely the normal law, we note that the tails of the sample distribution for $\left| \Delta_{\mathrm{norm}} V\right| \gtrapprox 2.3$ (the approximate abscissa where the left and right sample tails are above the normal law) are heavier than those of the normal law. The sample left and right tails are populated by \num{999} ($\approx \SI{6.8}{\percent}$) objects and about \num{580} objects are likely in excess compared to the normal law. For those \num{580} objects ($\approx \SI{4}{\percent}$ of the sample), the random uncertainty reported by the two experiment is not able to explain the radial velocity differences between the \Gaia and \GESsh datasets. This could be due to an incorrect estimate of the random uncertainty \revun{(\eg, see \citealp{2015A&A...580A..75J} for a discussion on the non-Gaussianity of the random uncertainty in \GESsh)} or a biased estimate of $v_{\mathrm{rad}}$. It could also be due to a still unidentified astrophysical variation (\eg stellar multiplicity, jitter, pulsations) of the radial velocity.

The statistical difference between the two sets of \num{14692} radial velocities can also be evaluated with a two-sample K-S test. Computed with the {\sc SciPy} module, for $v_{\mathrm{rad,GES}}$ and $v_{\mathrm{rad,Gaia}}$, the test returns a statistics $D = 0.004$ and a $p$-value of \num{0.999} under the null hypothesis $H_0$: "the two samples are drawn from the same unknown distribution". Therefore, we fail at rejecting the null hypothesis at the confidence level $\alpha = 0.05$: in other words, there is no strong evidence that the two datasets are statistically different. This result is in agreement with the discussion of the previous paragraph.

We remind the reader that \GESlg has observed their targets with a various choice of FLAMES/UVES and FLAMES/GIRAFFE setups, meaning that two given stars are not necessarily observed with the same setup, and so they are not observed at the same wavelengths and same resolution. During the homogenisation phase, \GESlg has selected the single setup or the combination of setups that will be used to publish the final (average) radial velocity of a given star. This choice can be traced back using the column \texttt{ORIGIN\_VRAD} of the \GESsh catalogue. The setup HR10 was used as a reference setup by \GESlg and velocity offsets have been computed and applied to put the radial velocities measured from other setups onto the HR10 radial velocity scale. A consequence of this observational strategy is that the agreement between the \GESsh and \Gaia radial velocities may vary between setups (\eg quality of the wavelength calibration of a given setup, efficiency of the cross-correlation technique depending on the absorption-line content of a given setup). Table~\ref{Table:Radial_velocity_differences_per_GES_setup} lists the mean, median and standard deviation of $\Delta_{\mathrm{norm}} v_{\mathrm{rad}} = \left(v_{\mathrm{rad,Gaia}} - v_{\mathrm{rad,GES}}\right) / \sqrt{\sigma[v_{\mathrm{rad,Gaia}}]^2 + \sigma[v_{\mathrm{rad,GES}}]^2}$ and of $\Delta v_{\mathrm{rad}}= v_{\mathrm{rad,Gaia}} - v_{\mathrm{rad,GES}}$ when only one \GESsh-setup combination is used for the \Gaia vs. \GESsh comparison. We note that indeed, the agreement between the \GESsh and \Gaia radial velocity scales depends on the \GESsh setup. An excellent agreement is obtained when the \GESsh radial velocity derives from observations with HR10, HR15N, HR3, U580: the mean of $\Delta_{\mathrm{norm}} v_{\mathrm{rad}}$ is below \num{0.1} in absolute value and its standard deviation is below \num{2}. For three setups (HR9B, HR14A, U520) used for warm stars, the standard deviation of $\Delta_{\mathrm{norm}} v_{\mathrm{rad}}$ is larger than \num{3} and the mean of $\Delta_{\mathrm{norm}} v_{\mathrm{rad}}$ is larger than $\approx \num{0.2}$ in absolute value. This larger bias and larger scatter of $\Delta_{\mathrm{norm}} v_{\mathrm{rad}}$ for these three setups are not correlated with the mean $G$ magnitude nor the mean $T_{\mathrm{eff,GES}}$ of the stars nor the mean \snr of \GESsh spectra. According to the last column `STD' of Table~3 in \citet{2023A&A...676A.129H}, the radial velocity homogenisation was less precise for HR9B, HR14A and U520 than for HR15N, HR21, U580 but it was not worse than for HR3 and WG13 combination.

\begin{table*}
  \centering
  \caption{\label{Table:Radial_velocity_differences_per_GES_setup} Mean, median and standard deviation of $\Delta_{\mathrm{norm}} v_{\mathrm{rad}} = \left(v_{\mathrm{rad,Gaia}} - v_{\mathrm{rad,GES}}\right) / \sqrt{\sigma[v_{\mathrm{rad,Gaia}}]^2 + \sigma[v_{\mathrm{rad,GES}}]^2}$ and $\Delta v_{\mathrm{rad}} = v_{\mathrm{rad,Gaia}} - v_{\mathrm{rad,GES}}$ when only one \GESsh-setup combination is used for the comparison. The first column gives the name of the setup, the second column gives the number of corresponding objects in the selection, the third to eighth columns give the aforementioned statistics.} 
  \begin{tabular}{lS[table-format=5.0]|S[table-format=5.2]S[table-format=5.2]S[table-format=5.2]|S[table-format=5.2]S[table-format=5.2]S[table-format=5.2]}
    \toprule
    \GESsh-setup combination & {\#} & \multicolumn{3}{c}{$\left(v_{\mathrm{rad,Gaia}} - v_{\mathrm{rad,GES}}\right) / \sqrt{\sigma[v_{\mathrm{rad,Gaia}}]^2 + \sigma[v_{\mathrm{rad,GES}}]^2}$} & \multicolumn{3}{c}{$v_{\mathrm{rad,Gaia}} - v_{\mathrm{rad,GES}}$}\\
    & & {mean} & {median} & {s.d.} & {mean} & {median} & {s.d.}\\
    \midrule
    CASU|HR15N                                    &  5630 &  -0.02 &  -0.02 &   1.74 &   0.19 &  -0.04 &   8.96\\
    CASU|HR10                                     &  3936 &  -0.01 &  -0.01 &   1.21 &   0.22 &  -0.02 &   5.67\\
    Arcetri|U580                                  &  3800 &  -0.00 &   0.02 &   1.39 &   0.03 &   0.03 &   6.30\\
    CASU|HR9B                                     &   605 &  -0.37 &  -0.20 &   3.49 &  -0.03 &  -0.31 &  29.63\\
    CASU|HR21                                     &   417 &  -0.09 &  -0.07 &   2.37 &  -0.85 &  -0.26 &  11.44\\
    WG13|combination                              &   166 &  -0.20 &   0.00 &   1.00 &  -2.83 &   0.02 &  16.57\\
    Arcetri|U520                                  &    76 &  -0.18 &   0.17 &   3.67 &  -0.51 &   0.26 &  22.29\\
    CASU|HR3                                      &    50 &   0.03 &  -0.03 &   1.55 &  -2.48 &  -0.00 &  15.69\\
    CASU|HR14A                                    &    12 &   0.56 &  -0.04 &   3.35 &   5.83 &   0.60 &  25.56\\

\bottomrule
\end{tabular}
\end{table*}

\begin{figure*}
  \centering
  \includegraphics[width=\textwidth,clip]{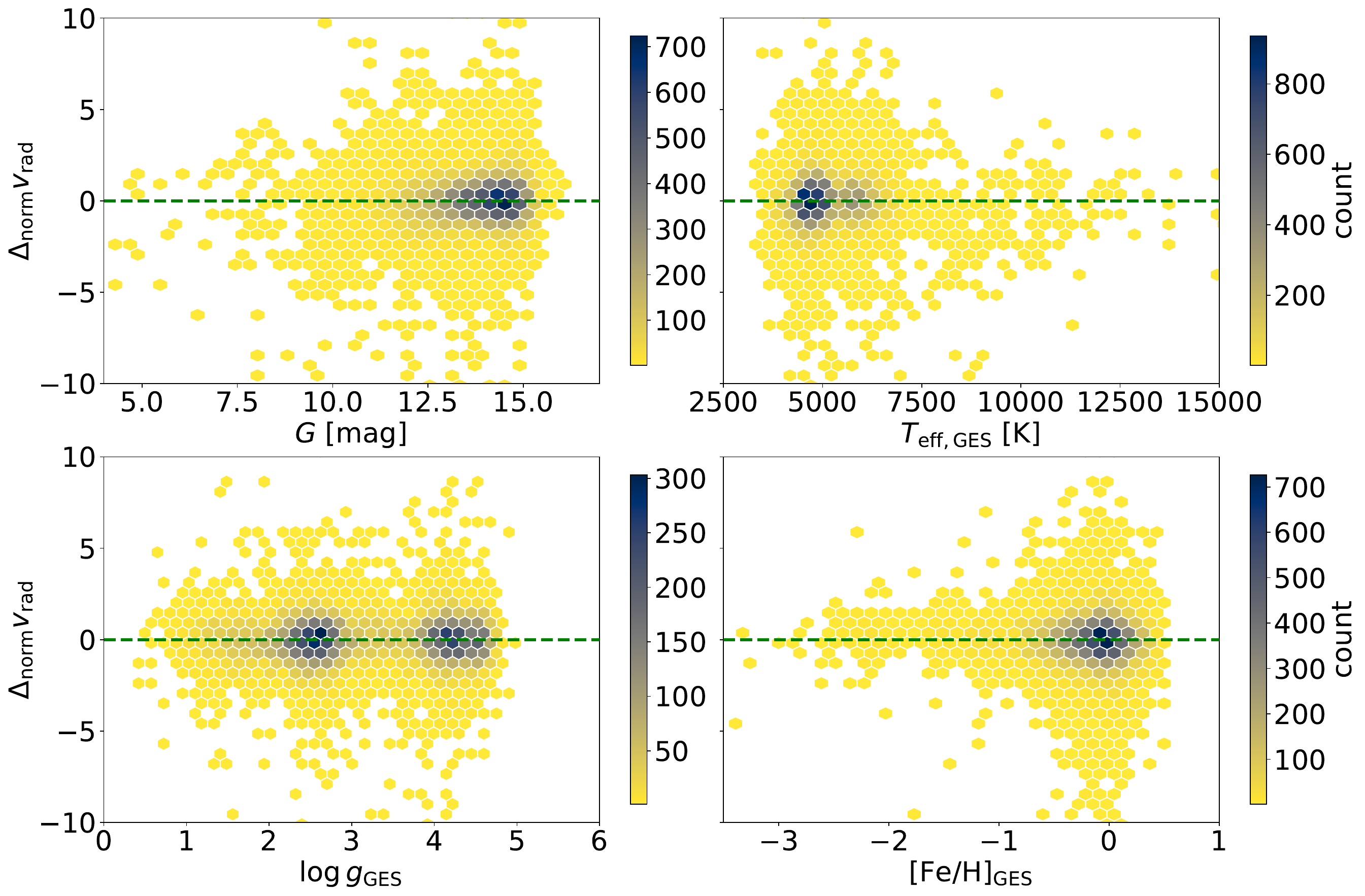}
  \caption{\label{Fig:Dependency_DeltaVrad_vs_four_parameters} From top to bottom, left to right: $\Delta_{\mathrm{norm}} v_{\mathrm{rad}}$ vs. $G$ magnitude, $T_{\mathrm{eff,GES}}$, $\log g_{\mathrm{GES}}$ and $\abratio{Fe}{H}_{\mathrm{GES}}$. The vertical axis is the same for the four panels. The colour scale changes from one plot to another.}
\end{figure*}

Figure~\ref{Fig:Dependency_DeltaVrad_vs_four_parameters} shows the dependency of $\Delta_{\mathrm{norm}} v_{\mathrm{rad}}$ with $G$ magnitude, $T_{\mathrm{eff,GES}}$, $\log g_{\mathrm{GES}}$ and $\abratio{Fe}{H}_{\mathrm{GES}}$. We note that the distributions are rather symmetrical around $\Delta_{\mathrm{norm}} v_{\mathrm{rad}} = 0$: the difference between the \GESsh and \Gaia radial velocities is not correlated with any of these four parameters. \revun{As a last remark, we note that \cite{2023A&A...674A...5K} recommend a correction of the \Gaia radial velocity in the form of calibration depending on $G_{\mathrm{RVS}}$. If we apply it, the correction is never larger than \SI{0.4}{\kilo\metre\per\second}, the sample is downsized to \num{14173} objects because of some unavailable $G_{\mathrm{RVS}}$ estimates; the mean of $\Delta_{\mathrm{norm}} v_{\mathrm{rad}}$ becomes \num{-0.09} and its standard deviation remains unchanged at \num{1.67}. The rest of the discussion remains true. We also note that \citet{2023A&A...674A..32B} show that the uncertainty on the \Gaia radial velocity is underestimated. They publish a calibration as a function of $G_{\mathrm{RVS}}$ to estimate a correcting factor. There calibration seems to be defined only on the $G_{\mathrm{RVS}}$ range $[8, 14$; we cannot compute the correcting factor for a small fraction of our selection brighter than $G_{\mathrm{RVS}} = 8$. Taking into account this correction does not change the above discussion. Figure~\ref{Fig:PDF_difference_normalised_velocities_corr_unc} is the same as Fig~\ref{Fig:PDF_difference_normalised_velocities} but it uses the corrected uncertainties for \Gaia radial velocities instead of the raw ones.}

In conclusion, after discarding objects with suspicious or variable radial velocities from the sample $\mathcal{S}_{1}$, we find an excellent agreement between the \GESsh and \Gaia radial velocity scales, given their respective uncertainties. The mean and median difference between the two datasets are respectively \num{0.07} and \SI{-0.02}{\kilo\meter\per\second}. We cannot explain the disagreement for about \SI{4}{\percent} of the analysed stars.

\subsection{Binarity}

The \GES is not designed to discover and monitor the variations of the radial velocities of a star with time. Nonetheless, thanks to the repeated observations needed to achieve a \snr sufficient for determining abundances, and thanks to the good resolving power of the FLAMES/GIRAFFE and FLAMES/UVES multi-object spectrographs, it is still possible to identify spectroscopic binaries with one visible component (SB1), discovered by looking for unaccountably scattered radial velocity series, and spectroscopic binaries with two or more visible components (SB$n \ge 2$), discovered by finding multi-peaked cross-correlation functions (CCFs).

A final census of the \GESsh SB1 and SB$n \ge 2$, based on the analysis of the final data release, is still under preparation (Van der Swaelmen et al., \emph{in prep.}): our preliminary analysis of \GESsh DR5.1 give a total of \num{2117} SB$n$ with \num{1216} SB1, \num{878} SB2, \num{20} SB3 and three SB4. However, three publications have made use of the previous internal \GESsh data releases. \citet{2017A&A...608A..95M} have listed \num{342} SB2, eleven SB3 and one SB4 after analysing the whole \GESsh iDR4; \citet{2020A&A...635A.155M} have found \num{803} SB1 among the HR10 and HR21 observations released in \GESsh iDR5; finally, \citet{2023arXiv231204721V} have found \num{322} SB2 (four of which being also SB3 candidates), ten SB3 and two SB4 among the HR10 and HR21 observations of field stars released in \GESsh iDR5. \revdeux{Once combined, these three publications give a list of \num{1113} unique SB$n$}. \citet{2017A&A...608A..95M} used the CCFs computed by the \GESsh WG, while \citet{2020A&A...635A.155M} and \citet{2023arXiv231204721V} are using the Nacre CCFs described and computed in \citet{2023arXiv231204721V}. A future publication will exploit the strength of the Nacre CCFs to provide the complete census of \GESsh SB1 and SB$n \ge 2$ among the GIRAFFE (HR10, HR21 and HR15N) and UVES observations but in the mean time, it is still possible to check how the \GESsh SBs are flagged by \Gaia. We point the reader that a detailed comparison of \GESsh iDR5 and \Gaia DR3 in terms of binarity is given in \citet{2023arXiv231204721V}. 

\Gaia DR3 provides various ways to identify confirmed or suspected non-single stars. The most direct way is to look at the column \texttt{non\_single\_star} of the \Gaia main catalogue to find the confirmed stellar multiples. Due to stringent filters, the \Gaia DR3 multiple-star census is mostly populated by bright object ($G \le 13$) and therefore, one can anticipate a small intersection with the \GESsh multiple-star census. Using the published (resp., new preliminary) census, we found \num{8} \revdeux{out of \num{1113}} (resp., \num{22} \revdeux{out of 2117}) SB$n$ among \GESsh DR5.1 targets that are \revdeux{also} flagged as non-single stars by \Gaia DR3: two (resp., twelve) astrometric binaries, one (resp., two) eclipsing binaries, five (resp., eight) spectroscopic binaries and zero binaries confirmed by a combination of techniques. \num{161} (resp., \num{414}) \GESsh SB$n$ have a \Gaia radial velocity. The median uncertainty on the \Gaia radial velocity is slightly larger for the \GESsh SB$n$ than for the non-SB$n$: \SI{3.91}{\kilo\meter\per\second} vs. \SI{3.24}{\kilo\meter\per\second} (resp., \SI{4.26}{\kilo\meter\per\second} vs. \SI{3.22}{\kilo\meter\per\second}). \revun{In other words,} the uncertainty on the \Gaia radial velocity tends to be slightly larger for the population of \GESsh SB$n$ candidates. \revdeux{The fact that the \GESsh SB$n$ population has a larger median uncertainty on their \Gaia radial velocity may indicate that, in future \Gaia releases, these faint objects will also be seen as non-single stars by \Gaia.}

The quantity \texttt{RUWE} (Renormalised Unit Weight Error) can be used to identify objects for which the single-star model does not permit a good fit of the astrometric observations. The \Gaia documentation indicates that a \texttt{RUWE} larger than \num{1.4} should be treated as unusual and this may or may not point at a hidden stellar companion. \num{7440} objects of $\mathcal{S}_{0}$ have $\texttt{RUWE} \ge 1.4$ but only \num{66} (resp., \num{168}) are flagged has SB$n$ in the published (resp., preliminary) \GESsh census: we confirm that $\texttt{RUWE} \ge 1.4$ is not a necessary nor a sufficient condition to identify spectroscopic binaries.

The \Gaia `Astrophysical parameters' tables provide the community with a series of columns that are intended to help in tracking down potential binaries. \citet{2023arXiv231204721V} discuss the use of the columns \texttt{classprob\_dsc\_combmod\_binarystar} and \texttt{classprob\_dsc\_specmod\_binarystar} from the Discrete Source Classifier \citep[DSC;][]{2023A&A...674A..31D} and \texttt{flags\_msc} from the Multiple Star Classifier \citep[MSC;][]{2023A&A...674A..26C}. We find zero (resp., six and five for the preliminary final census) \GESsh SB2 with a probability \texttt{classprob\_dsc\_combmod\_binarystar} and \texttt{classprob\_dsc\_specmod\_binarystar} larger than 0.5. Among the preliminary census, there are six and two \GESsh SB2 with a probability \texttt{classprob\_dsc\_combmod\_binarystar} and \texttt{classprob\_dsc\_specmod\_binarystar} larger than 0.9. According to \citet{2023A&A...674A..31D}, only \SI{0.2}{\percent} of the unresolved binaries of their validation data-set are recovered (see their Table~3) by the the two DSC classifiers. Therefore, we do not expect more than a couple SB2 to be correctly flagged by DSC: our findings seem to be compatible with their prediction: \num{303} (resp., \num{748}) \GESsh SB2 have \texttt{flags\_msc} set to 0, which indicate that the inference of atmospheric parameters of each component cannot be rejected a priori.

\section{Rotational velocities}
\label{Sec:Rotational_velocities}

\begin{figure}
  \centering
  \includegraphics[width=\columnwidth,clip]{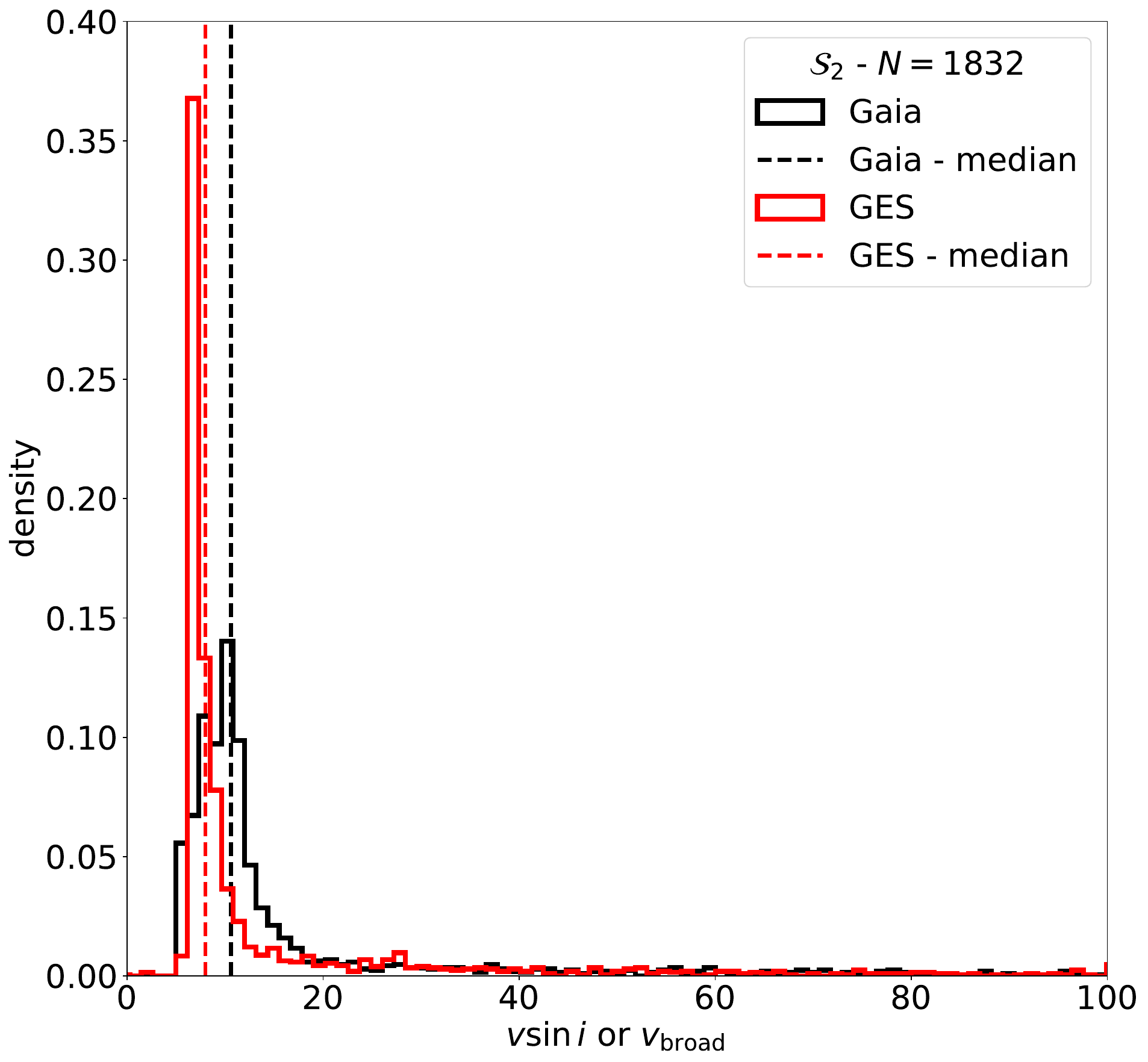}
  \caption{\label{Fig:Histogram_GES_vsini_and_Gaia_vbroad} Normalised histogram of $v \sin i_{\mathrm{GES}}$ (red) and the broadening parameter $v_{\mathrm{broad,Gaia}}$ (black).}
\end{figure}

\begin{figure}
  \centering
  \includegraphics[width=\columnwidth,clip]{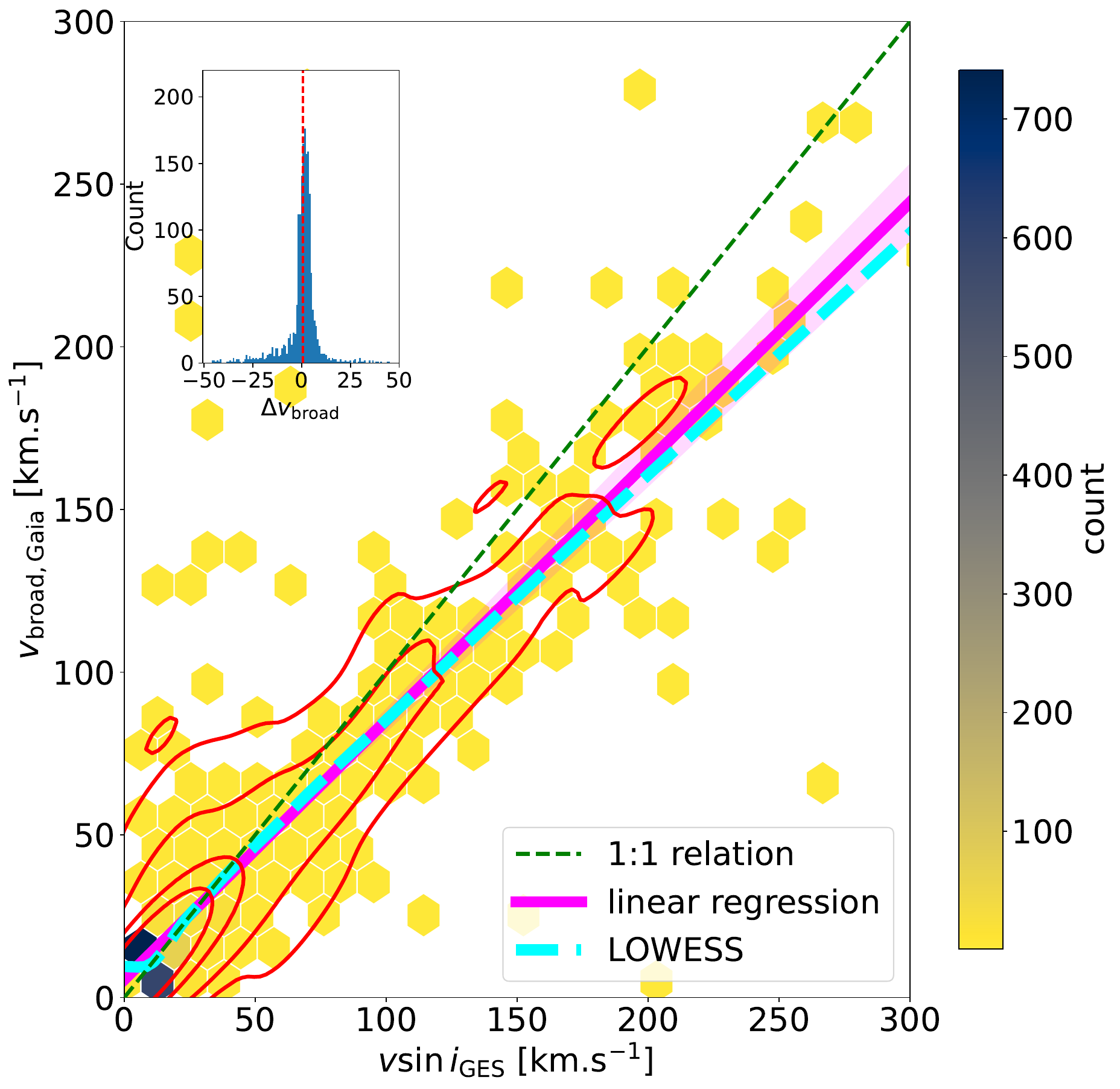}
  \caption{\label{Fig:Comparison_GES_vsini_vs_Gaia_vbroad} $v_{\mathrm{broad,Gaia}}$ vs $v \sin i_{\mathrm{GES}}$ (2D histogram). The green dashed line is the 1-to-1 relation. The pink thick line is the linear regression (see text for the parametrisation). The cyan thick dashed line is  \revdeux{a non-parametric LOWESS model (locally-weighted linear regression) implemented using the Python module StatsModels}. The inset shows the distribution of $\Delta v_{\mathrm{broad}} = v_{\mathrm{broad,Gaia}} - v \sin i_{\mathrm{GES}}$. The red dashed line in the inset indicates the location of the mean difference.}
\end{figure}

In this section, we compare the projected rotational velocity $v \sin i_{\mathrm{GES}}$ and the broadening parameter $v_{\mathrm{broad,Gaia}}$ \citep{2023A&A...674A...8F} for the \num{2094} stars of the sample $\mathcal{S}_{2}$. \GESlg provides no estimate of $v \sin i$ for \num{262} out of \num{2094} stars with a valid $v_{\mathrm{broad,Gaia}}$. The sample $\mathrm{S}_{2}$ is therefore downsized to \num{1832} objects. Figure~\ref{Fig:Histogram_GES_vsini_and_Gaia_vbroad} shows the (normalised) distributions of $v \sin i_{\mathrm{GES}}$ and $v_{\mathrm{broad,Gaia}}$. We note that the two distributions are quite different: the mode and median of the distribution of \GES $v \sin i$ are about \SI{8}{\kilo\metre\per\second}, while the mode and median of the distribution of $v_{\mathrm{broad,Gaia}}$ are around \SI{11}{\kilo\metre\per\second}. \revun{It is expected since $v \sin i_{\mathrm{GES}}$ and $v_{\mathrm{broad,Gaia}}$ do not measure the same quantity. \revdeux{The \GESlg quantity named $v \sin i$ in the final public release comes from one of three possible sources according to \citet{2023A&A...676A.129H}: the global fitting code ROTFIT by the OACT node for GIRAFFE spectra \citep[\eg][]{ 2015A&A...575A...4F}, the CCF width -- $v \sin i$ calibration by the Arcetri node for UVES spectra \citep{2014A&A...565A.113S} and one estimate or a combination of estimates provided by the WG13 for stars in young clusters. All the used techniques account for the GIRAFFE or UVES instrumental resolution such that $v \sin i_{\mathrm{GES}}$ takes into account the spectral broadening due to the stellar rotation and the macroturbulence.} On the other hand, for \Gaia RVS, $v_{\mathrm{broad,Gaia}}$ measures any source of broadening: instrumental, rotation, turbulence.} We therefore do not expect a strong correlation between these two parameters: indeed, the linear regression shown in Fig.~\ref{Fig:Comparison_GES_vsini_vs_Gaia_vbroad} gives a slope of $\num{0.796} \pm \num{0.011}$, a $y$-intercept of $\num{5.559} \pm \num{0.571}$ and the coefficient $r^2 = \num{0.72495}$ indicates a loose correlation. This selection is essentially FGK dwarf and giant stars with a $G$ magnitude in the range $[10, 14]$: such stars have in general a small rotational velocity that causes a line-broadening smaller or comparable to the one caused by the instrumental resolution. In their Table~3, \cite{2023A&A...674A...8F} give a $v_{\mathrm{broad,Gaia}}$ range as a function of $G$ and $T_{\mathrm{eff}}$ where $v_{\mathrm{broad,Gaia}}$ has a probability higher than \SI{90}{\percent} to be within $2\sigma$ of $v \sin i$. For FGK stars ($T_{\mathrm{eff}} = \SI{4000}{\kelvin}$ or \SI{5500}{\kelvin} or \SI{7500}{\kelvin} in their table) and for $G \ge 10$, the validity range is for $v_{\mathrm{broad,Gaia}} \gtrapprox \SI{12}{\kilo\meter\per\second}$ (conservative value). If we restrict the sample to stars fainter than $G = 10$ and with $v_{\mathrm{broad,Gaia}} \gtrapprox \SI{12}{\kilo\meter\per\second}$, the correlation coefficient $r^2$ marginally increases and remains below \num{0.8}, still indicating a loose correlation. \revun{In other words, $v_{\mathrm{broad,Gaia}}$ is not a reliable proxy for $v \sin i_{\mathrm{GES}}$, especially for slow rotators.} Our results are compatible with the results obtained by \cite{2023A&A...674A...8F} when they compare the \Gaia broadening parameter to the rotational velocities published by APOGEE and GALAH (and also, by RAVE and LAMOST, but the resolution of these two surveys is much lower than those of \GESlg).

\section{Temperatures, gravities and abundances}
\label{Sec:comparison_stars_in_common}

In this section, we compare the spectroscopic parameters of the stars in common between the two surveys for the \num{2079} stars of the sample $\mathcal{S}_{3}$. As stated in Sec.~\ref{Sec:gaia_ges_intersection}, the initial selection $\mathcal{S}_{3}$ of \num{2079} stars is obtained by requesting the availability of the three \Gaia \gspspec parameters $\left\{T_{\mathrm{eff}}, \log g, \abratio{Fe}{H}\right\}$ for a star of the parent sample. Then, we check in the \GESsh DR5.1 catalogue if some stellar parameters and some abundances have been derived by the \GESsh consortium. The result of such counting is summarised in the Table~\ref{Table:Intersections_summary}. We note that some of the stars of $\mathcal{S}_{3}$ are missing the corresponding \GESsh stellar parameters. Thus, \num{1939} stars have both \Gaia \gspspec and \GESsh $T_{\mathrm{eff}}$, \num{1575} stars both \Gaia \gspspec and \GESsh $\log g$, and \num{1904} stars both \Gaia \gspspec and \GESsh $\abratio{Fe}{H}$. Table~\ref{Table:Intersections_summary} also counts the published abundances in \GESsh and \Gaia for the stars in $\mathcal{S}_{3}$. As expected, this time, the \GESsh catalogue is more complete than the \Gaia DR3 catalogue when it comes to individual abundances. Four $\alpha$ elements -- namely Mg, Si, Ca, and Ti -- possess individual abundances in both \Gaia DR3 and \GES DR5.1 for more than \num{100} stars. Since the \Gaia RVS spectra are centred around the strong lines of the near-infrared \ion{Ca}{II} triplet, Ca is logically the element most-often measured by \Gaia \gspspec and \num{503} stars have a Ca abundance in both surveys. In the next subsections, we present statistical tests done on the atmospheric parameters and abundances of the selected stars to discuss the agreement between the \GESsh and \Gaia catalogues.

\citet{2023A&A...674A..29R} use three external heterogeneous (different instruments, spectral coverage, analysis methods) catalogues, namely APOGEE DR17, GALAH DR3 and RAVE DR6, to validate the \gspspec parametrisation of the \Gaia RVS spectra. In short, after filtering using the uncertainties and quality flags provided in these external catalogues and the \gspspec flags, \citet{2023A&A...674A..29R} define a best-quality subset (\num{170000} unique stars) and a medium-quality subset (\num{750000} unique stars) to investigate the possible differences between the \gspspec parameters ($\left\{T_{\mathrm{eff}}, \log g, \abratio{M}{H}\right\}$) and the literature-compilation ones. The authors find no biases for $T_{\mathrm{eff}}$ but provide the reader with three calibrations in the form of low-order ($n \le 4$) polynomials to correct for the identified biases for $\log g$ and $\abratio{M}{H}$: a calibration for $\log g$ (hereafter, `calibrated $\log g$'), a general calibration for $\abratio{M}{H}$ (hereafter, `calibrated $\abratio{M}{H}$') and a specific calibration for $\abratio{M}{H}$ in open clusters (hereafter, `OC-calibrated $\abratio{M}{H}$'). The authors check also the dependency of individual abundances with the surface gravity using a subset of \gspspec-parametrised \Gaia sources. A series of calibrations for the \gspspec abundance ratios are derived by forcing stars of the solar neighbourhood with near-solar metallicities and on near-circular orbits to have $\abratio{X}{Fe}$ close to zero. It is worth noting here that a) the aforementioned calibrations are not guaranteed to work for any science case or for any volume of the parameter space; b) \gspspec atmospheric parameters are calibrated against external catalogues while \gspspec individual abundances are calibrated against a subset of the \gspspec catalogue; c) since each spectroscopic survey comes with its own biases induced by the choice of analysis techniques and tools, the need for recalibrating the \gspspec parameters is to be expected. \revun{\citet{2023A&A...674A..29R} publish also a set of quality flags for the atmospheric parameters and the chemical abundances. Out of the thirteen flags qualifying the atmospheric parameters and for the \num{2079} stars of the sample $\mathcal{S}_{2}$, all are set to 0 except for the flag \texttt{fluxNoise} (\texttt{flag07}): for \num{889} stars, \texttt{fluxNoise} is set to 0; for \num{809} stars, it is set to 1; for \num{324} stars, it is set to 2; for \num{57} stars, it is set to 3.} The present study is interesting in that it will \emph{independently} test the recommended \gspspec calibrations \revun{and quality flags} against another external catalogue \revun{not really used by \citet{2023A&A...674A..29R} in their validation}, namely \GESsh DR5.1, and furthermore, these tests are carried out in the faint-magnitude regime.

\subsection{Effective temperature}

\begin{figure}
  \centering
  \includegraphics[width=\columnwidth,clip]{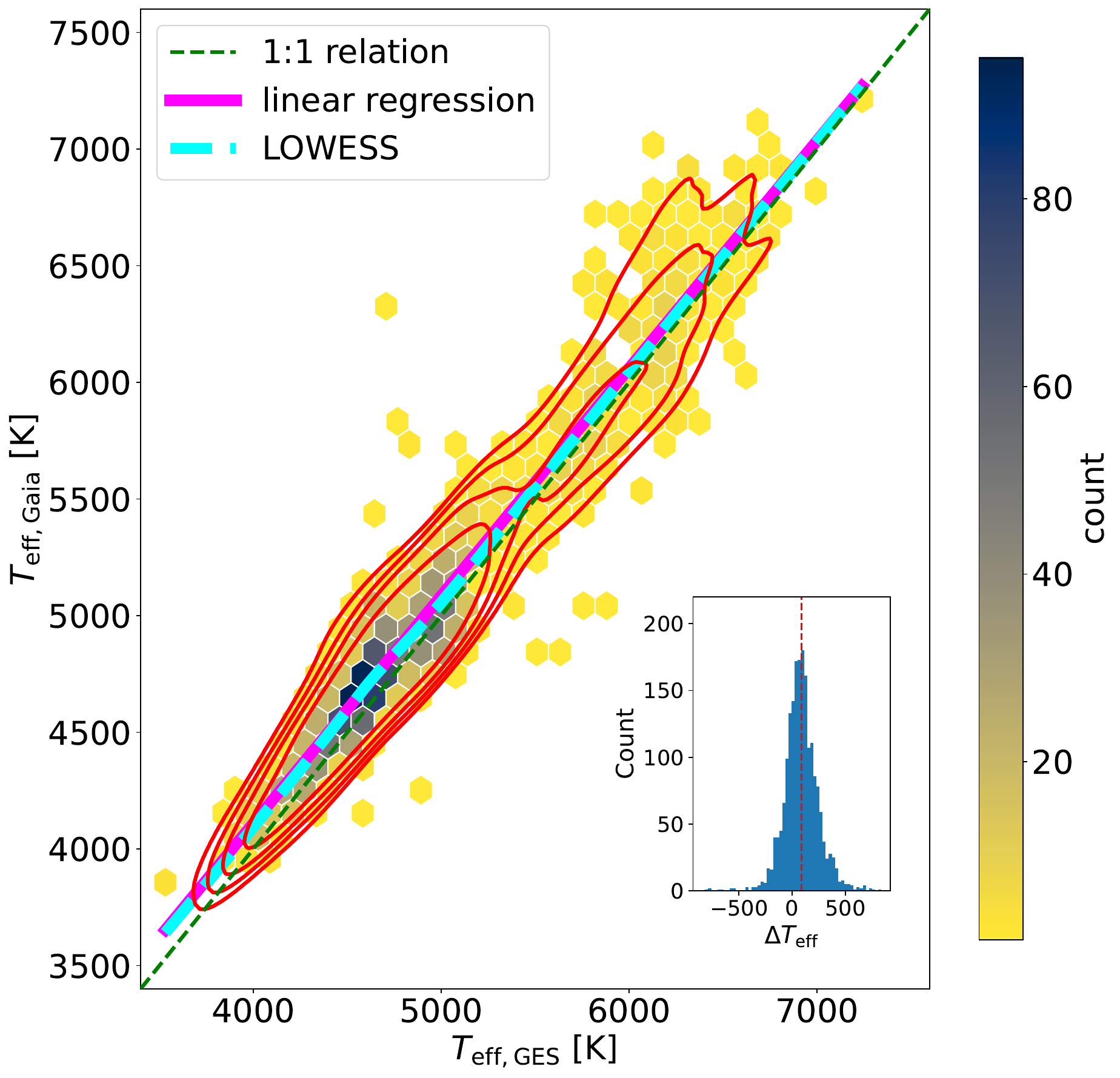}
  \caption{\label{Fig:Comparison_effective_temperatures} Comparison of the effective temperatures for the \num{1939} stars of $\mathcal{S}_{3}$ with both a \Gaia \gspspec and \GESsh temperature estimate. The 2D hexagonal bins are colour-coded by the number of stars. The red lines show density levels containing 68, 80, 90 and \SI{95}{\percent} of the population. The dashed green line shows the 1-to-1 relation, the pink thick line is the linear regression (see text for the parametrisation). The cyan thick dashed line is the LOWESS line. The inset shows the distribution of $\Delta T_{\mathrm{eff}} = T_{\mathrm{eff,Gaia}} - T_{\mathrm{eff,GES}}$.}
\end{figure}

\begin{figure}
  \centering
  \includegraphics[width=\columnwidth,clip]{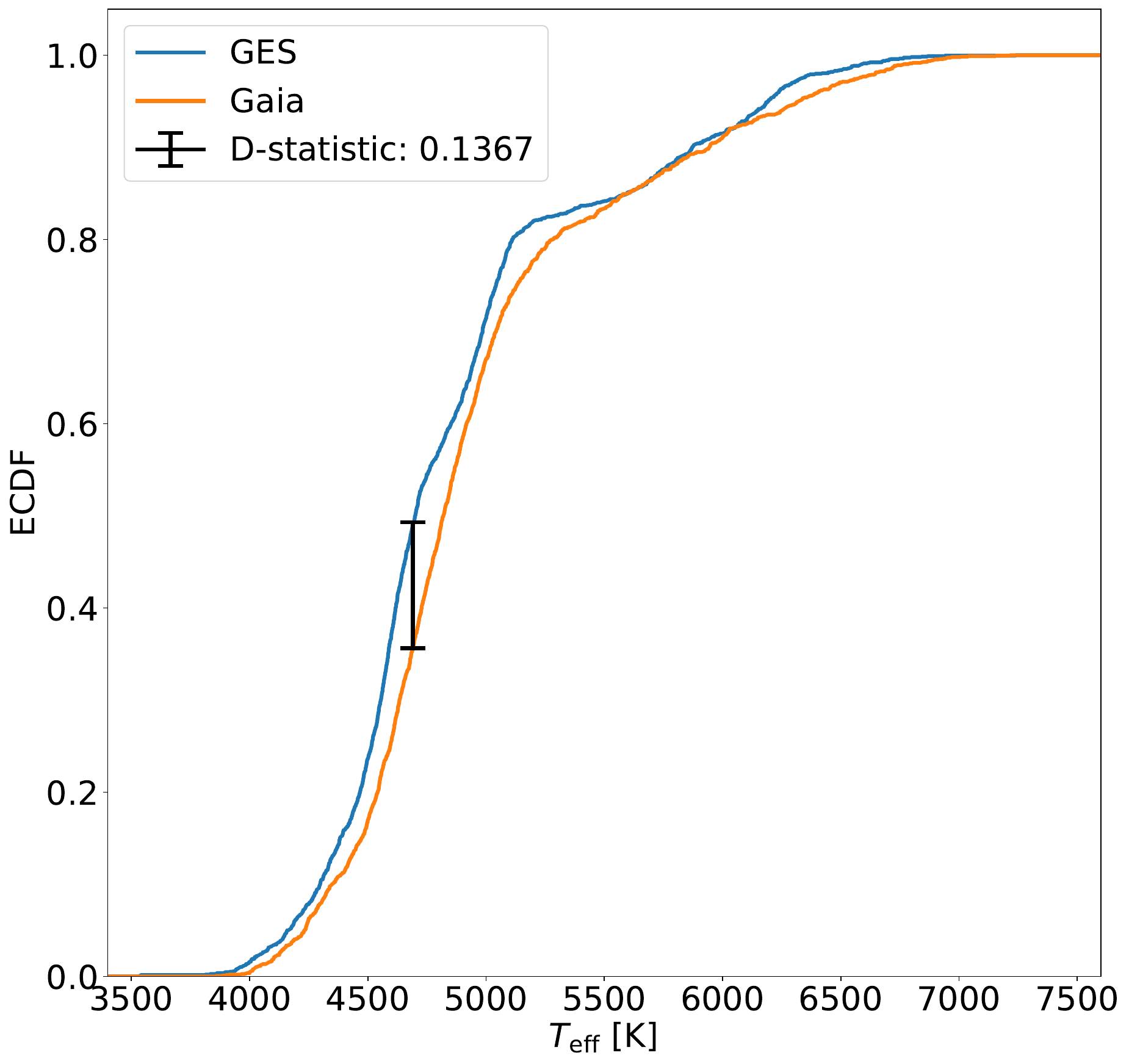}
  \caption{\label{Fig:KS_statistic_temperatures} Empirical cumulative distribution function (ECDF; continuous line) and $D$-statistic (vertical line) of the two-sample KS test for $T_{\mathrm{eff,GES}}$ (blue) and $T_{\mathrm{eff,GES}}$ (orange).}
\end{figure}

In Figure~\ref{Fig:Comparison_effective_temperatures}, we show the comparison between the effective temperatures for \num{1939} stars of the subsample $\mathcal{S}_{3}$. The distribution lies along the 1-to-1 relation and the linear regression give a slope of $\num{0.975} \pm \num{0.007}$, a $y$-intercept of $\SI{210}{\kelvin} \pm \SI{32}{\kelvin}$ and a coefficient $r^2 = \num{0.921}$. We conclude that $T_{\mathrm{eff,GES}}$ and $T_{\mathrm{eff,Gaia}}$ are strongly correlated. Figure~\ref{Fig:KS_statistic_temperatures} shows the empirical cumulated distribution functions (ECDF) and the $D$-statistic of the two-sample Kolmogorov-Smirnov (KS) test. We find $D = \num{0.1367}$ and a $p$-value of $\num{3e-16}$, which indicates that we should reject the null hypothesis that the two distributions are drawn from the same distribution: it is not surprising given the degeneracies that hamper the determination of atmospheric parameters and the different methods adopted by \GESlg and \gspspec to get these parameters. We find a mean and standard deviation for $\Delta T_{\mathrm{eff}} = T_{\mathrm{eff,Gaia}} - T_{\mathrm{eff,GES}}$ of \SI{89}{\kelvin} and \SI{170}{\kelvin}. \revun{Allowing only $\texttt{fluxNoise} = 0$ or 1 does not significantly change the comparison. If we keep only the objects with \texttt{fluxNoise} set to 0, the mean and standard deviation become \SI{38}{\kelvin} and \SI{112}{\kelvin}, but the sample size is divided by more than 2 (\num{829} objects left).}

\subsection{Surface gravity}

In Figure~\ref{Fig:Comparison_gravities}, we compare the \GESsh surface gravity $\log g$ of \num{1575} stars to the \Gaia uncalibrated (top panel) and calibrated (bottom panel) $\log g$. We find a mean and standard deviation for $\Delta \log g = \log g_{\mathrm{Gaia,uncal}} - \log g_{\mathrm{GES}}$ of \num{-0.19} and \num{0.39}, indicating that $\log g_{\mathrm{Gaia,uncal}}$ is underestimated on average. The behaviour is slightly different if we split the population in dwarf and giant stars: the mean and standard deviation become \num{-0.09} and \num{0.30} for dwarf stars (\num{329} objects with $\log g_{\mathrm{GES}} \ge \num{3.5}$), and \num{-0.22} and \num{0.40} for giant stars (\num{1246} objects). The bias affecting $\log g_{\mathrm{Gaia,uncal}}$ is a bit larger, in absolute value, for the giant subpopulation than for the dwarf subsample.

The \gspspec calibrated gravity improves the situation for both dwarf and giant stars. \revun{For the full sample, the mean and standard deviation for $\Delta \log g = \log g_{\mathrm{Gaia,cal}} - \log g_{\mathrm{GES}}$ become \num{0.08} and \num{0.37}, respectively}; for the dwarf subsample, they are equal to \num{-0.03} and \num{0.26}, respectively; for the giant subsample, they are equal to \num{0.11} and \num{0.37}, respectively. The parameters (slope, $y$-intercept and $r^2$) of the linear regressions shown in Fig.~\ref{Fig:Comparison_gravities} are: $\left( 0.998, -0.185, 0.8161 \right)$ for the $\log g_{\mathrm{Gaia,uncal}}$ and $\left( 0.866, 0.455, 0.8137 \right)$ for $\log g_{\mathrm{Gaia,cal}}$. \revun{If we keep only the objects with \texttt{fluxNoise} set to 0, the mean and standard deviation become \num{-0.02} and \num{0.28}, but again the sample size is divided by more than 2 (\num{710} objects left).} We conclude that it is better to use the calibrated \gspspec gravity and \revun{to clean the selection with the help of the quality flags, in line with the prescriptions from \citet{2023A&A...674A..29R}}.

\begin{figure}
  \centering
  \includegraphics[width=\columnwidth,clip]{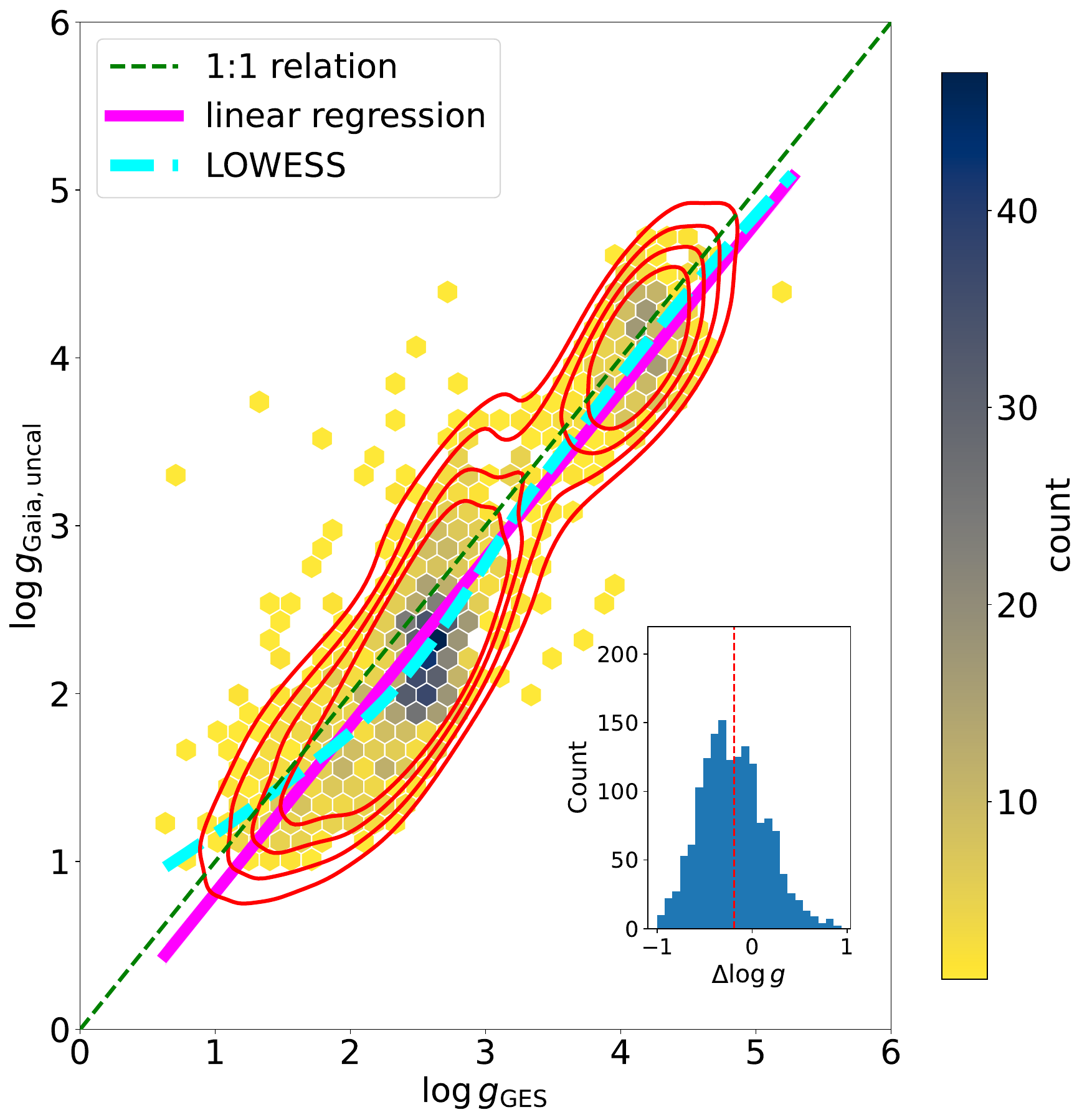}

  \includegraphics[width=\columnwidth,clip]{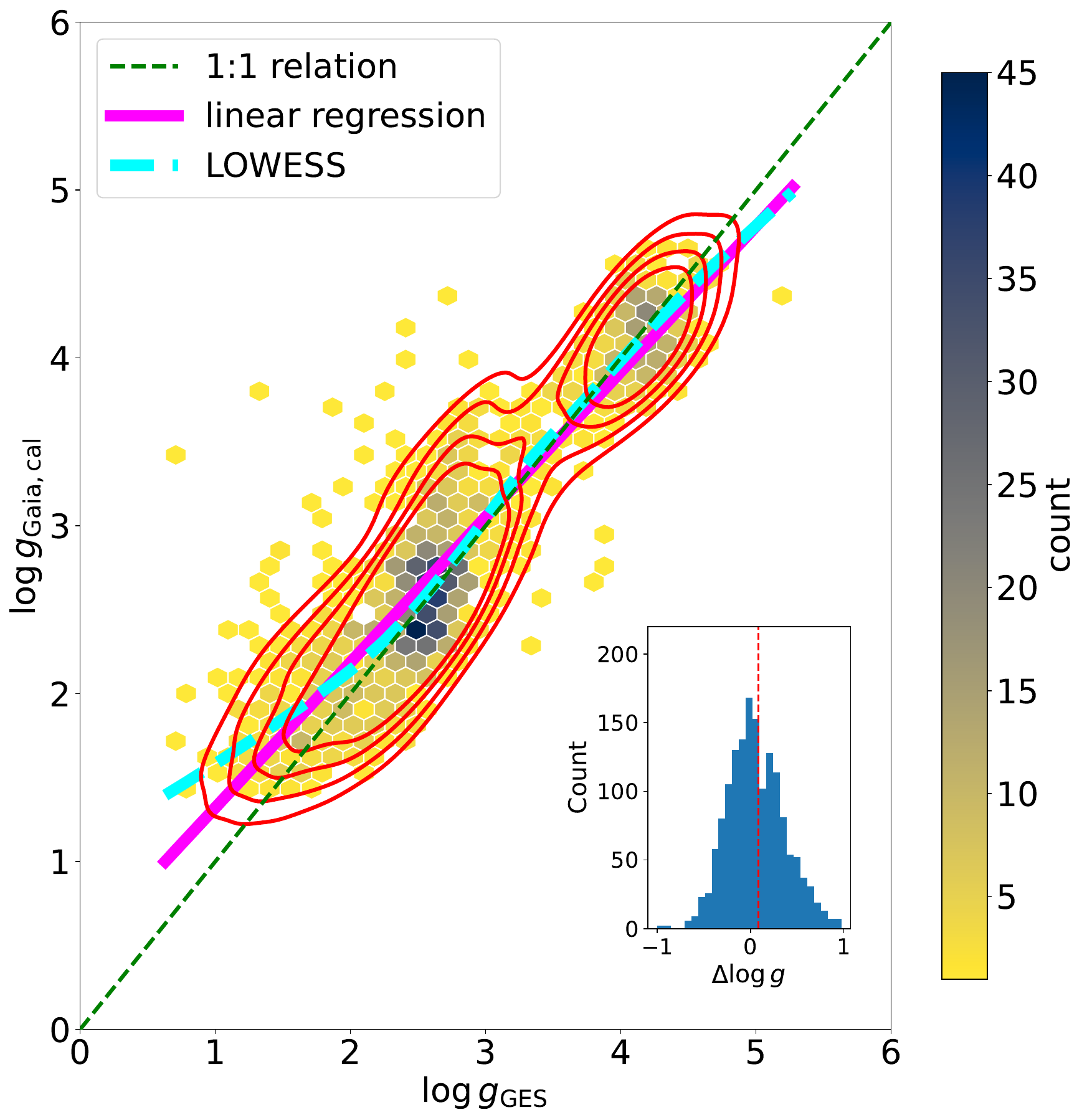}
  \caption{\label{Fig:Comparison_gravities} Comparison of the \GESlg surface gravity with that of \Gaia uncalibrated (top) and calibrated (bottom) ones. Symbols and colours are as in the Fig.~\ref{Fig:Comparison_effective_temperatures}.}
\end{figure}

\subsection{Metallicity}

In this section, we compare the \num{1904} objects with both a \GESsh and a \Gaia $\abratio{Fe}{H}$ estimates. Figure~\ref{Fig:Comparison_gravities} shows the comparison for the uncalibrated, the calibrated and the OC-calibrated \Gaia $\abratio{M}{H}$. The metallicity range goes from \SI{-2.2}{\dex} to \SI{0.5}{\dex} in terms of \GESsh $\abratio{Fe}{H}$ but \SI{94}{\percent} of the \num{1904} stars have a $\abratio{Fe}{H}_{\mathrm{GES}}$ in $[-0.5, 0.5]$. The mean and standard deviation of $\Delta \abratio{Fe}{H} = \abratio{Fe}{H}_{\mathrm{Gaia}} - \abratio{Fe}{H}_{\mathrm{GES}}$ are: \SI{0.03}{\dex} and \SI{0.17}{\dex} for the uncalibrated metallicity, \SI{0.03}{\dex} and \SI{0.16}{\dex} for the calibrated metallicity, and \SI{0.02}{\dex} and \SI{0.16}{\dex} for the OC-calibrated metallicity. The parameters (slope, $y$-intercept, $r^2$) of the linear regressions are: $\left( 0.887, 0.020, 0.6744 \right)$ for the uncalibrated metallicity, $\left( 0.855, 0.014, 0.6896 \right)$ for the calibrated metallicity, $\left( 0.835, 0.008, 0.6759 \right)$ for the OC-calibrated metallicity. \revun{If we keep only the objects with \texttt{fluxNoise} set to 0, the mean and standard deviation become \num{-0.03} and \num{0.13} for the uncalibrated case, and \num{-0.01} and \num{0.12} for the calibrated case, but again the sample size is divided by more than 2 (\num{822} objects left).} We conclude from these comparisons that the two calibrations and a flag-based selection \revun{appear to} marginally improve the agreement between the \GESsh and the \Gaia metallicity scales, \revun{in agreement with Sec.~4.3 of \citet{2023A&A...674A..32B}}.

\begin{figure}
  \centering
  \includegraphics[width=0.91\columnwidth,clip]{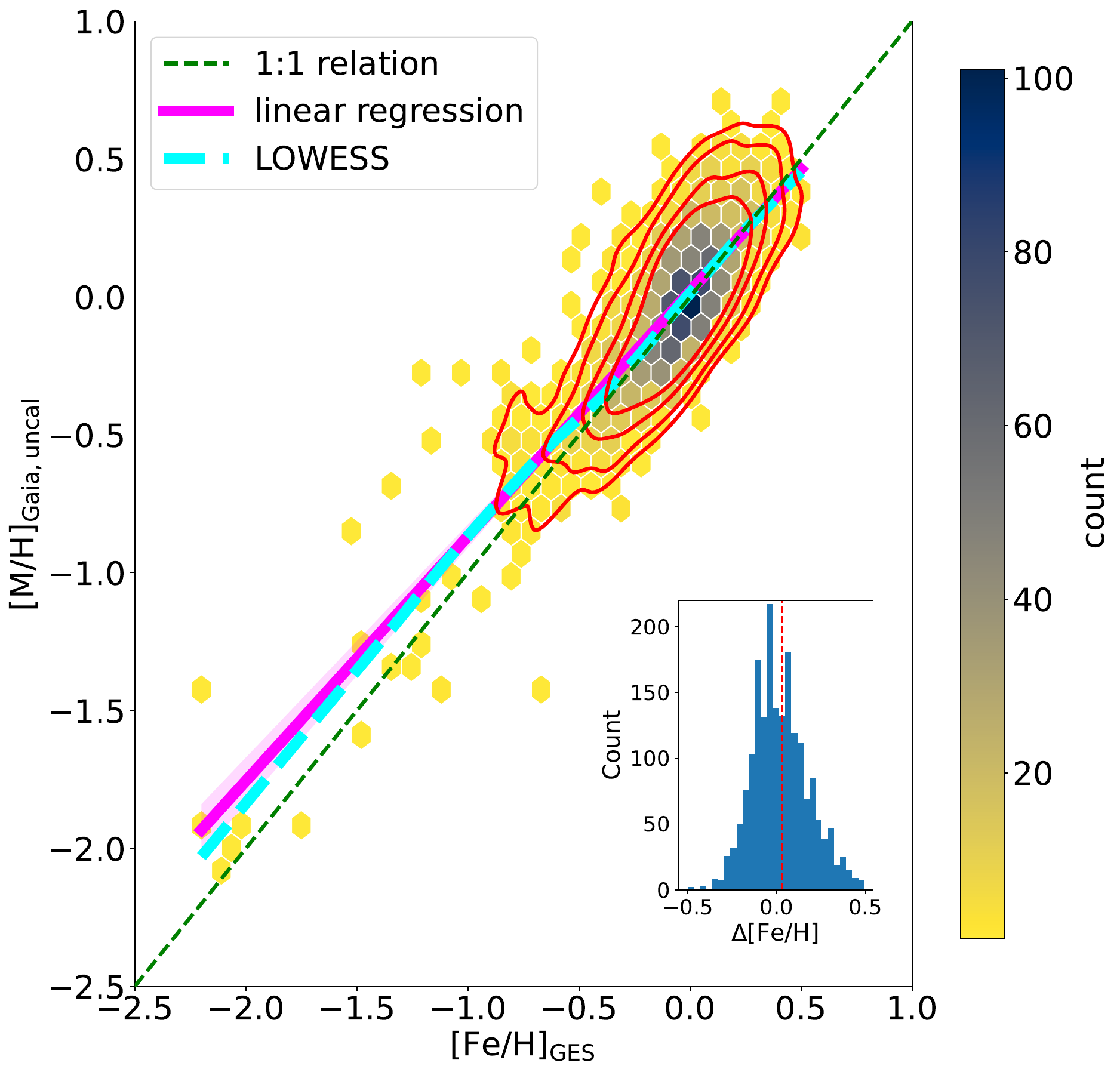}

  \includegraphics[width=0.91\columnwidth,clip]{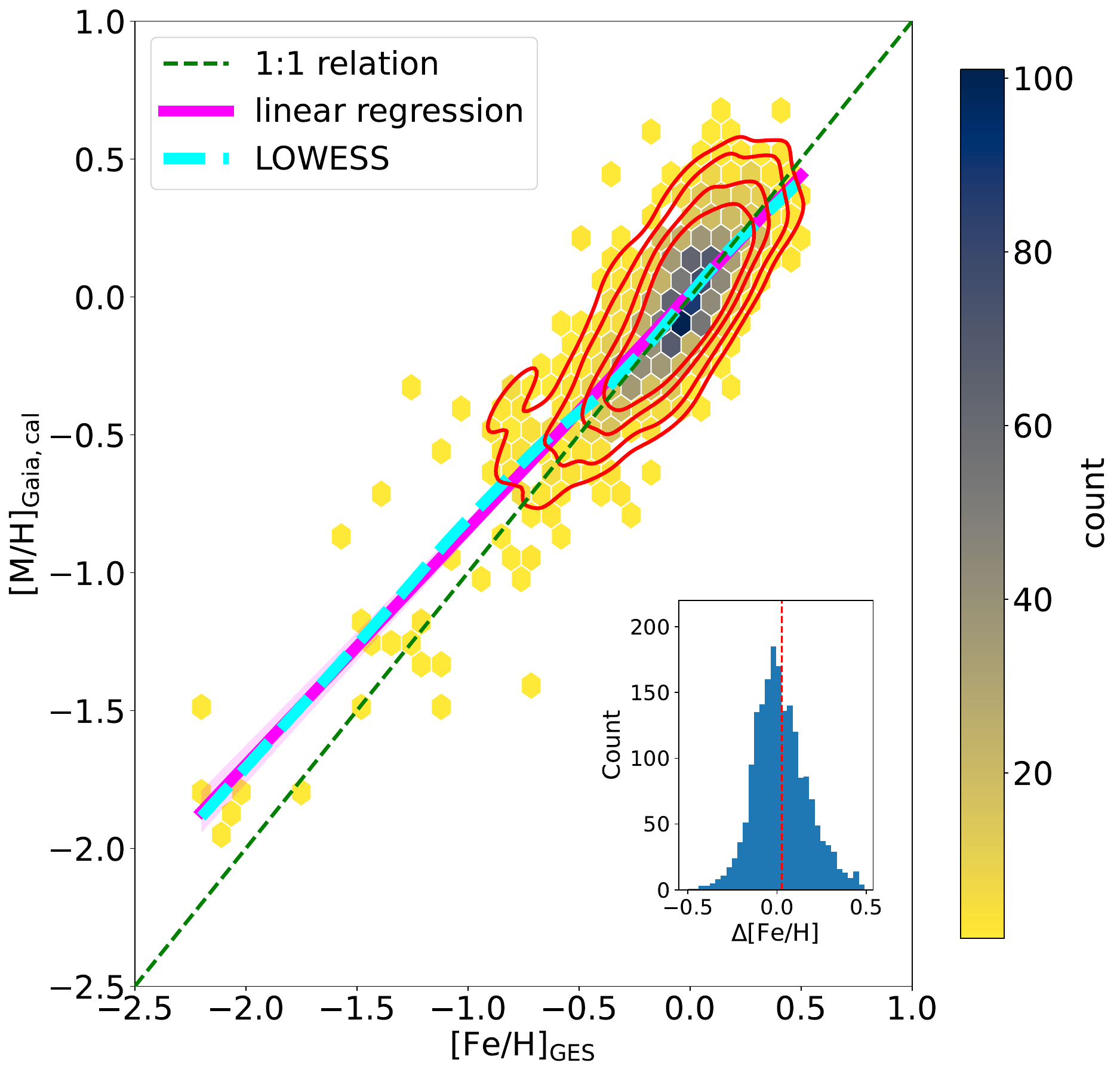}

  \includegraphics[width=0.91\columnwidth,clip]{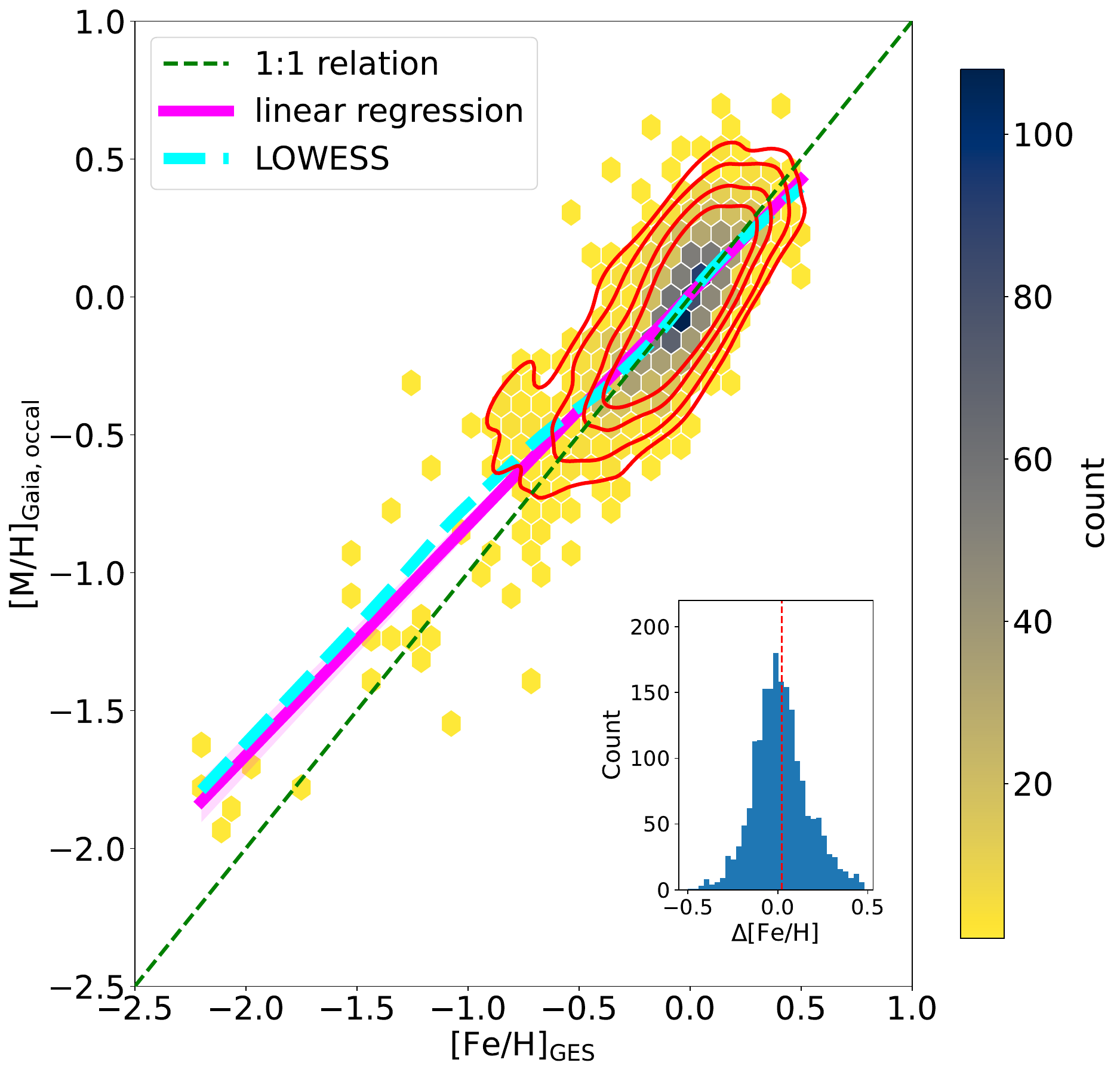}
  \caption{\label{Fig:Comparison_gravities} Comparison of the \GESlg metallicity with that of \Gaia uncalibrated (top), calibrated (middle) and OC-calibrated (bottom) ones. Symbols and colours are as in the Fig.~\ref{Fig:Comparison_effective_temperatures}.}
\end{figure}

\subsection{$\alpha$ abundance}

\begin{figure*}
  \centering
  \includegraphics[width=\columnwidth,clip]{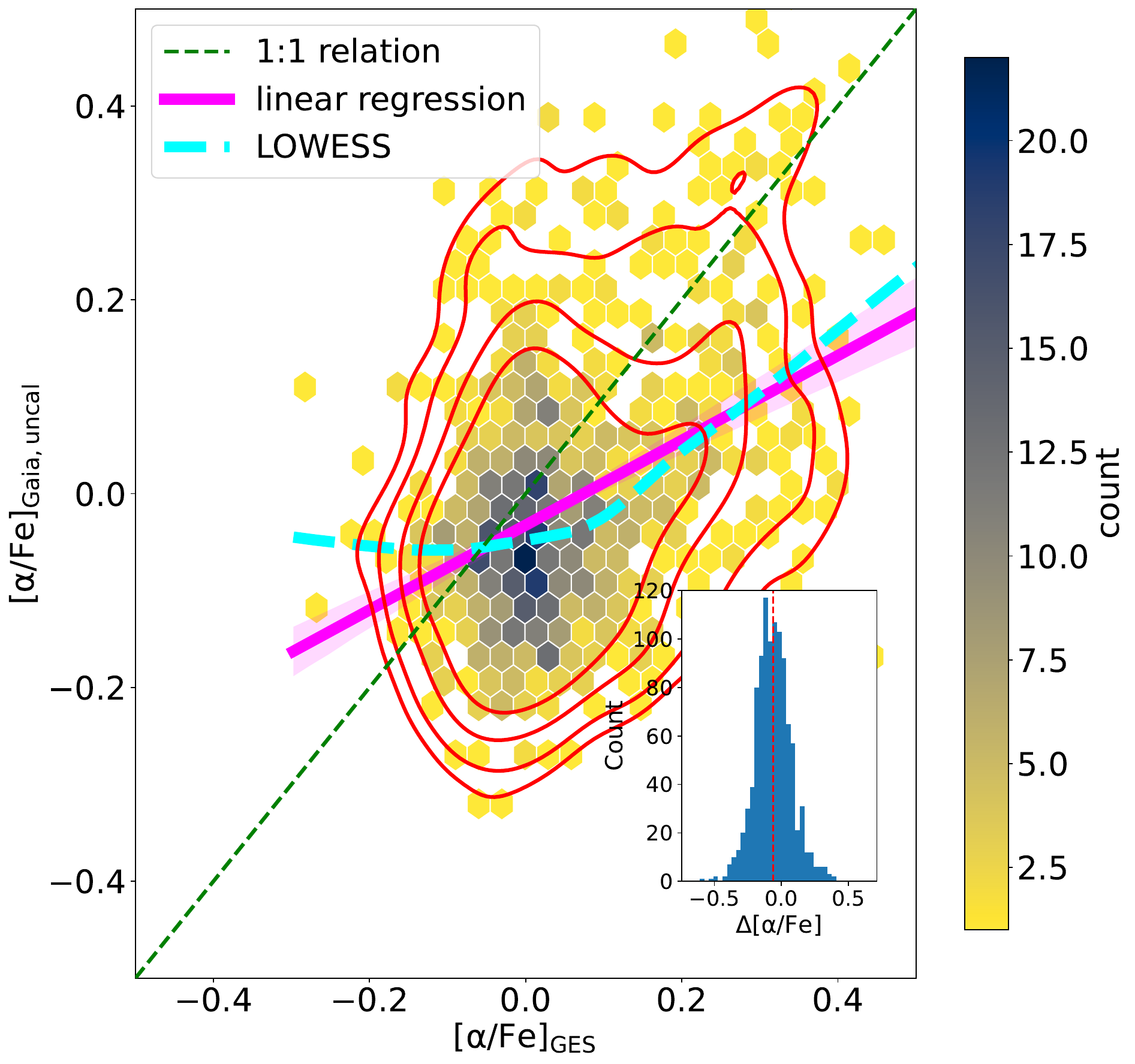}
  \includegraphics[width=\columnwidth,clip]{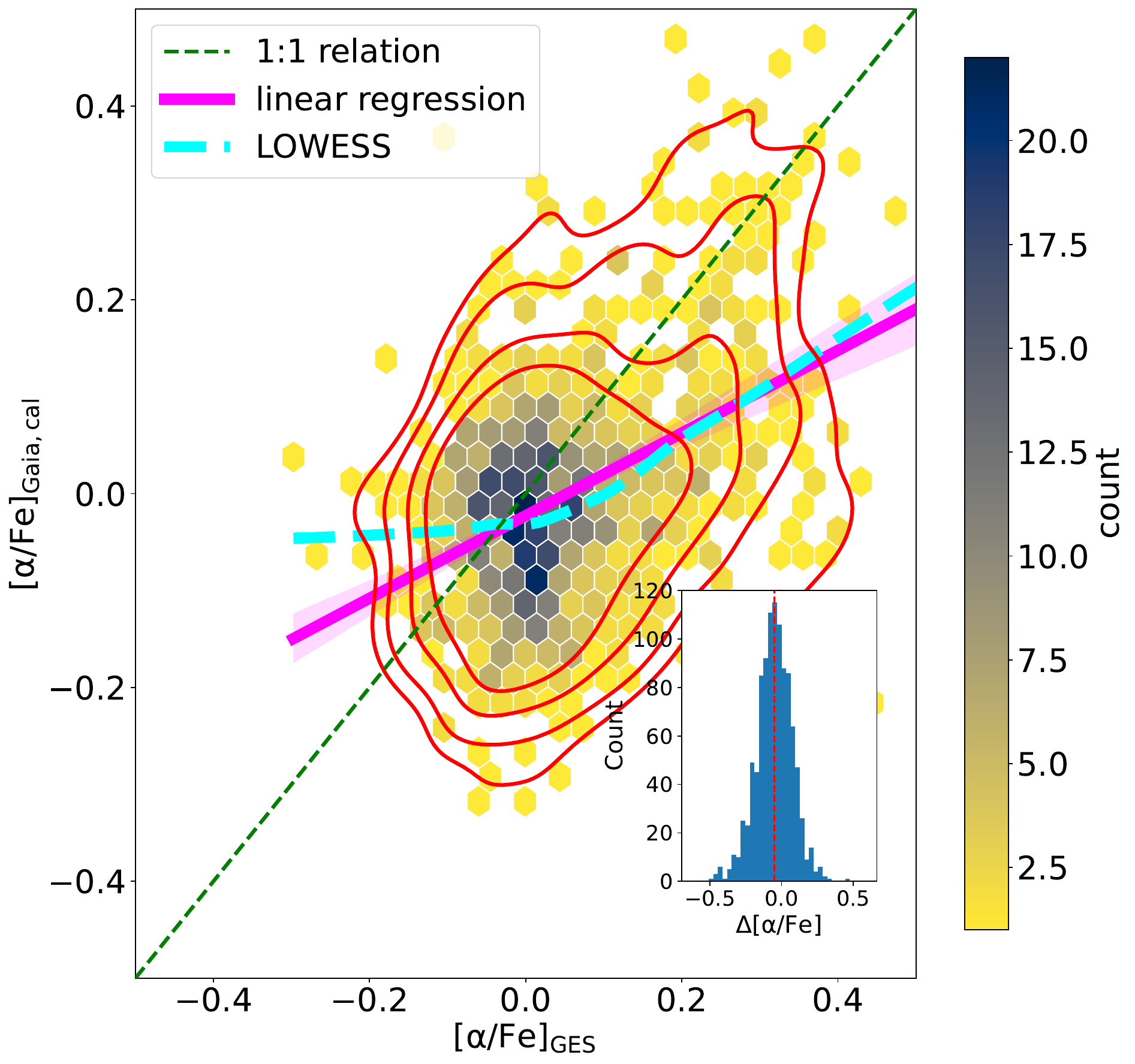}

  \includegraphics[width=\columnwidth,clip]{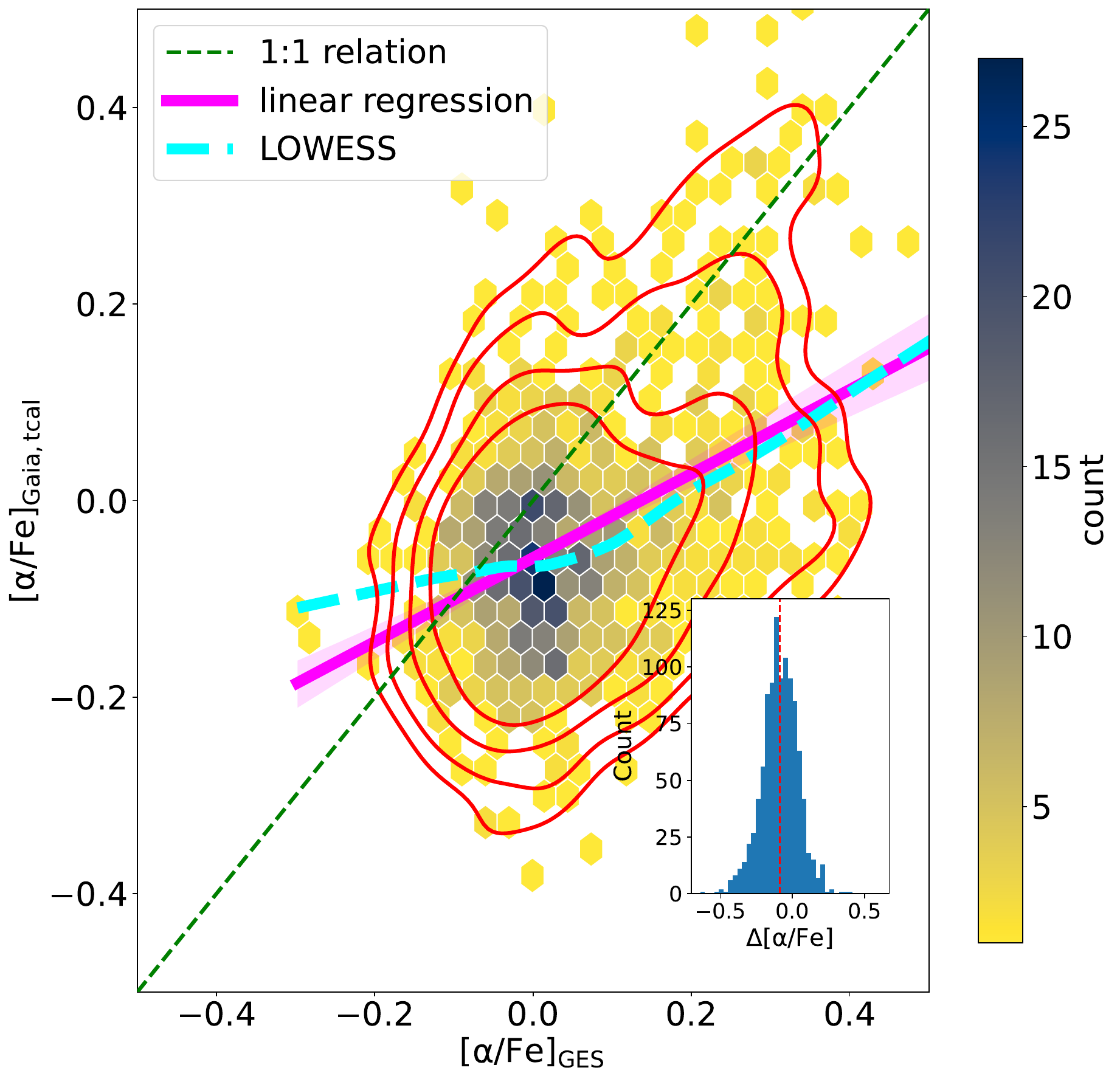}
  \includegraphics[width=\columnwidth,clip]{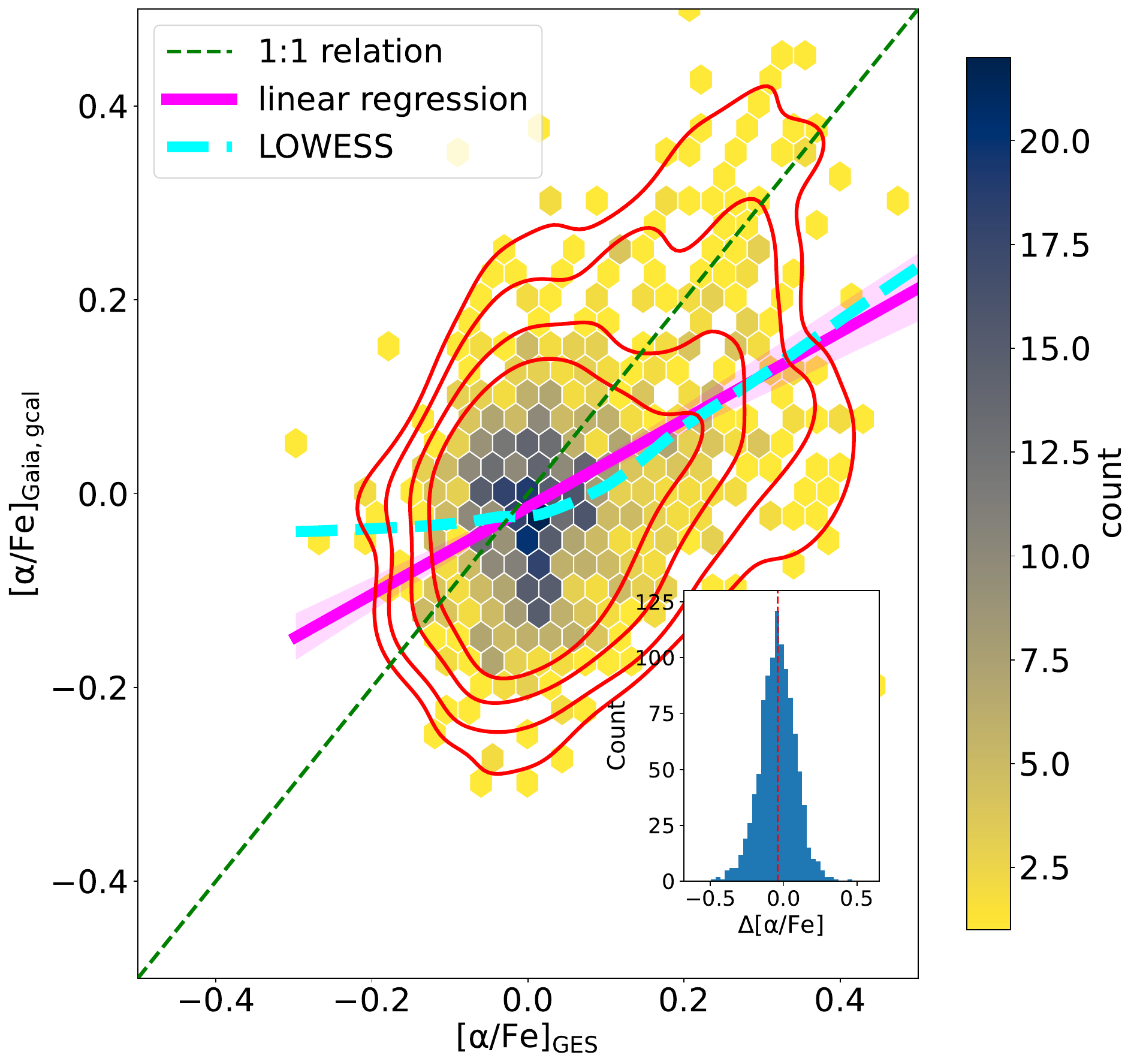}
  \caption{\label{Fig:Comparison_alpha} Comparison of the \GESlg $\abratio{\alpha}{Fe}$ with that of \Gaia uncalibrated (top left), calibrated (top right), $T_{\mathrm{eff}}$-calibrated (bottom left) and $\log g$-calibrated (bottom right) ones. Symbols and colours are as in the Fig.~\ref{Fig:Comparison_effective_temperatures}.}
\end{figure*}

We now compare the \num{1037} objects of $\mathcal{S}_{3}$ with a valid measurement of the $\alpha$ content in the \GESsh DR5.1 and the \Gaia DR3 catalogues. For \GESlg, $\abratio{\alpha}{Fe}$ is obtained by averaging the abundances of Mg, Si, Ca and Ti; for \Gaia, $\abratio{\alpha}{Fe}$ is directly parametrised. Figure~\ref{Fig:Comparison_alpha} shows the comparison between \GESsh and \Gaia for the \gspspec uncalibrated, calibrated, $T_{\mathrm{eff}}$-calibrated and $\log g$-calibrated values. The sample is made of Milky Way disc stars and therefore, $\abratio{\alpha}{Fe}$ approximately ranges from \num{0} to \num{0.4} in terms of \GESsh $\abratio{\alpha}{Fe}$. The mean and standard deviation of $\Delta \abratio{\alpha}{Fe} = \abratio{\alpha}{Fe}_{\mathrm{Gaia}} - \abratio{\alpha}{Fe}_{\mathrm{GES}}$ are: \SI{-0.06}{\dex} and \SI{0.14}{\dex} for the uncalibrated $\abratio{\alpha}{Fe}$, \SI{-0.05}{\dex} and \SI{0.13}{\dex} for the calibrated $\abratio{\alpha}{Fe}$, \SI{-0.09}{\dex} and \SI{0.13}{\dex} for the $T_{\mathrm{eff}}$-calibrated $\abratio{\alpha}{Fe}$, and \SI{-0.04}{\dex} and \SI{0.12}{\dex} for the $\log g$-calibrated $\abratio{\alpha}{Fe}$. \revun{Imposing \texttt{fluxNoise} equal to 0 does not significantly improve the agreement. For instance, the mean and standard deviation become \SI{-0.05}{\dex} and \SI{0.11}{\dex} for the calibrated case.} These numbers indicate that the calibrated $\abratio{\alpha}{Fe}$ and $\log g$-calibrated $\abratio{\alpha}{Fe}$ \revun{and the use of the quality flags, though preferable, offer a marginal improvement for the sample under study. It is in agreement with \citet{2023A&A...674A..32B} who note that biases remain after applying one of the above calibrations for $\abratio{\alpha}{Fe}$.}

\begin{figure*}
  \centering
  \includegraphics[width=\columnwidth,clip]{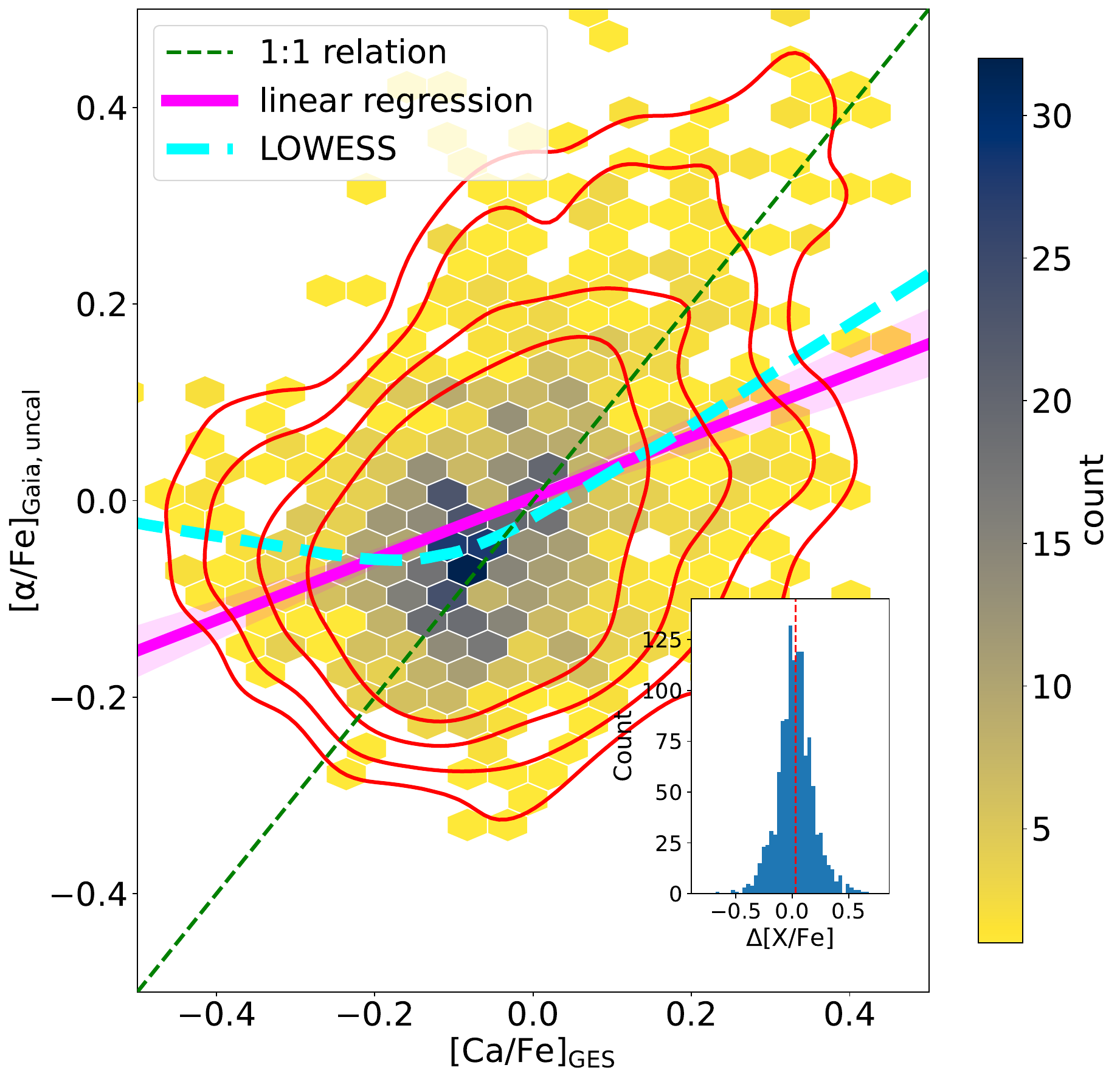}
  \includegraphics[width=\columnwidth,clip]{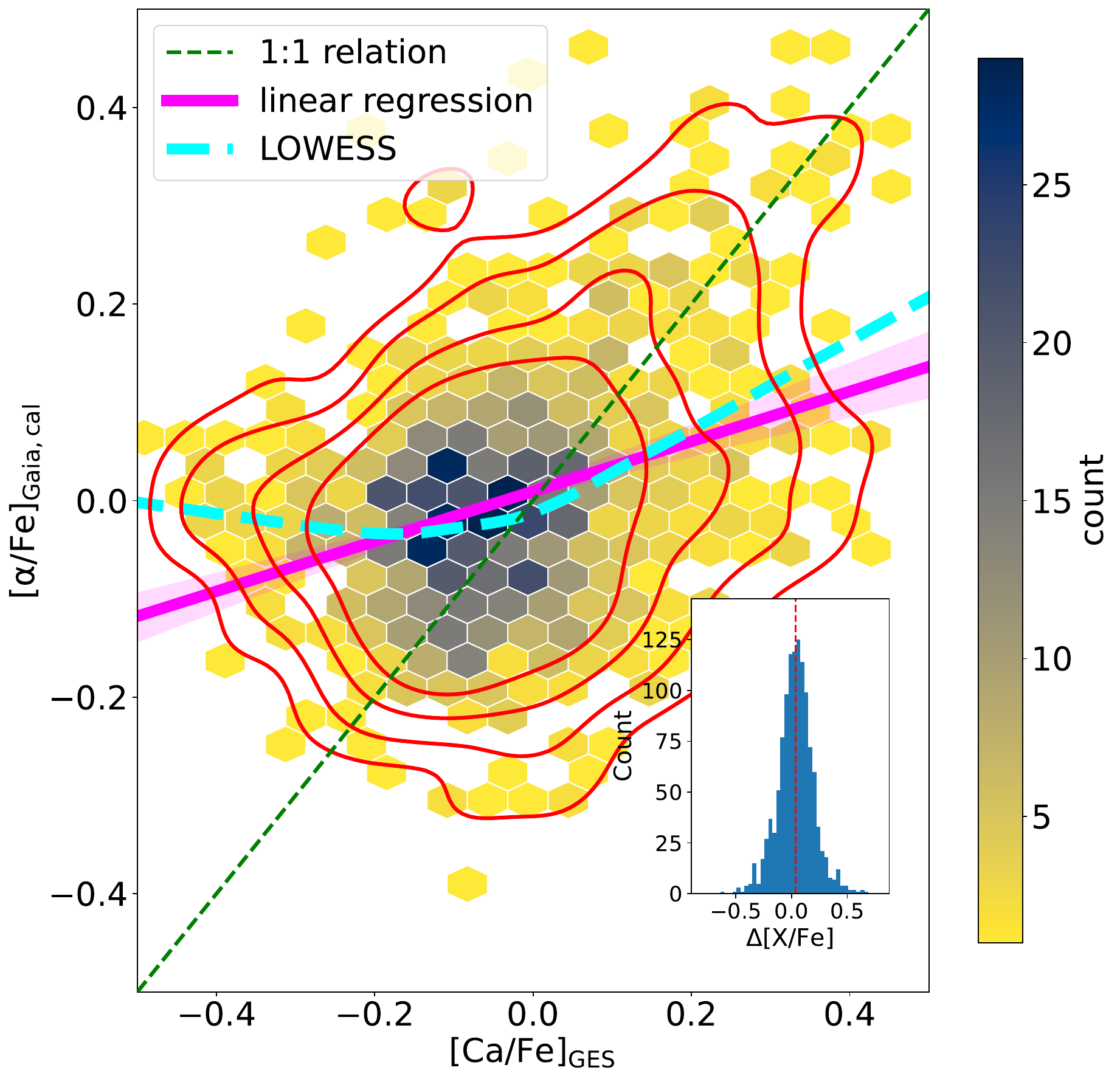}

  \includegraphics[width=\columnwidth,clip]{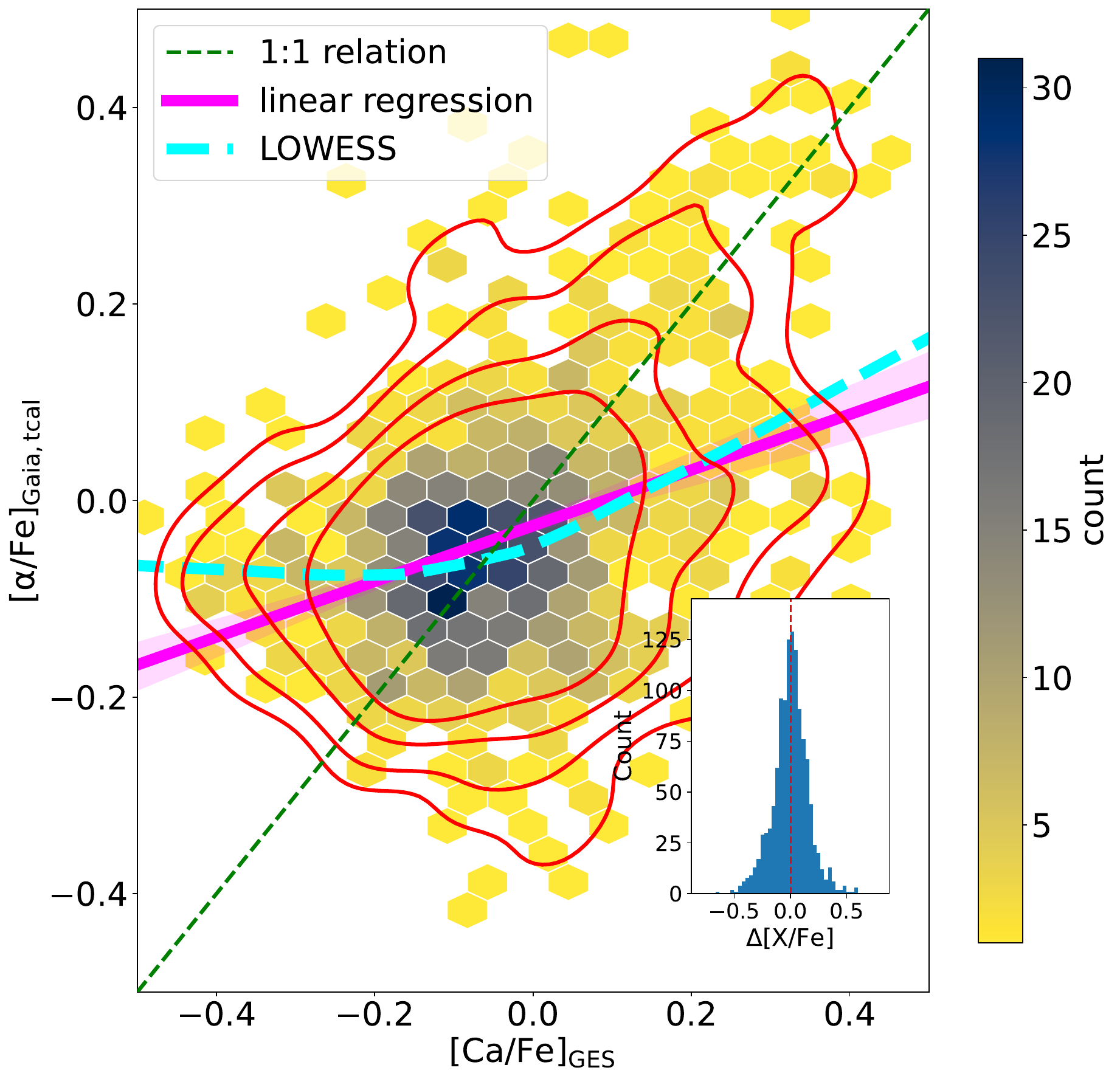}
  \includegraphics[width=\columnwidth,clip]{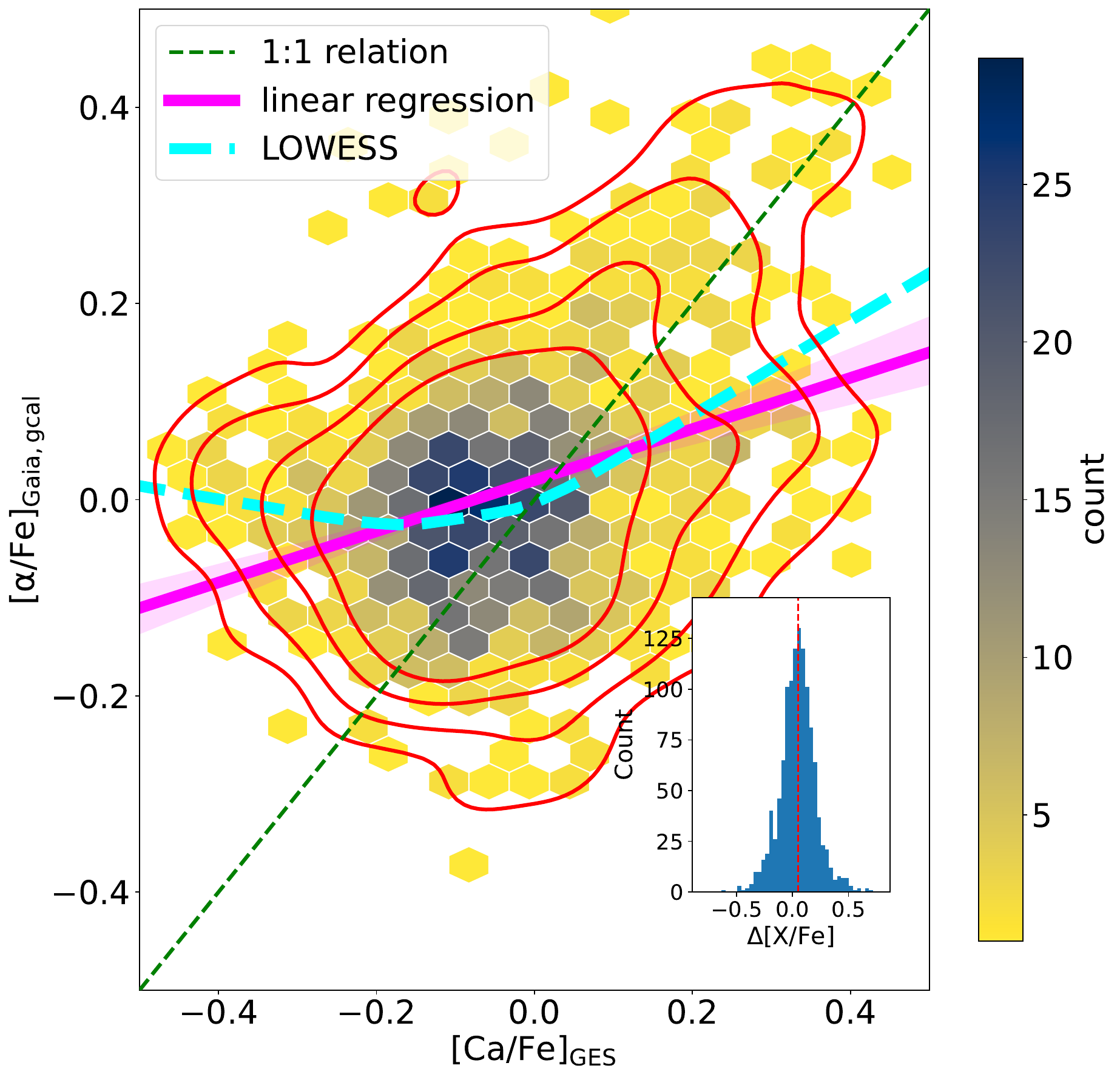}
  \caption{\label{Fig:Comparison_alpha_cafe} Comparison of the \GESlg $\abratio{Ca}{Fe}$ with the \Gaia uncalibrated (top left), calibrated (top right), $T_{\mathrm{eff}}$-calibrated (bottom left) and $\log g$-calibrated (bottom right) $\abratio{\alpha}{Fe}$. Symbols and colours are as in the Fig.~\ref{Fig:Comparison_effective_temperatures}.}
\end{figure*}

In late-type stars, the RVS spectrum will be dominated by the \ion{Ca}{II} triplet lines, and therefore, calcium will weigh more in the estimation of the \Gaia $\abratio{\alpha}{Fe}$ parameter. Figure~\ref{Fig:Comparison_alpha_cafe} compares the \Gaia $\abratio{\alpha}{Fe}$ to the \GESsh $\abratio{Ca}{Fe}$. The mean and standard deviation of $\Delta \abratio{\alpha}{Fe} = \abratio{\alpha}{Fe}_{\mathrm{Gaia}} - \abratio{Ca}{Fe}_{\mathrm{GES}}$ are: \SI{0.03}{\dex} and \SI{0.16}{\dex} for the uncalibrated $\abratio{\alpha}{Fe}$, \SI{0.04}{\dex} and \SI{0.16}{\dex} for the calibrated $\abratio{\alpha}{Fe}$, \SI{0}{\dex} and \SI{0.16}{\dex} for the $T_{\mathrm{eff}}$-calibrated $\abratio{\alpha}{Fe}$, and \SI{0.05}{\dex} and \SI{0.16}{\dex} for the $\log g$-calibrated $\abratio{\alpha}{Fe}$. The plots and these numbers show that there is indeed a similar agreement between \GESsh $\abratio{Ca}{Fe}$ and \Gaia $\abratio{\alpha}{Fe}$ as there is between \GESsh $\abratio{\alpha}{Fe}$ and \Gaia $\abratio{\alpha}{Fe}$. \revun{In other words, for the sample under study, the agreement between \Gaia $\abratio{\alpha}{Fe}$ and the \GESsh $\abratio{Ca}{Fe}$ is as good as the agreement between \Gaia $\abratio{\alpha}{Fe}$ and the \GESsh $\abratio{\alpha}{Fe}$.}

\subsection{Individual abundances}

\begin{figure*}
  \centering
  \includegraphics[width=0.99\columnwidth,clip]{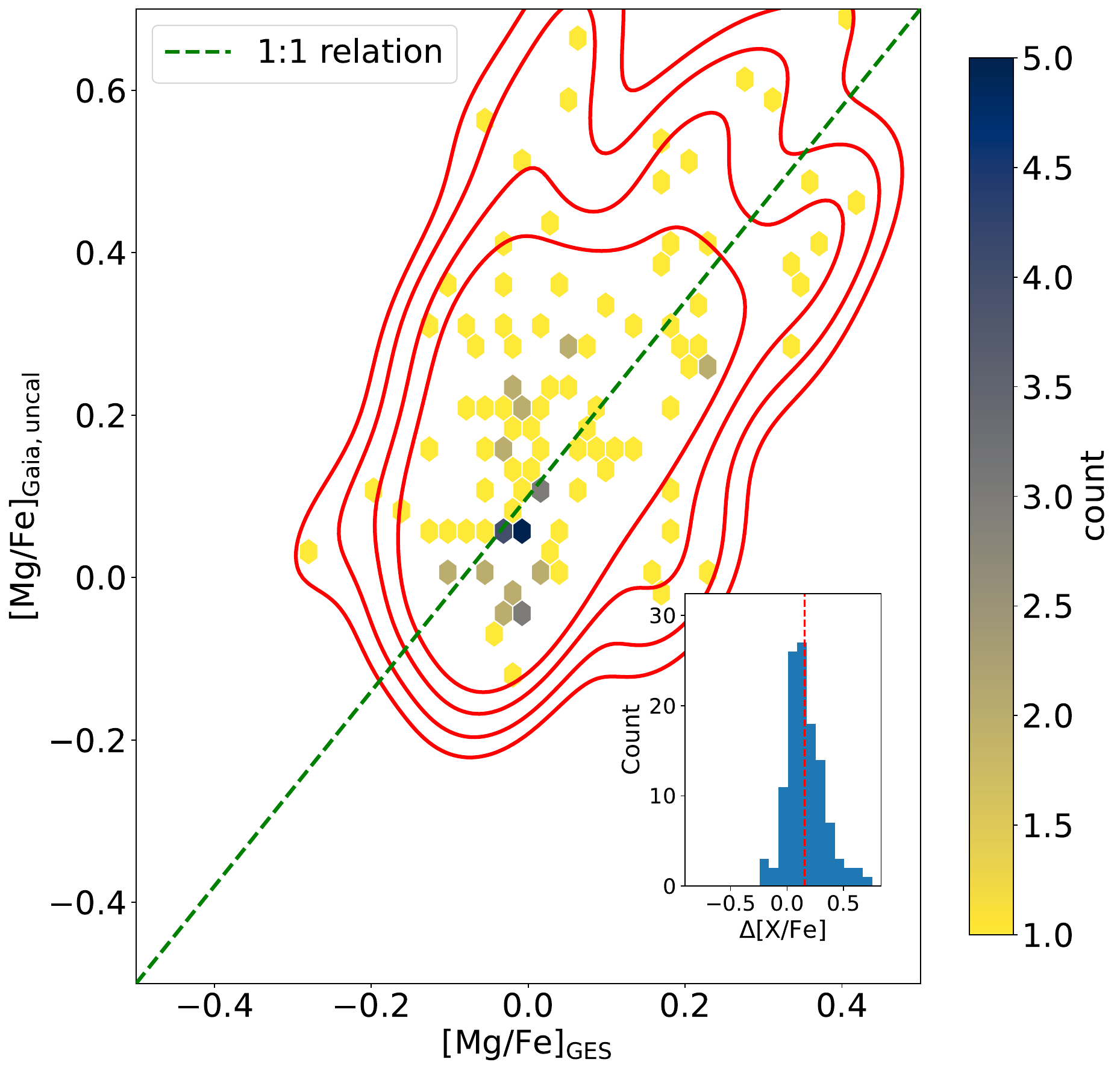}
  \includegraphics[width=0.99\columnwidth,clip]{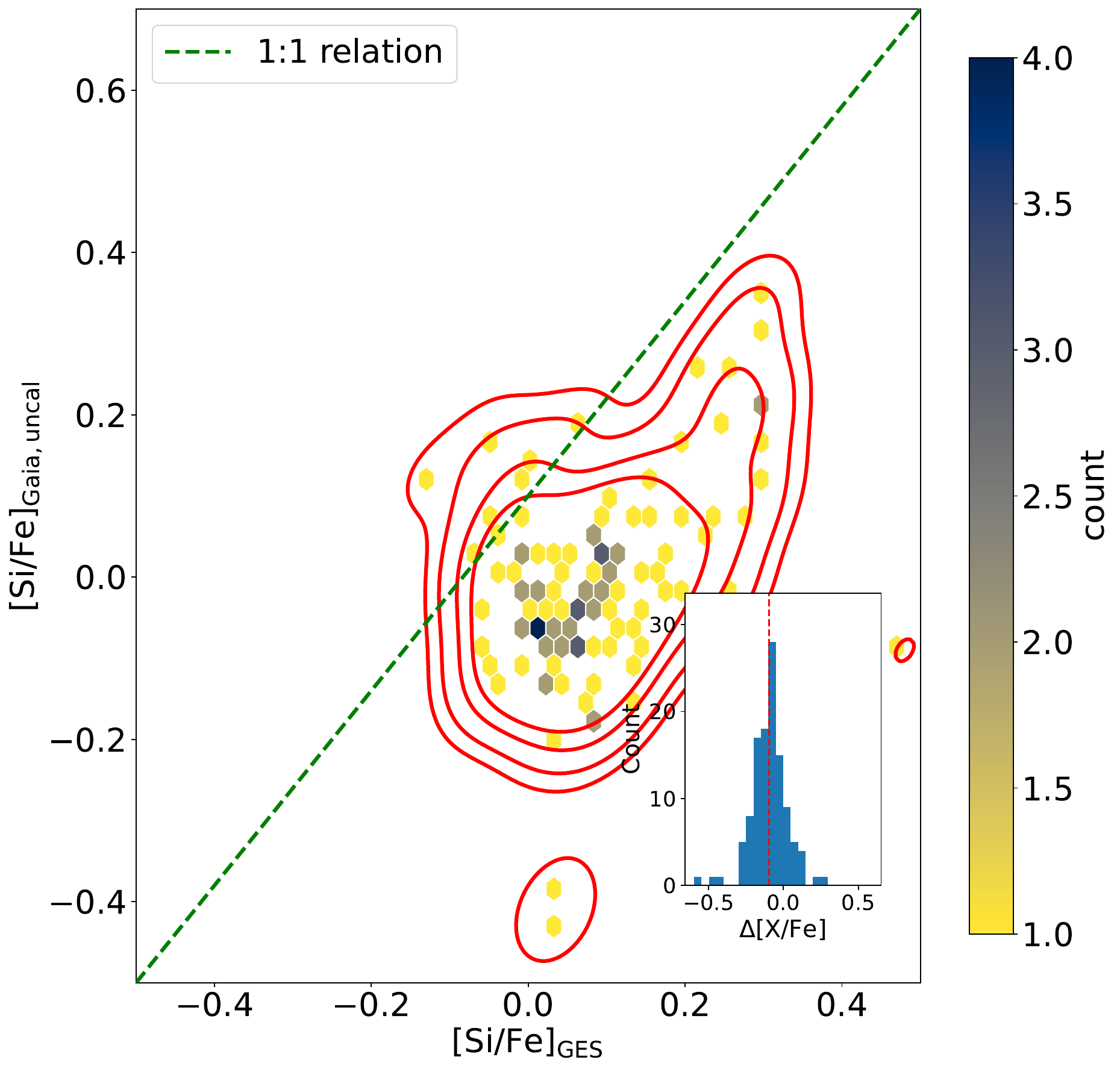}
  
  \includegraphics[width=0.99\columnwidth,clip]{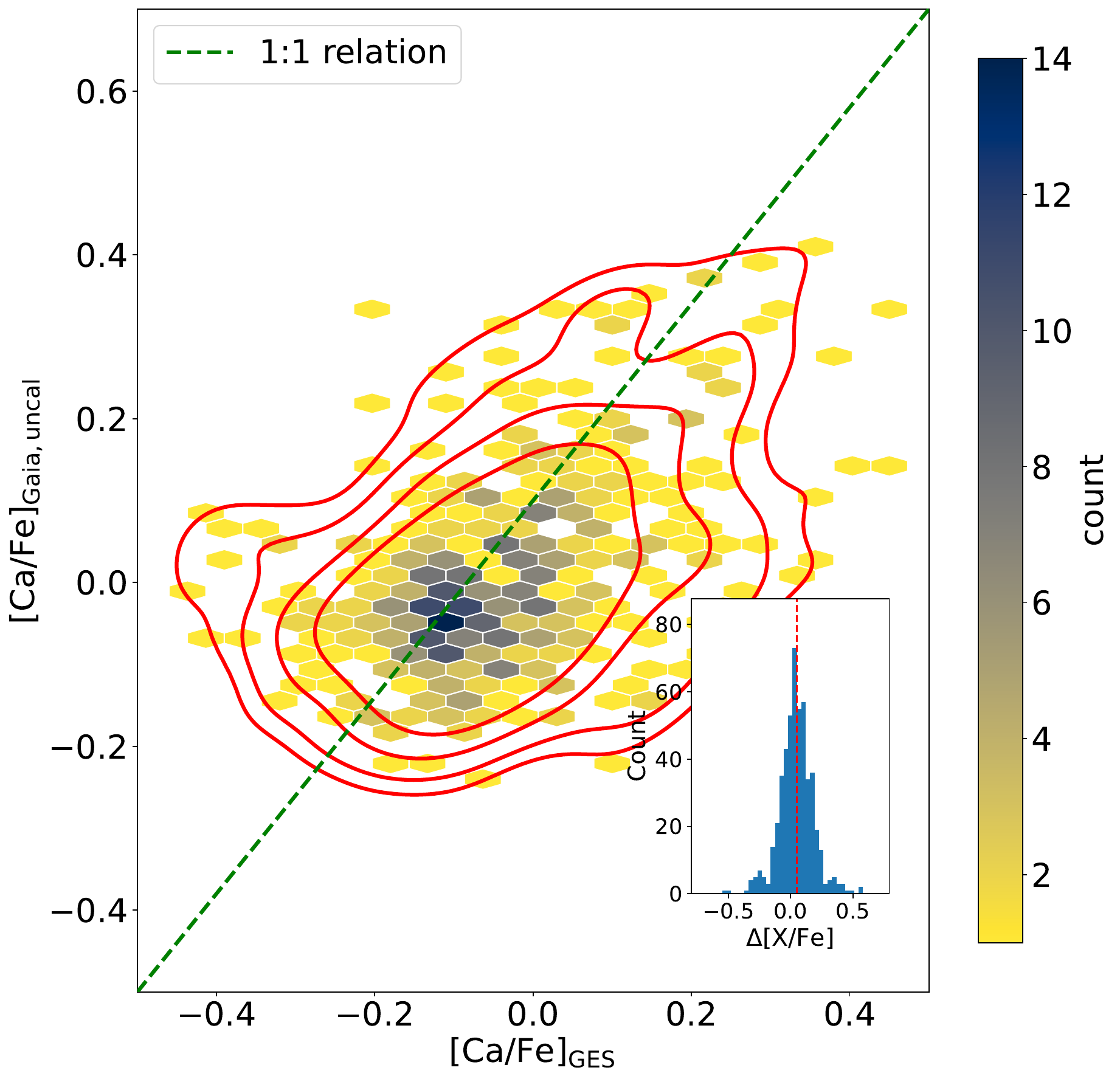}
  \includegraphics[width=0.99\columnwidth,clip]{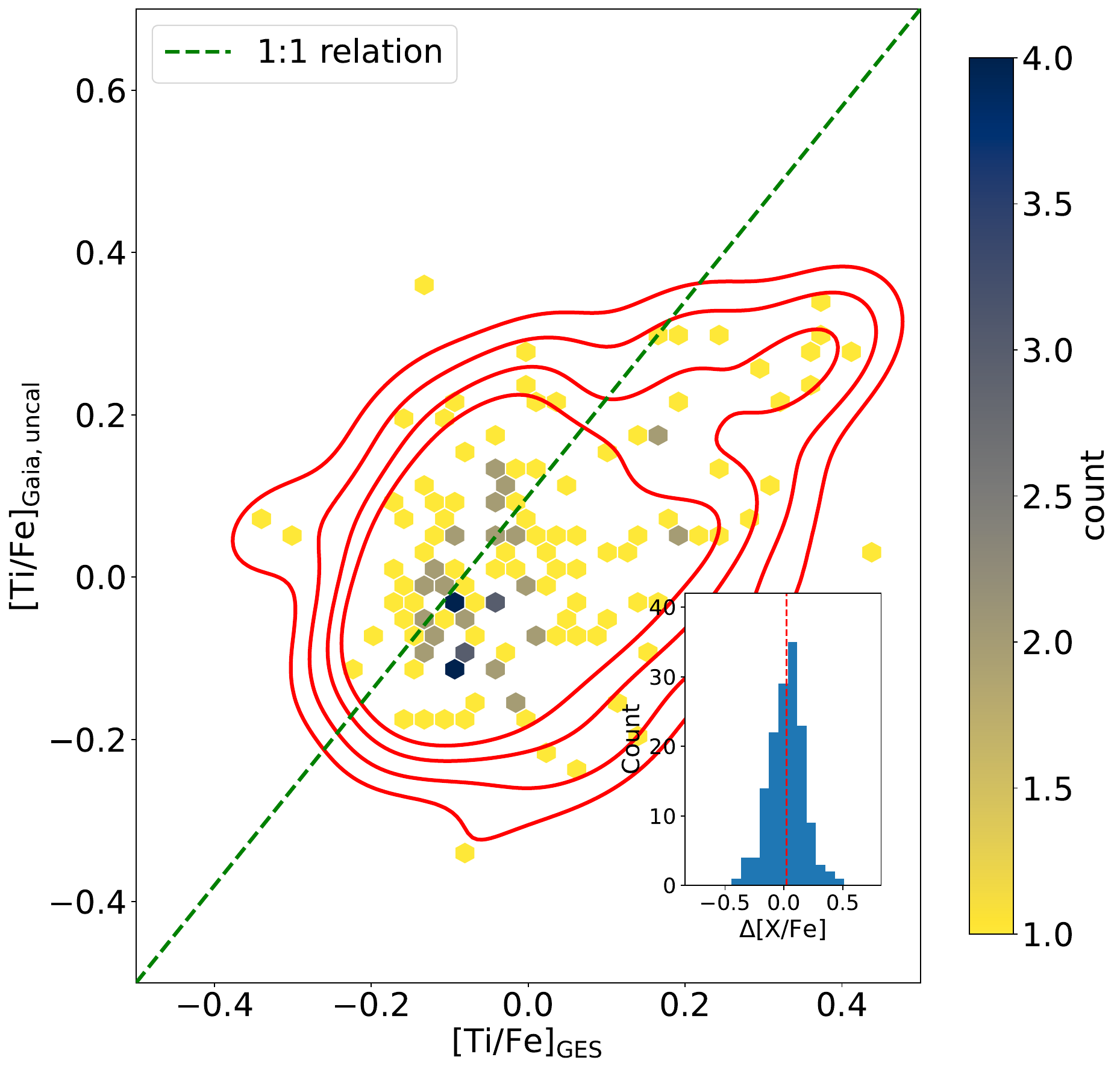}
  \caption{\label{Fig:Comparison_other_elements_p1} Comparison of the \GESlg $\abratio{X}{Fe}$ with the \Gaia uncalibrated $\abratio{X}{Fe}$ for the following chemical species (from top to bottom, left to right): Mg, Si, Ca, and Ti. Symbols and colours are as in the Fig.~\ref{Fig:Comparison_effective_temperatures}.}
\end{figure*}

\begin{figure*}
  \centering
  \includegraphics[width=0.99\columnwidth,clip]{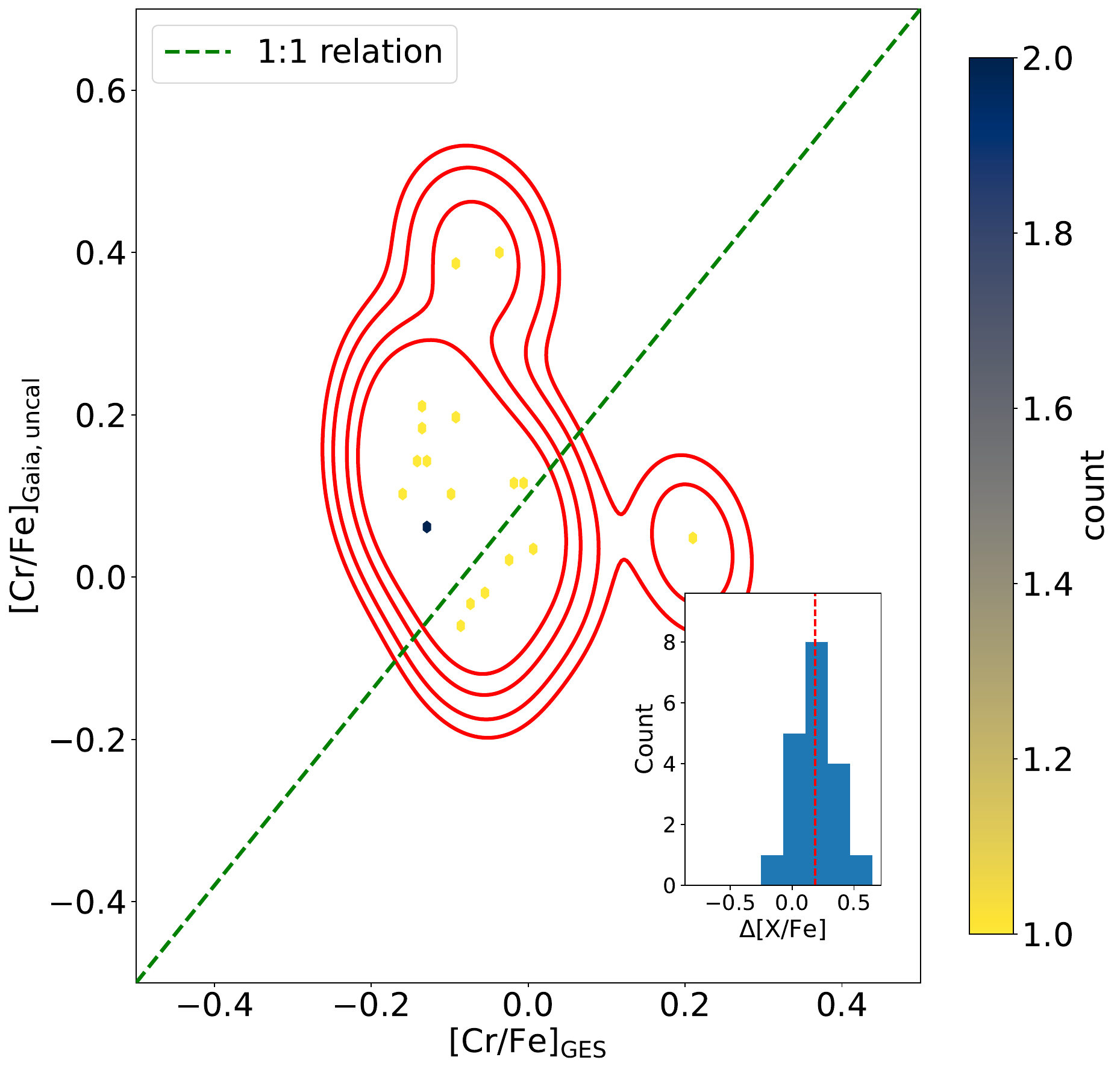}
  \includegraphics[width=0.99\columnwidth,clip]{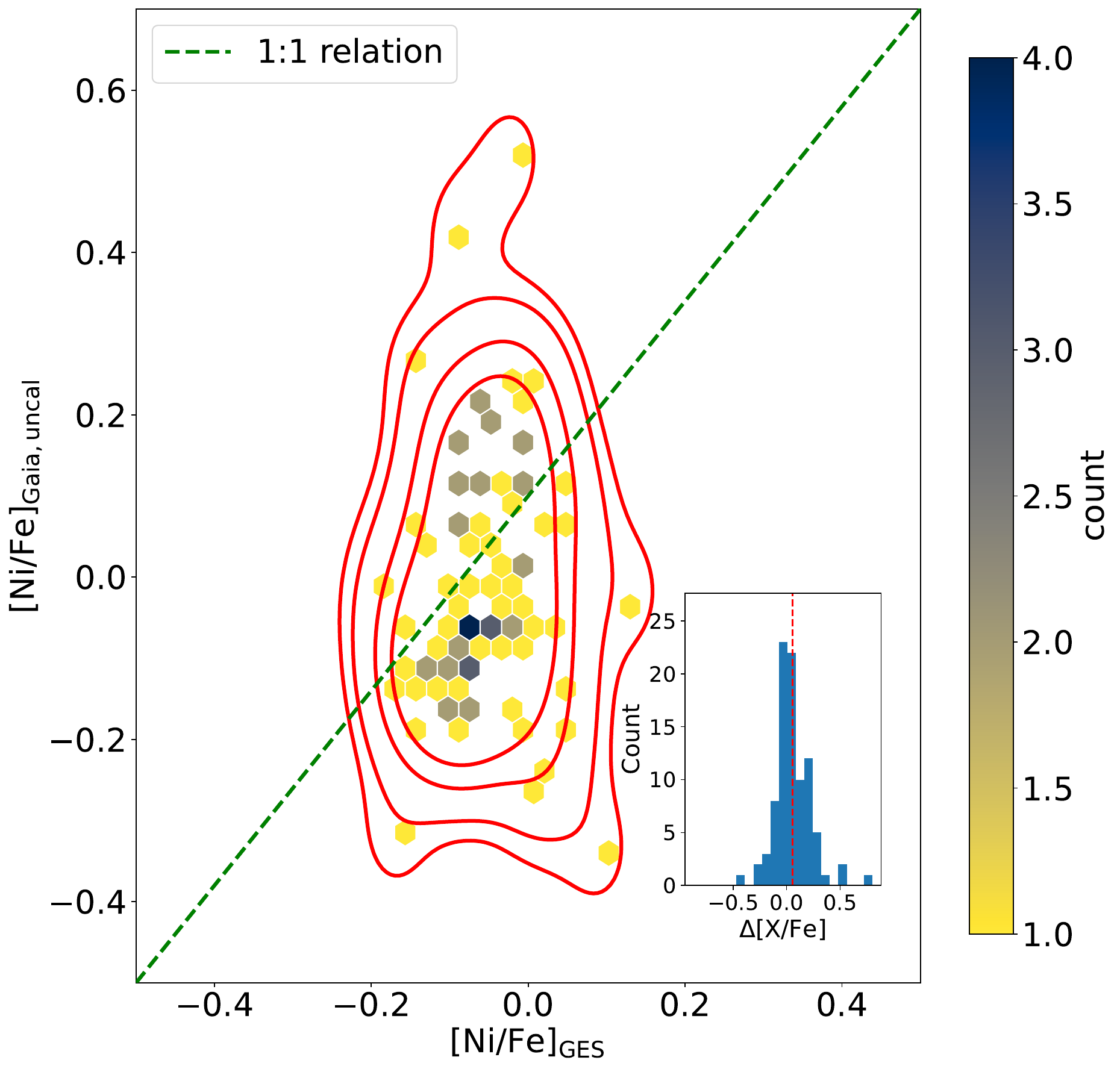}
  
  \includegraphics[width=0.99\columnwidth,clip]{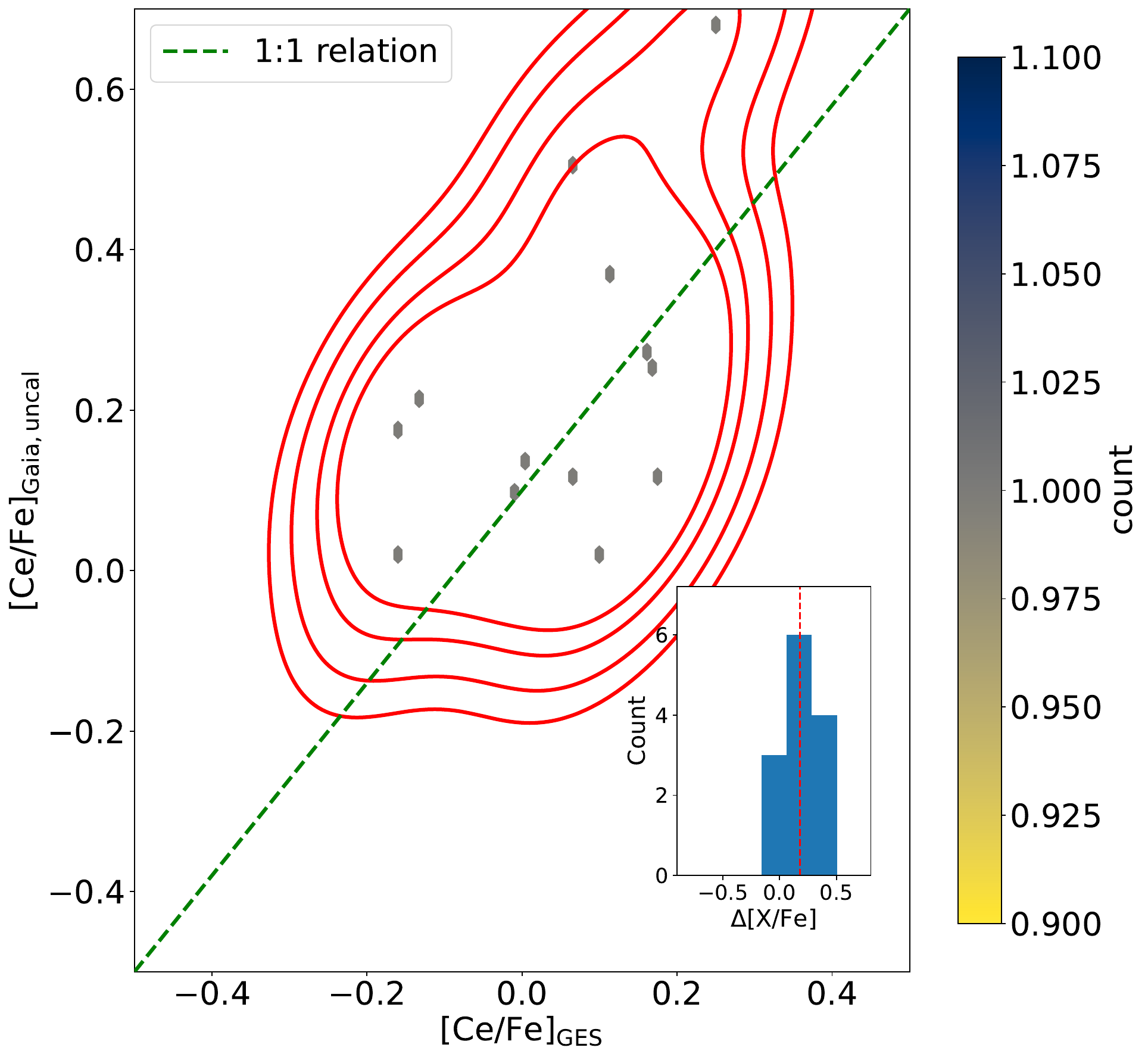}
  \caption{\label{Fig:Comparison_other_elements_p2} Comparison of the \GESlg $\abratio{X}{Fe}$ with the \Gaia uncalibrated $\abratio{X}{Fe}$ for the following chemical species (from top to bottom, left to right): Cr, Ni, and Ce. Symbols and colours are as in the Fig.~\ref{Fig:Comparison_effective_temperatures}.}
\end{figure*}

In Figures~\ref{Fig:Comparison_other_elements_p1} and \ref{Fig:Comparison_other_elements_p2}, we compare $\abratio{X}{Fe}$ for seven of the eight elements available both in \GESsh and \Gaia: four $\alpha$ elements (Mg, Si, Ca, Ti), two iron peak elements (Cr and Ni) and one neutron-capture element (Ce). We note that the \GESsh -- \Gaia intersection leads to small to very small samples (less than twenty datapoints for Cr and Ce) when it comes to comparing individual abundances. Nevertheless, Figure~\ref{Fig:Comparison_other_elements_p1} and \ref{Fig:Comparison_other_elements_p2} suggests that the best agreement is obtained for Ca, Ti and Ni (smallest biases) and the agreement is a bit worse for Mg and Si. We cannot conclude for Cr and Ce because of the paucity of data. \revun{For Mg, Si, Ti and Ni, we note that the \gspspec quality flags take either the values 0, 1 or 2 (rarely for Si and Ti). For Ca, all of the \num{503} stars have their Ca abundance quality flags set to 0. In particular, for these five species, none of the stars has an abundance quality flag set to 9, \ie a value that should be absolutely discarded according to the prescriptions from \citet{2023A&A...674A..29R}.}

\subsection{Dependency with $G$ magnitude}

\begin{figure*}
  \centering
  \includegraphics[width=\textwidth,clip]{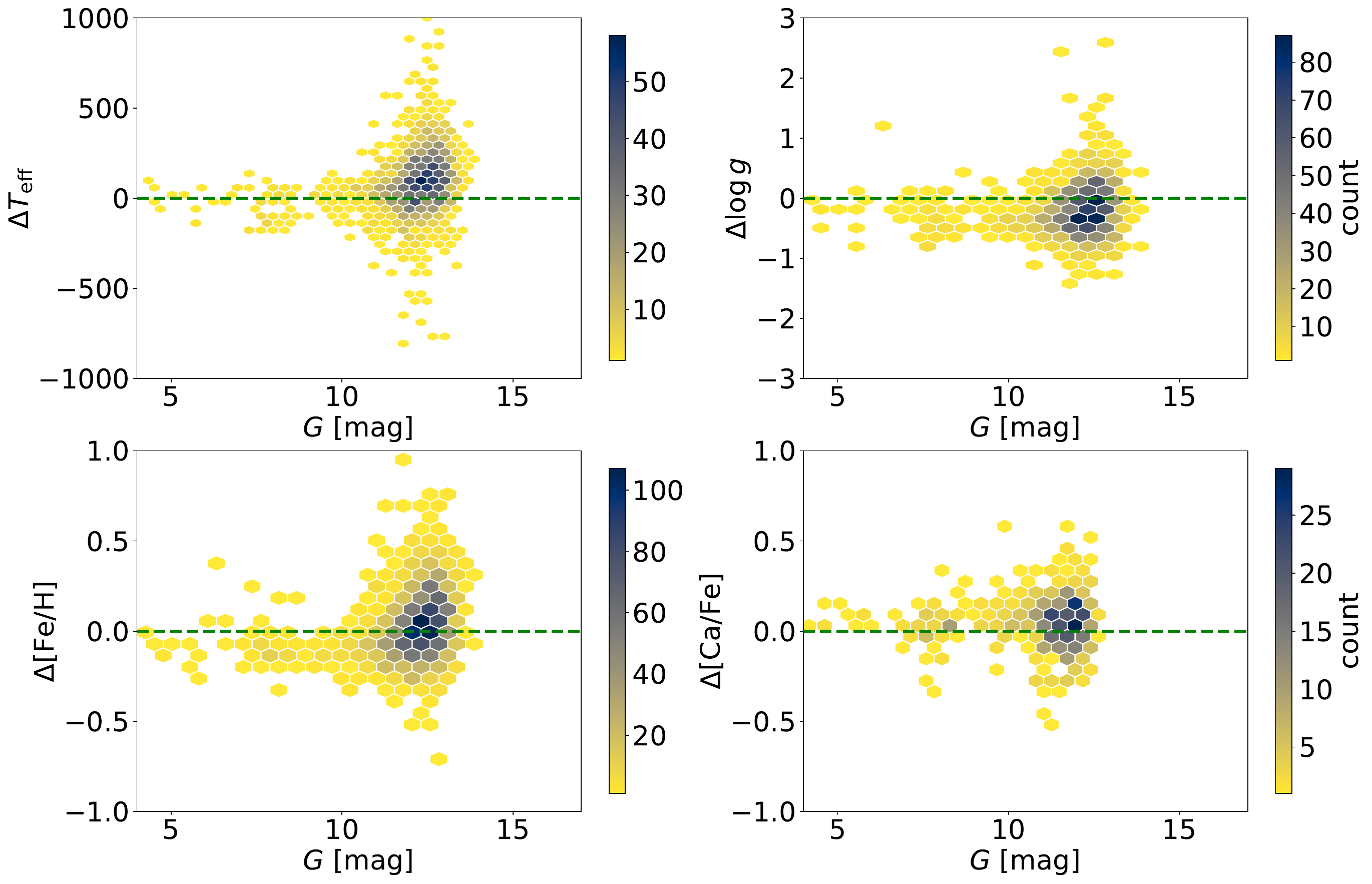}

  \caption{\label{Fig:parameter_difference_vs_G_magnitude} Difference between \Gaia and \GESlg parameters $\Delta \mathcal{P} = \mathcal{P}_{\mathrm{Gaia}} - \mathcal{P}_{\mathrm{GES}}$ as a function of the $G$ magnitude (blue dots) where $\mathcal{P}$ is, from left to right and top to bottom, $T_{\mathrm{eff}}$, $\log g$, $\abratio{Fe}{H}$, $\abratio{Ca}{Fe}$. The dashed black horizontal line has equation $\Delta \mathcal{P} = 0$.}
\end{figure*}

Figure~\ref{Fig:parameter_difference_vs_G_magnitude} investigates the dependency of the difference between \Gaia and \GESlg parameters $\Delta \mathcal{P} = \mathcal{P}_{\mathrm{Gaia}} - \mathcal{P}_{\mathrm{GES}}$ as a function of the $G$ magnitude where $\mathcal{P}$ is $T_{\mathrm{eff}}$, $\log g$ ($\log g_{\mathrm{Gaia,uncal}}$ or $\log g_{\mathrm{Gaia,cal}}$ for \gspspec), $\abratio{Fe}{H}$ ($\abratio{M}{H}_{\mathrm{Gaia}}$ or $\abratio{M}{H}_{\mathrm{Gaia,cal}}$ for \gspspec) or $\abratio{Ca}{Fe}$. Table~\ref{Tab:parameter_difference_stats} gives the mean and standard deviation of $\Delta \mathcal{P}$ for two $G$-magnitude ranges $[3.47, 11[$ and $[11, 13.87]$ as well as the $p$-value of the Kolmogorov-Smirnov two-sample test for these two subsamples. Table~\ref{Tab:parameter_difference_stats} also investigates different selections based on the setup used in \GESlg to observe a given star (high resolution UVES U580 vs. medium resolution GIRAFFE HR15N) or the evolutionary stage of a given star (dwarf vs. giant).

\begin{figure*}
  \centering
  \includegraphics[width=0.32\textwidth,clip]{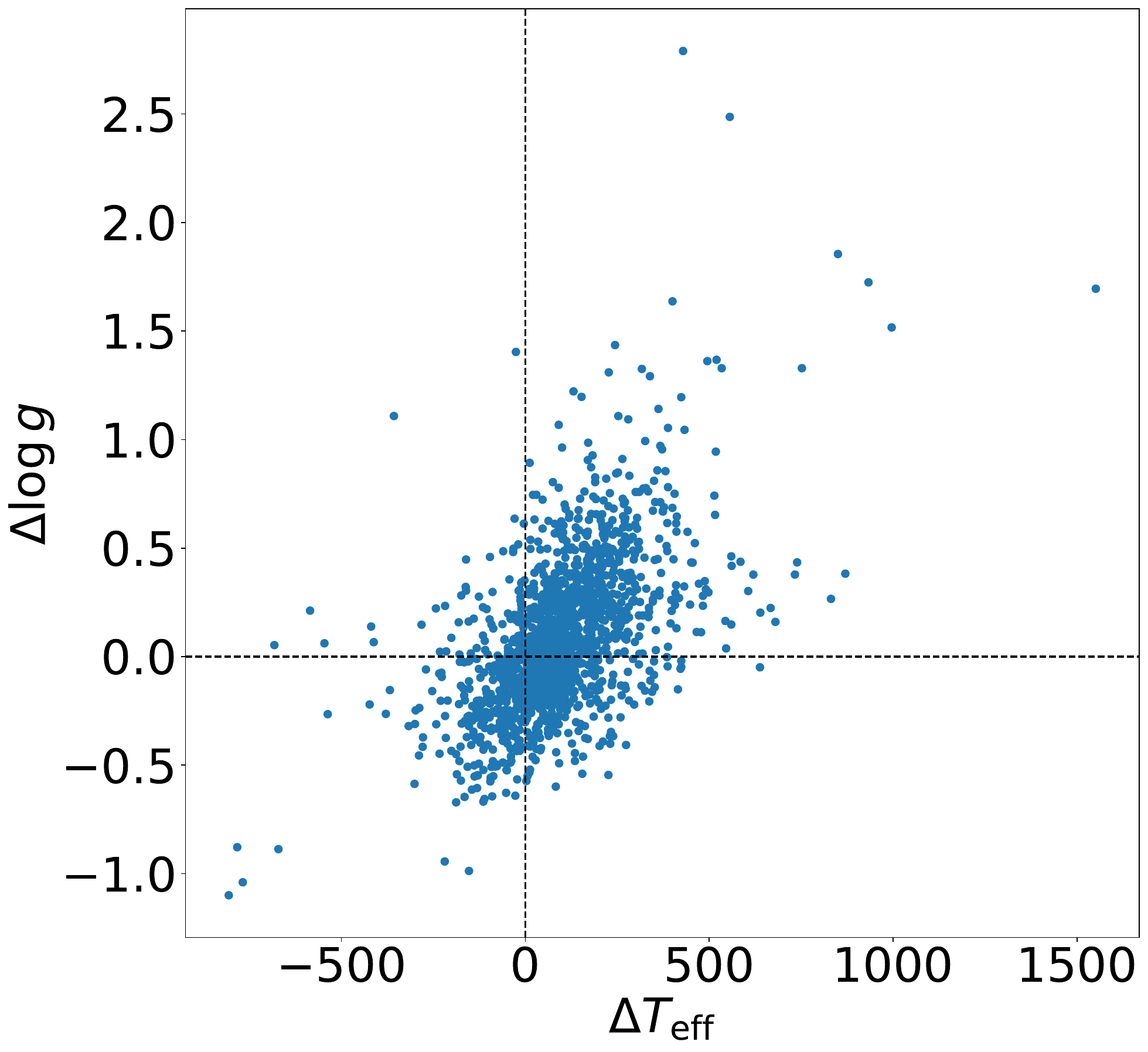}
  \includegraphics[width=0.32\textwidth,clip]{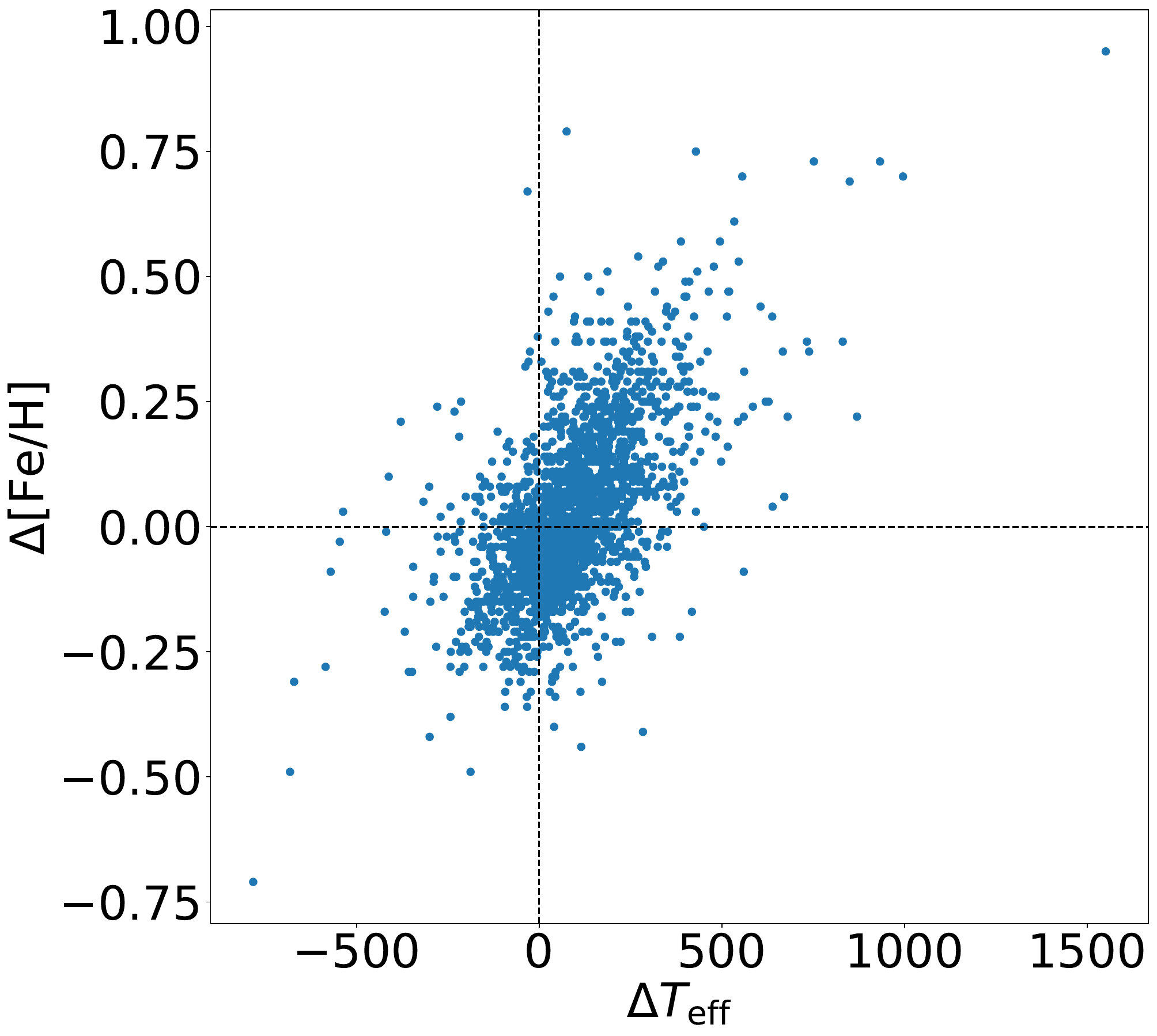}
  \includegraphics[width=0.32\textwidth,clip]{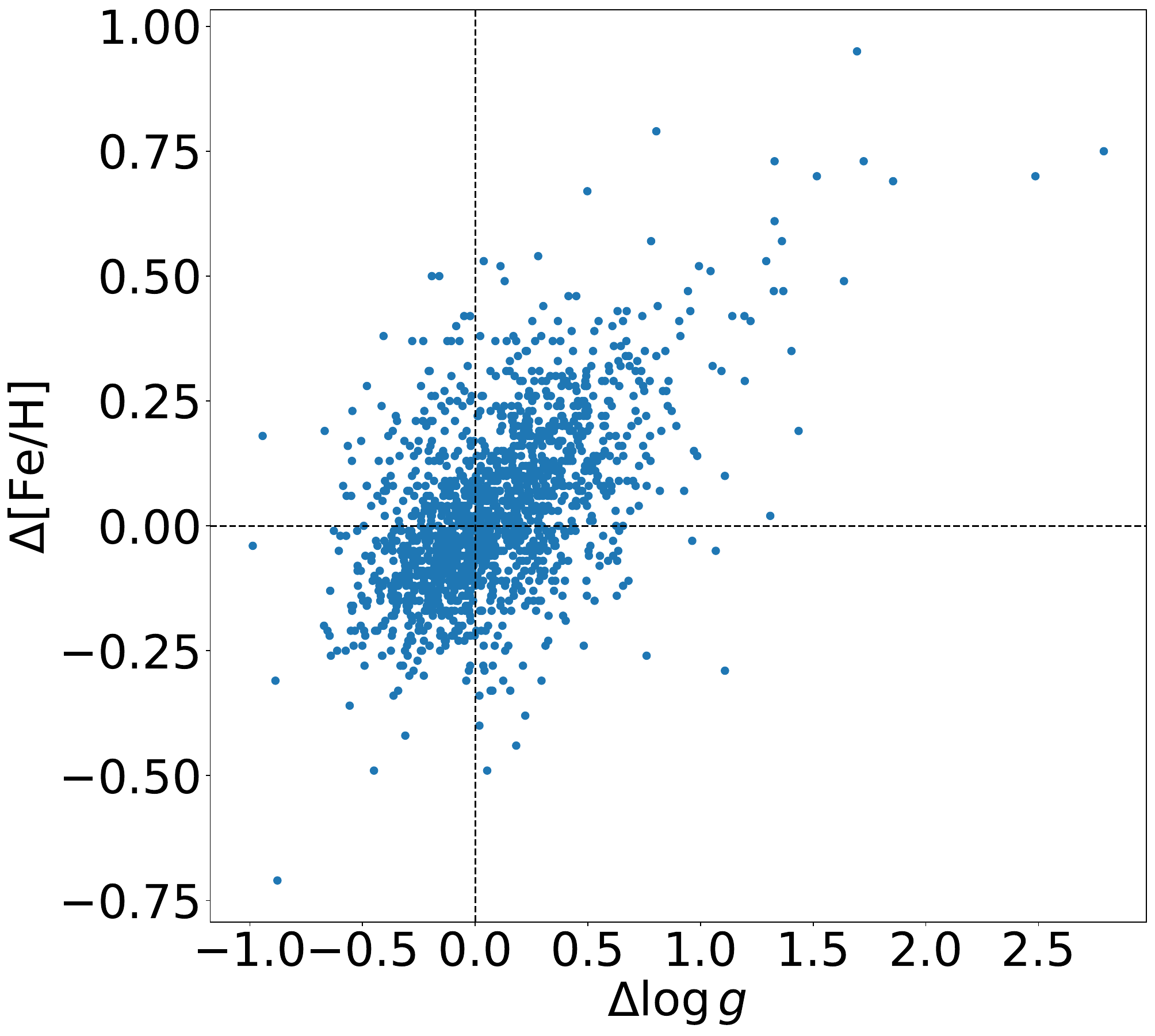}
  \caption{\label{Fig:correlation_parameter_difference} $\Delta \mathcal{P}'$ vs. $\Delta \mathcal{P}$ where $\mathcal{P}$ and $\mathcal{P}'$ are chosen among $T_{\mathrm{eff}}$, $\log g$, $\abratio{Fe}{H}$. The dashed black horizontal and vertical line have equation $\Delta \mathcal{P} = 0$ and $\Delta \mathcal{P}' = 0$. Here we use the \Gaia calibrated $\log g$ and the uncalibrated $\abratio{M}{H}$.}
\end{figure*}

As already noted in Figure~\ref{Fig:histogram_magnitude}, faint objects are more numerous than bright objects in our sample: \SI{91}{\percent} ($1885 / 2079$) of the \Gaia\,--\,\GESsh \revun{subset} intersection lie in the $G$ range $[11, 13.87]$. This is a consequence of the \GESsh selection function: \GESlg is indeed designed to sample the faint part of the \Gaia catalogue and \SI{79}{\percent} of the full \GESsh DR5.1 catalogue has a $G$ magnitude bigger than \num{15}. For the full intersection $\mathcal{S}_{3}$, we note that for the temperature, surface gravity and metallicity, the two subsamples obtained for $G < \SI{11}{\mag}$ and $G \ge \SI{11}{\mag}$ behave differently: a) the mean and standard deviation of the $\Delta T_{\mathrm{eff}}$, $\Delta \log g$ and $\Delta \abratio{Fe}{H}$ increase with $G$ (so when stars get fainter); b) often, the $p$-values are extremely small indicating that we can reject the null hypothesis that the two subsamples are drawn from the same underlying distribution. As already noted in the previous subsections, the use of the \gspspec calibrated gravity instead of the uncalibrated gravity allows us to get the centre of the distribution closer to the line $\Delta \log g = 0$ but a small offset of \num{0.08} still exists and the difference between the mean $\Delta \log g$ of the bright and the faint subsamples remains \num{0.15} in absolute value. The same behaviour is observed when we breaks the initial \Gaia\,--\,\GESsh intersection according to the \GESsh setup (U580 or HR15N) or the evolutionary stage (dwarf vs. giant). An exception may exist for the gravity in the case of dwarf stars: the different behaviour between the bright and the faint subsamples tends to vanish, in particular when we use the calibrated gravity. One knows that the determination of the atmospheric parameters is a degenerate problem, and indeed, we observe positive correlations between two $\Delta \mathcal{P}$ as shown in Fig.~\ref{Fig:correlation_parameter_difference}: in other words, $T_{\mathrm{eff}}$, $\log g$ and $\abratio{Fe}{H}$ tend to be simultaneously overestimated.

On the other hand, $\Delta \abratio{Ca}{Fe}$ shows little dependency with $G$: the distribution is flat around $\Delta \abratio{Ca}{Fe} \approx 0.05$ and we cannot reject the null hypothesis that the two subsamples corresponding to the bright and the faint ranges are drawn from the same underlying distribution. These remarks hold when we break the full intersection according to the \GESsh setups or the stellar evolutionary stage. This is a frequent observation in stellar spectroscopic studies where systematic effects tend to cancel out for abundance ratios in the form $\abratio{X}{Fe}$. Indeed, we do not find a correlation between $\Delta T_{\mathrm{eff}}$ and $\Delta \abratio{Ca}{Fe}$ or between $\Delta \log g$ and $\Delta \abratio{Ca}{Fe}$.

\begin{figure*}
  \centering
  \includegraphics[width=\textwidth,clip]{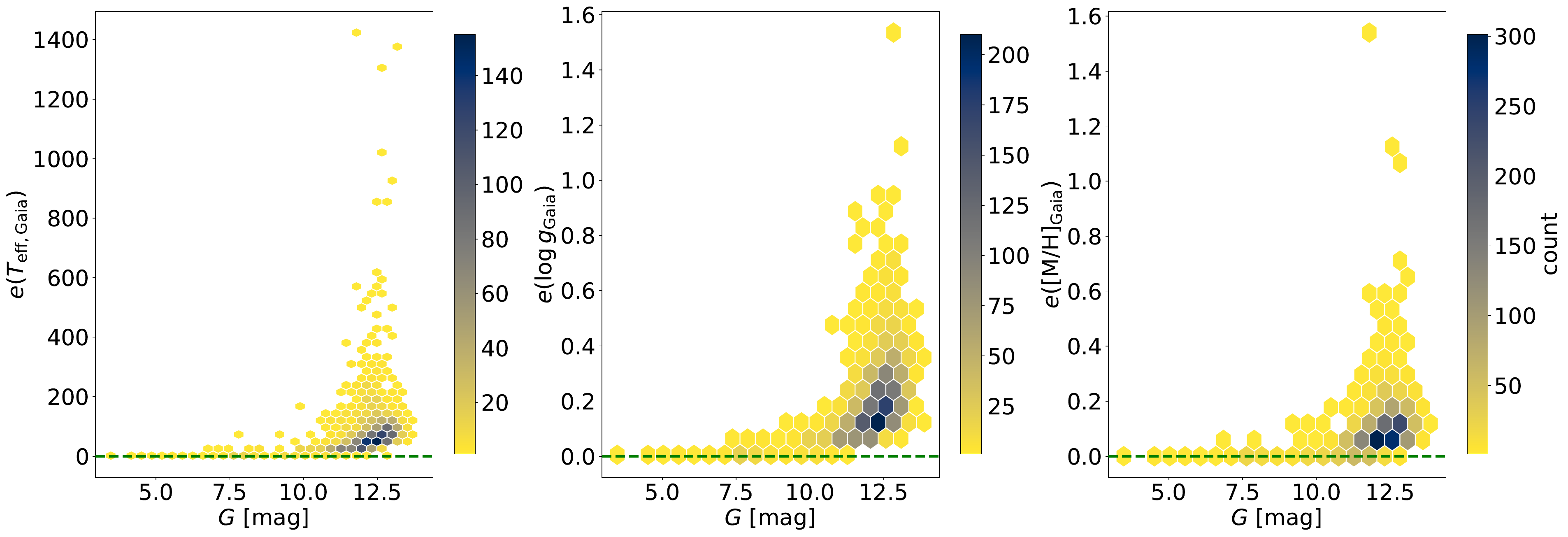}
  \caption{\label{Fig:GSPSpec_error_vs_G_magnitude} Correlation between $e(\mathcal{P}_{\mathrm{Gaia}})$ and $G$ where $\mathcal{P}$ if either $T_{\mathrm{eff}}$, $\log g$ or $\abratio{M}{H}$ (from left to right).}
\end{figure*}

The cut at $G \approx \SI{11}{\mag}$ adopted above is empirically chosen from the plots in Fig.~\ref{Fig:parameter_difference_vs_G_magnitude}: one could argue that the bright sample covers about seven magnitudes, while the faint sample covers only three magnitudes, and that the number of data-points is significantly different between the two magnitude ranges. However, we see that this bifurcation around $G \approx \SI{11}{\mag}$ is seen in Fig.~\ref{Fig:GSPSpec_error_vs_G_magnitude} displaying the change of the error on a given \gspspec parameter $e \left( \mathcal{P}_{\mathrm{Gaia}} \right)$ as a function of the $G$ magnitude. Since \gspspec gives for each quantity a lower and upper uncertainty, not necessarily symmetrical, we define $e \left( \mathcal{P}_{\mathrm{Gaia}} \right)$ as the arithmetic average of the lower and upper uncertainties. The Figure~\ref{Fig:GSPSpec_error_vs_G_magnitude} relies only on \Gaia data and does not include \GESsh data. It shows clearly that a change happens around $G \gtrapprox \SI{11}{\mag}$: $e \left( \mathcal{P}_{\mathrm{Gaia}} \right)$ becomes significantly scattered for $G \gtrapprox \SI{11}{\mag}$ compared to its typical scatter for $G \lessapprox \SI{11}{\mag}$.

\begin{figure}
  \centering
  \includegraphics[width=\columnwidth,clip]{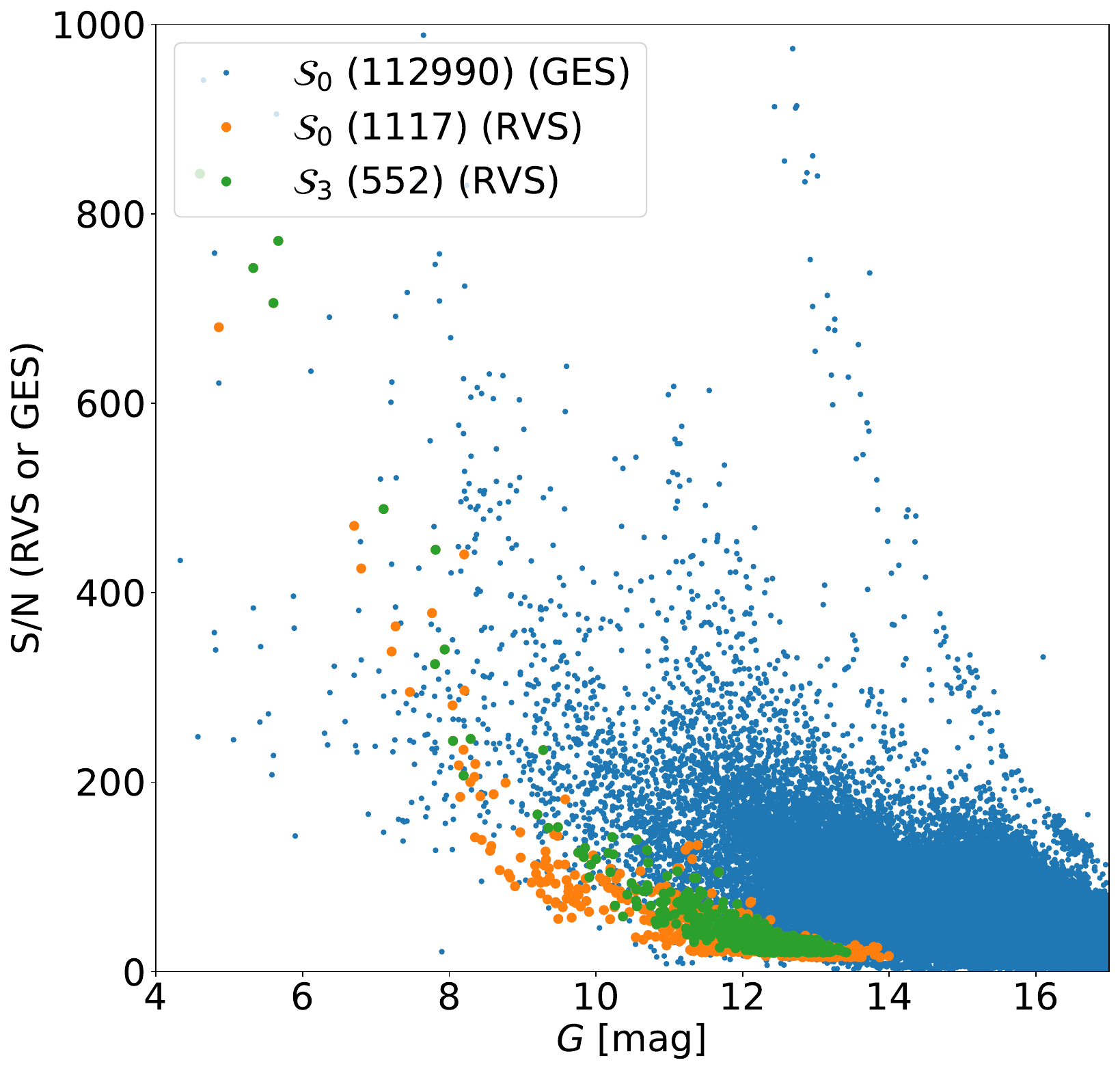}
  \caption{\label{Fig:G_magnitude_vs_snr} \revdeux{\GESsh \snr and RVS \snr as a function of $G$ for respectively \num{112990} (blue) and \num{1117} (orange) objects of $\mathcal{S}_{0}$ with a valid \GESsh (resp., RVS) \snr and RVS \snr as a function of $G$ for the \num{552} (green) objects of $\mathcal{S}_{3}$ with a valid RVS \snr.}}
\end{figure}

Unsurprisingly, as shown in Fig.~\ref{Fig:G_magnitude_vs_snr}, the $G$ magnitude and the \Gaia RVS \snr is strongly correlated. More specifically, $G$ and $\log \mathrm{SNR}_{\mathrm{RVS}}$ are linearly dependent. \revdeux{On the other hand, the relation between the $G$ magnitude and the \GESsh \snr is not straightforward since, within \GESlg, the exposure time has been adjusted depending on the magnitude regime in which the target lies; still, the \GESsh \snr tends to be higher towards lower $G$ mag.} If we combine the two \snr through, for instance, a geometric mean $\sqrt{\mathrm{SNR}_{\mathrm{GES}} \mathrm{SNR}_{\mathrm{RVS}}}$, the correlation between the two quantity remains \revdeux{tight}: the logarithm of the geometric average of the two \snr approximately varies linearly with $G$ \revdeux{(based on only \num{552} objects of $\mathcal{S}_{3}$ with published \snr for the RVS spectra)}. \revdeux{No sharp drop of} the \snr is to be noted at $G \approx \SI{11}{\mag}$, \revdeux{it still follows the relation observed at brighter regime. However, the mean RVS \snr is \num{83} in the $G$ range $[10, 11]$, \num{51} in the $G$ range $[11, 12]$ and \num{29} in the $G$ range $[12, 14]$. While a \snr of \num{30} is still enough in high-resolution spectroscopy of faint objects to estimate parameters with, for example, typical uncertainties lower than \SI{150}{\kelvin} for $T_{\mathrm{eff}}$ or lower than \SI{0.15}{\dex} for the $\alpha$ abundances, it appears that at the RVS resolution and sampling, and for the RVS wavelength window, a \snr lower than $\approx \num{50}-\num{70}$ is not enough to reach such a precision for faint objects.} Finally, we note that the \gspspec flags \texttt{flag01} to \texttt{flag13} but not \texttt{flag07} are equal to 0 for all of the \num{2079} stars in the \Gaia\,--\,\GESsh intersection. The flag 7 can be equal to 0, 1, 2 or 3 but forcing \texttt{flag07} to be equal to 0 does not make the scatter of $\Delta \mathcal{P}$ for the faint range similar to that of the bright range.

\revun{It would be more homogeneous to compare the spectroscopic quantities to $G_{\mathrm{RVS}}$ instead of to the broad-band $G$ magnitude. In the above discussion, we used $G$ since this photometric quantity is the most used in the literature. For the sake of completeness, we have checked that the conclusions are not changed when considering $G_{\mathrm{RVS}}$: the break at $G \approx 11$ translates into a break at $G_{\mathrm{RVS}} \approx 10$. Fig~\ref{Fig:GSPSpec_error_vs_Grvs_magnitude} and \ref{Fig:Grvs_magnitude_vs_snr} are similar to Fig~\ref{Fig:GSPSpec_error_vs_G_magnitude} and \ref{Fig:G_magnitude_vs_snr} but they use $G_{\mathrm{RVS}}$.} 

From the above discussion, we conclude that the statistical differences between the bright and the faint ranges are: a) not explained by a \GESsh setup that would preferentially populate one of the two magnitude range; b) not explained by a luminosity class that would preferentially populate one of the two magnitude ranges; c) not explained by a significant drop of \snr for $G \gtrapprox \SI{11}{\mag}$. However, we do note a bifurcation at $G \gtrapprox 11$, whose origin remains elusive, in the quality (accuracy, precision) of the \gspspec atmospheric parameters ($T_{\mathrm{eff}}$, $\log g$, $\abratio{M}{H}$), and which approximately corresponds to a \snr of $\num{70} \pm \num{20}$. \revdeux{This finding probably indicates that a \snr lower than $\approx 70$ at the RVS resolution, sampling and short wavelength domain is not yet enough to obtain the precision commonly seen for a \snr of $\approx \num{30}$ at higher resolution performed on wider wavelength windows.} Unfortunately, most of the \Gaia -- \GESsh intersection lie in a range of $G$ magnitudes that is unfavourable for the determination of accurate and precise \gspspec atmospheric parameters. The situation can be marginally improved with the use of the published calibrations. On the other hand, the abundance ratios in the form $\abratio{X}{Fe}$ or $\abratio{A}{B}$ (with element `B' other than hydrogen) are probably not significantly affected by this ``magnitude effect'', which means that it is possible to use the \gspspec abundances of sources fainter than $G \approx 11$. \revdeux{We show that this statement is at least true for $\alpha$ and Ca.} We note that \citet{2023A&A...674A..29R} use the criterion $\mathrm{SNR}_{\mathrm{RVS}} \ge 150$ to define their high-quality sample for $\left\{ T_{\mathrm{eff}}, \log g, \abratio{M}{H} \right\}$: it is \revdeux{comparable} to the threshold we independently find in this section.

\begin{table*}
  \centering
  \caption{\label{Tab:parameter_difference_stats} Statistical quantities for the quantities $\Delta \mathcal{P}$ where $\mathcal{P}$ is $T_{\mathrm{eff}}$, $\log g$, $\abratio{Fe}{H}$, $\abratio{Ca}{Fe}$ computed for two $G$-magnitude ranges. Columns 2, 3 and 4 (resp., 5, 6 and 7) gives the mean, standard deviation and number of stars for the first (resp., second) $G$ range. The last column gives the $p$-value of the Kolmogorov-Smirnov two-sample test. The different blocks of the table give the statistical quantities for respectively the full intersection, the intersection restricted to stars observed by \GESlg only with UVES U580 or only with GIRAFFE HR15N, the intersection restricted to giant or dwarf stars based on the \GESsh $\log g$. For the selection restricted to stars observed with HR15N, only eleven stars are found in the bright $G$ range; only the statistical quantities for the faint range are provided.}
  \begin{tabular}{lS[table-format=5.2]S[table-format=5.2]S[table-format=3.0]S[table-format=5.2]S[table-format=5.2]S[table-format=5.0]S}
    \toprule
    $\Delta \mathcal{P}$ & {mean} & {s.d.} & {\#} & {mean} & {s.d.} & {\#} & {$p$-value}\\
    \cmidrule{2-4}
    \cmidrule{5-7}
    & \multicolumn{3}{c}{$[3.47, 11[$} & \multicolumn{3}{c}{$[11, 13.87]$} & \\
    \midrule
    \midrule
    \multicolumn{8}{c}{Full intersection (2079 stars)}\\
    \midrule
    $\Delta T_{\mathrm{eff}}$ [K] & -10 & 103 & 171 & 98 & 171 & 1768 & 6.0e-26\\
    $\Delta \log g_{\mathrm{Gaia,uncal}}$ & -0.32 & 0.30 & 167 & -0.17 & 0.39 & 1408\tablefootmark{a} & 2.6e-7\\
    $\Delta \log g_{\mathrm{Gaia,cal}}$ & -0.05 & 0.26 & 167 & 0.10 & 0.36 & 1408\tablefootmark{a} & 1.0e-11\\
    $\Delta \abratio{Fe}{H}$ & -0.07 & 0.12 & 171 & 0.04 & 0.17 & 1733\tablefootmark{a} & 6.1e-23\\
    $\Delta \abratio{Ca}{Fe}$ & 0.04 & 0.13 & 157 & 0.05 & 0.15 & 346 & 0.18\\

    \midrule
    \multicolumn{8}{c}{U580 only (806 stars)}\\
    \midrule
    $\Delta T_{\mathrm{eff}}$ [K] & -12 & 106 & 137 & 104 & 160 & 669 & 2.6e-20\\
    $\Delta \log g_{\mathrm{Gaia,uncal}}$ & -0.35 & 0.27 & 137 & -0.14 & 0.38 & 669 & 6.9e-9\\
    $\Delta \log g_{\mathrm{Gaia,cal}}$ & -0.07 & 0.23 & 137 & 0.12 & 0.35 & 669 & 7.2e-13\\
    $\Delta \abratio{Fe}{H}$ & -0.07 & 0.11 & 137 & 0.04 & 0.17 & 669 & 3.0e-19\\
    $\Delta \abratio{Ca}{Fe}$ & 0.04 & 0.12 & 136 & 0.05 & 0.11 & 253 & 0.44\\
    \midrule
    \multicolumn{8}{c}{HR15N only (818 stars)}\\
    \midrule
    $\Delta T_{\mathrm{eff}}$ [K] &  &  & & 89 & 185 & 818 & \\
    $\Delta \log g_{\mathrm{Gaia,uncal}}$ & & & & -0.21 & 0.39 & 445\tablefootmark{a} & \\
    $\Delta \log g_{\mathrm{Gaia,cal}}$ &  &  & & 0.05 & 0.36 & 445\tablefootmark{a} & \\
    $\Delta \abratio{Fe}{H}$ &  &  & & 0.04 & 0.18 & 770\tablefootmark{a} & \\
    $\Delta \abratio{Ca}{Fe}$ &  &  & & 0.01 & 0.22 & 25 & \\
    \midrule
    \multicolumn{8}{c}{Giants only, based on $\log g_{\mathrm{GES}} < 3.5$ (1246 stars)}\\
    \midrule
    $\Delta T_{\mathrm{eff}}$ [K] & -5 & 87 & 117 & 107 & 145 & 1129 & 4.3e-20\\
    $\Delta \log g_{\mathrm{Gaia,uncal}}$ & -0.41 & 0.29 & 117 & -0.20 & 0.40 & 1129 & 1.1e-12\\
    $\Delta \log g_{\mathrm{Gaia,cal}}$ & -0.06 & 0.29 & 117 & 0.13 & 0.38 & 1129 & 9.3e-11\\
    $\Delta \abratio{Fe}{H}$ & -0.08 & 0.12 & 117 & 0.04 & 0.17 & 1126 & 4.9e-20\\
    $\Delta \abratio{Ca}{Fe}$ & 0.03 & 0.13 & 108 & 0.04 & 0.15 & 293 & 0.33\\
    \midrule
    \multicolumn{8}{c}{Dwarfs only, based on $\log g_{\mathrm{GES}} \ge 3.5$ (329 stars)}\\
    \midrule
    $\Delta T_{\mathrm{eff}}$ [K] & -17 & 134 & 50 & 103 & 258 & 279 & 8.4e-8\\
    $\Delta \log g_{\mathrm{Gaia,uncal}}$ & -0.12 & 0.21 & 50 & -0.09 & 0.31 & 279 & 0.06\\
    $\Delta \log g_{\mathrm{Gaia,cal}}$ & -0.03 & 0.27 & 50 & -0.02 & 0.27 & 279 & 0.12\\
    $\Delta \abratio{Fe}{H}$ & -0.05 & 0.13 & 50 & 0.07 & 0.19 & 273 & 1.6e-7\\
    $\Delta \abratio{Ca}{Fe}$ & 0.09 & 0.12 & 49 & 0.09 & 0.13 & 53 & 0.61\\
    \bottomrule
  \end{tabular}
  \tablefoot{For some stars lacking an estimate of the surface gravity $\log g$, \GESlg could derive a gravity index. Used with the effective temperature, it made possible the determination of a metallicity estimate $\abratio{Fe}{H}$. This explains why there are more stars with an estimate of $\abratio{Fe}{H}$ than with an estimate of $\log g$.}
\end{table*}

\section{Asteroseismic targets}
\label{Sec:astero}

Part of the legacy of \GESsh has been the creation of new reference sets of stellar parameters. In particular, collaborations between asteroseismology and spectroscopy aim at providing atmospheric parameters using both spectroscopic and asteroseismic data, derived iteratively and converging on $T_{\mathrm{eff}}$, $\log g$ and $\abratio{Fe}{H}$ for stars targeted by \GESsh selected from the {\em K2} and {\em CoRoT} projects. This resulted in two reference sets: 90 stars from the {\em K2@Gaia-ESO} project \citep{Worley20}, and 1599 stars from the {\em CoRoT@Gaia-ESO} project (Masseron et al, in prep). These samples were observed in \GESlg as either high resolution (hereafter, the K2 or CoRot `HR' sample) with UVES or medium resolution (hereafter, the K2 or CoRoT `MR' sample) with GIRAFFE. 

The results obtained from the HR and MR reference samples reflect the different quality, \eg the \snr, and the wavelength coverage of the spectra that was used to derive them. The MR spectra are typically obtained for fainter stars and they cover a small wavelength range, while HR spectra correspond to the brightest targets and cover a spectral range of about \SI{2000}{\angstrom}. Further discussion about the comparisons of the parameters between HR and MR for the K2 sample are given in \citet{Worley20}. Note that none of the MR K2 stars are present in \Gaia DR3.

\begin{figure*}
  \centering
  \includegraphics[width=\textwidth,clip]{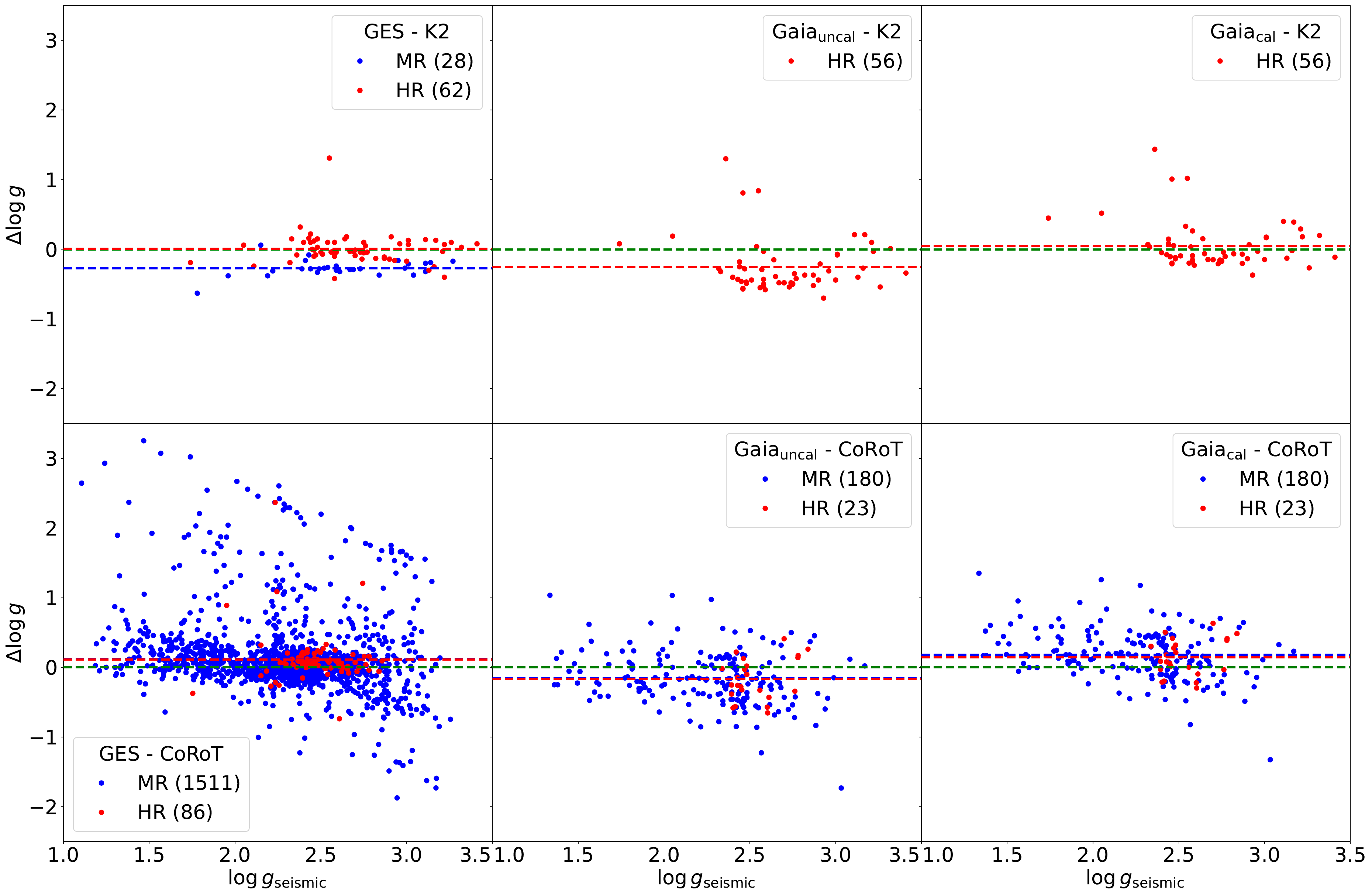}
  \caption{\label{Fig:Comparison_gravities_spec_vs_seismic} Comparison of $\log g$ values as a difference to the reference seismic values (K2 or CoRoT) for each of \GESlg, \Gaia uncalibrated and \Gaia calibrated. Top row: comparison with K2; bottom row: comparison with CoRoT. The HR (red) and MR (blue) reference samples are shown.}
\end{figure*}

Figure~\ref{Fig:Comparison_gravities_spec_vs_seismic} compares the $\log g$ for the K2 and CoRoT reference sets with the values obtained by \GESlg and the two sets of values generated by \gspspec, $\log g$ uncalibrated and $\log g$ calibrated. We note that a very good agreement is obtained between the \GESsh and seismic $\log g$, on the one hand, and between the \Gaia calibrated and seismic $\log g$ for the K2 HR sample with a mean difference less than \num{0.05} in absolute value. An offset of \num{-0.27} exists between the \GESsh and seismic $\log g$ for the K2 MR sample, while an offset of \num{-0.25} is found between the \Gaia uncalibrated and seismic $\log g$ for the K2 HR sample. An offset larger than \num{0.1} is found for the six comparisons with CoRot reference stars. For eight out of the ten comparisons shown in Fig.\ref{Fig:Comparison_gravities_spec_vs_seismic}, the standard deviation of the difference is larger than \num{0.25}.

\begin{figure*}
  \centering
  \includegraphics[width=\textwidth,clip]{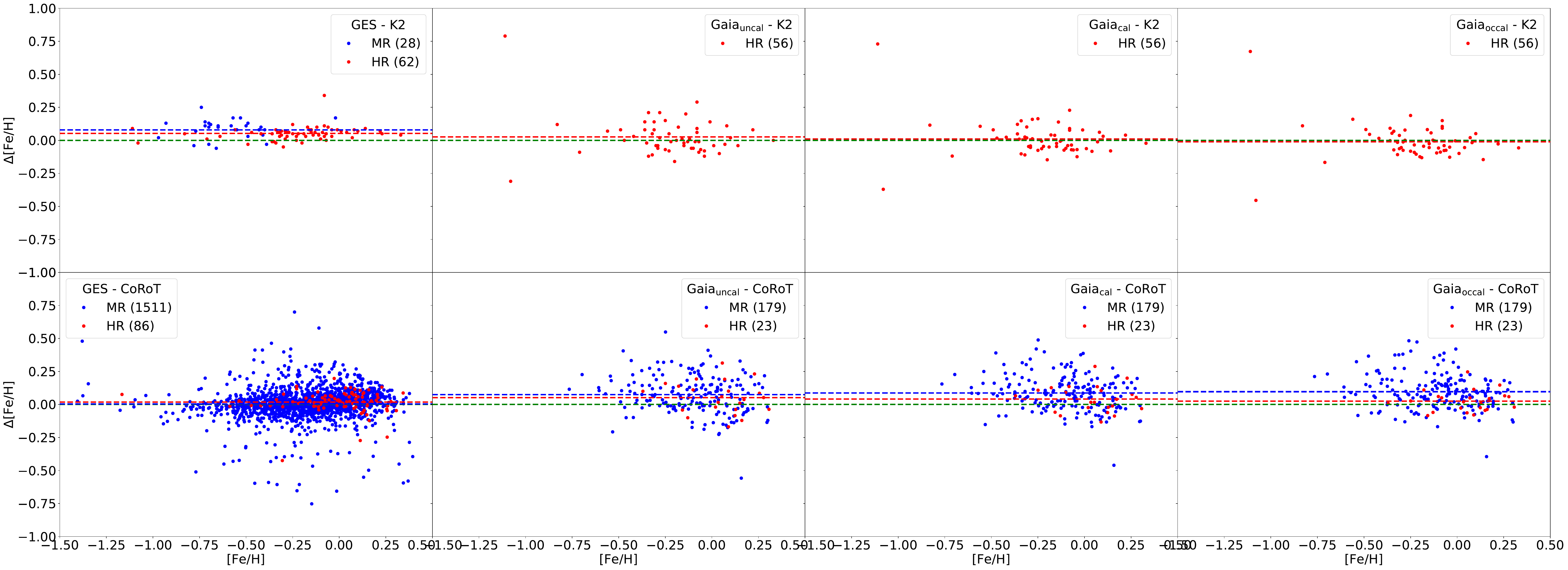}
  \caption{\label{Fig:Comparison_metallicities_spec_vs_seismic} Comparison of $\abratio{Fe}{H}$ values as a difference to the reference seismic values (K2 or CoRoT) for \GESlg, \Gaia, \Gaia calibrated, and \Gaia OC-calibrated. Top row: comparison with K2; bottom row: comparison with CoRoT. The HR (red) and MR (blue) reference samples are shown.}
\end{figure*}

Figure~\ref{Fig:Comparison_metallicities_spec_vs_seismic} shows the same kind of comparison but for the metallicity. Four estimates of $\abratio{Fe}{H}$ or $\abratio{M}{H}$ are compared to the seismic estimate, namely the \GESsh metallicity, and the \Gaia uncalibrated, calibrated and OC-calibrated ones. In all cases, the offsets are below \num{0.1} in absolute value, with a standard deviation of the difference between \num{0.05} and \num{0.15}. We can therefore conclude in a good agreement between the spectroscopic estimates of the metallicity and those based on spectral analysis adopting seismic surface gravities.  

In summary, the comparison of spectroscopic and seismic surface gravities reveals significant offsets, of different sign, for most tested estimates. At this stage, it is impossible to use the seismic data to argue in favour of one of the two \gspspec gravity scales. On the other hand, the comparison of spectroscopic and seismic metallicities let us think that all scales are more or less equivalent. This comparison shows that the multi-messenger approach to building reference sets of stellar parameters provides a useful validation of survey results, but yet there is no unanimous agreement between \Gaia data and spectroscopic data combined with asteroseismology. Further development and expansion of these sets is required.

\section{Stars in open star clusters: individual abundances and average properties}
\label{Sec:open_clusters}

\subsection{Open cluster member stars in common}

As shown in Fig.~\ref{Fig:pie_chart_ges_type}, a large fraction of stars in common between the two surveys belong to open star clusters. Among them, we have selected those which are highly probable members of clusters in order to compare both the metallicity and the abundances of individual members, as well as the average properties of the open clusters. For \GESsh, we used the membership analysis of \citet{jackson22} available for most clusters, and of \citet{viscasillas22a} for the remaining ones. In both cases, the membership probability was calculated considering, at the same time, the \GESsh radial velocities and the \Gaia proper motions and parallaxes. We cross-matched the member stars from \GESsh with the \Gaia database, finding that there are \num{136} member stars of open clusters which have \GESsh and \gspspec stellar parameters. They belong to \num{34} different open clusters. In the following analysis, we only consider member stars observed by \GESsh with the high-resolution setups, \ie observed with an UVES setup.

\begin{figure*}
  \centering
  \includegraphics[width=\textwidth,clip]{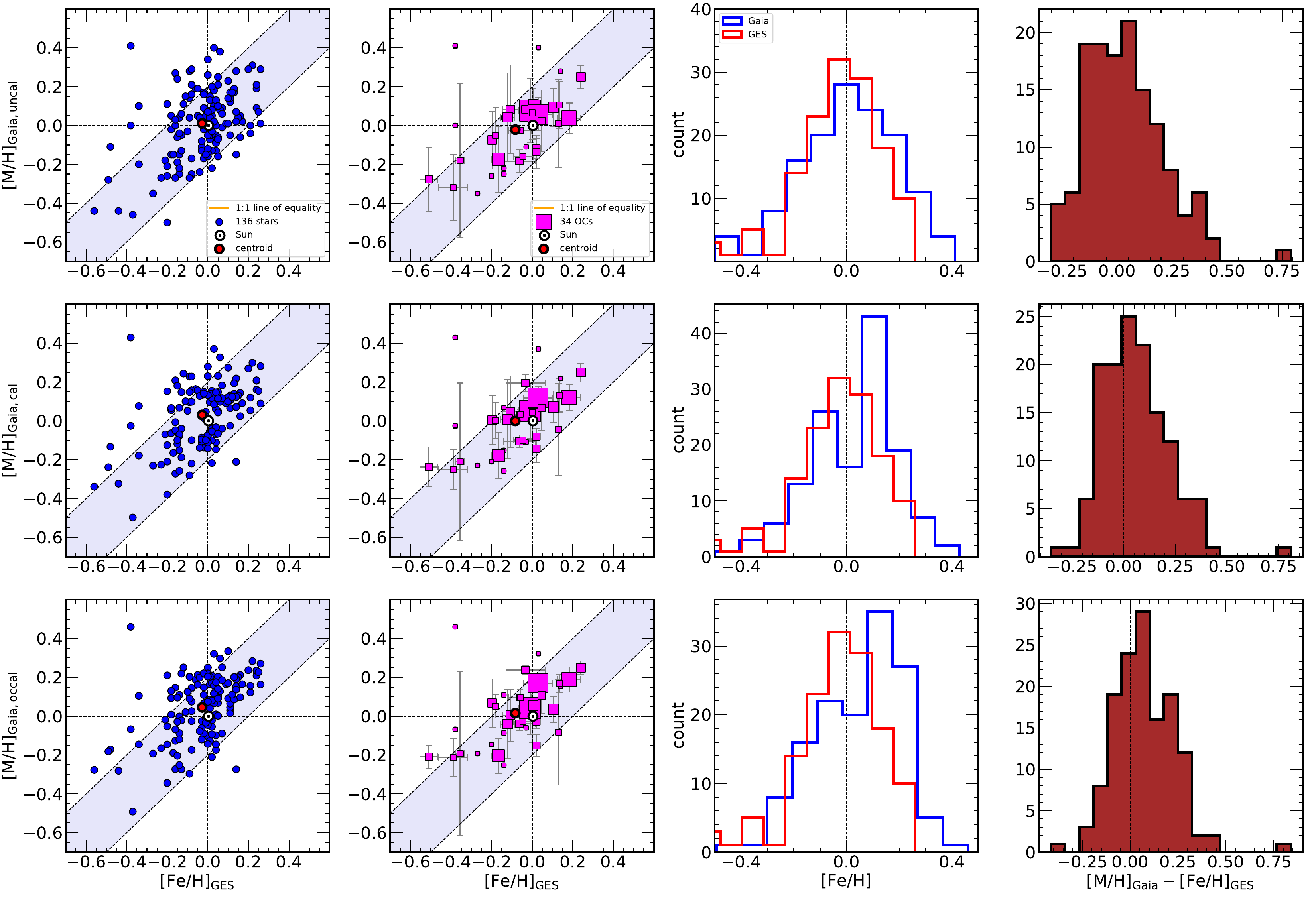}
  \caption{\label{fig:met_clusters} \GESsh vs. \Gaia metallicity (uncalibrated, calibrated, and calibrated with open clusters) for \num{136} stars belonging to \num{34} open clusters in common. The panels on the left show the member stars individually and the average metallicities per cluster, and the panels on the right show the histograms of the metallicity distributions of both samples and their difference $\Delta$. The size of the symbols is proportional to the number of members. The light-blue bands indicate an agreement between $\pm$ 0.2.}
\end{figure*}

\begin{figure}
  \centering
  \includegraphics[width=\columnwidth,clip]{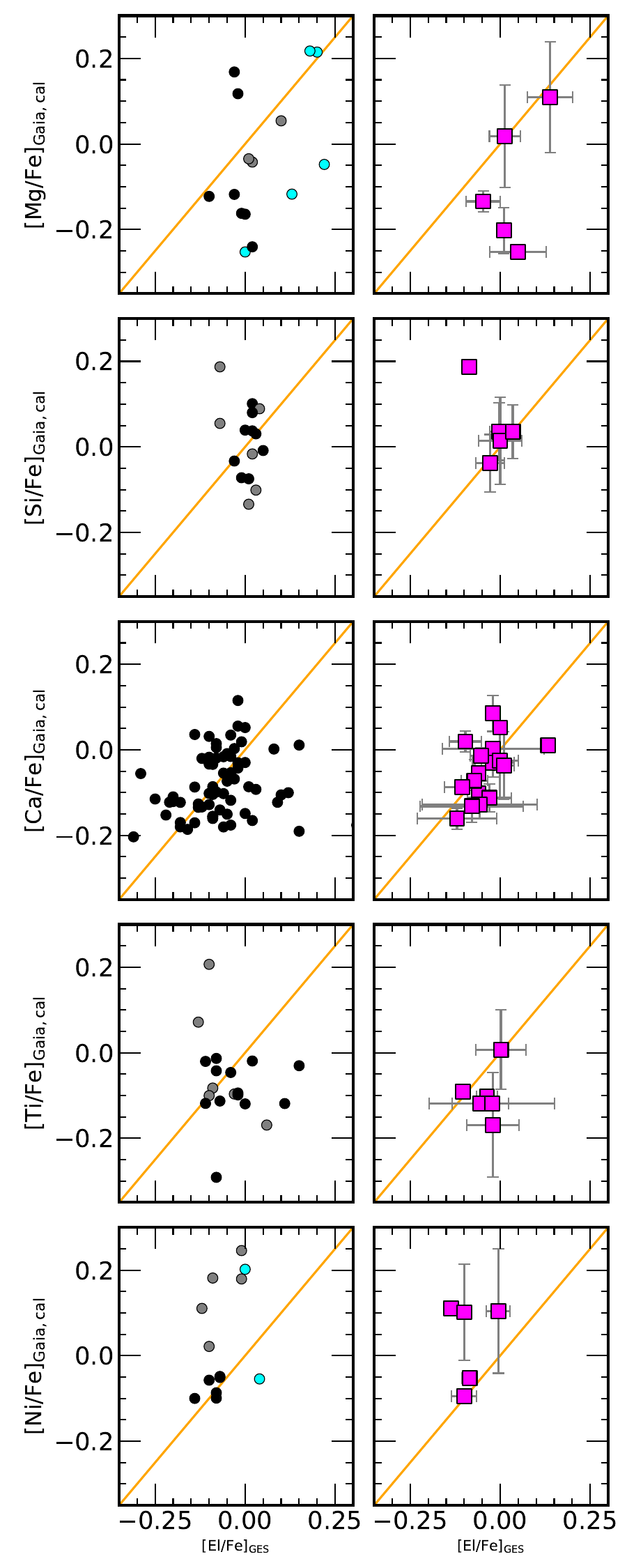}
 \caption{\label{fig:clusters_individual} Calibrated abundance ratios $\abratio{X}{Fe}$ from \Gaia versus those from \GESlg for individual members of open clusters in common (left) and their average abundance ratios (right). Each symbol represents a single member star coloured according to its quality flag (black: 0, grey: 1, cyan:2).}
\end{figure}

In Figure~\ref{fig:met_clusters} we show the Gaia metallicities $\abratio{M}{H}$ (uncalibrated, calibrated, and OC-calibrated) as a function of \GESsh $\abratio{Fe}{H}$. As seen earlier for the selection $\mathcal{S}_{3}$, we find a good agreement between the \GESsh metallicity scale and the three different \Gaia metallicity scale. The mean difference is not null but negligible given the cumulated uncertainty of the metallicity estimates. For a lower number of stars, which varies from element to element, we also have some individual elemental abundances. They are shown in Fig.~\ref{fig:clusters_individual} \revun{for Mg, Si, Ca, Ti and Ni} in which both abundances of individual member stars and averaged values per clusters are shown. The agreement between \GESsh and \Gaia abundances is in general difficult to judge given the low statistics \revun{per cluster}. \revun{We note that in general keeping individual measurements with a \gspspec abundance quality flag set to 0 (best case) removes most of the discordant values; this is not true for Ti where this filtering is not enough to reduce the scatter. If one looks at the per-cluster averaged quantities (right column), then the agreement for Ca is good. This shows that the \revdeux{somewhat imprecise} \gspspec abundances due to its medium-resolution spectroscopy, and the faint $G$ regime of the current sample can be counterbalanced by working with averaged quantities. Stellar clusters are an example of science case where averaging abundances is suitable.}

\subsection{Average metallicity and $\abratio{\alpha}{\mathrm{Fe}}$ of clusters in common}

\begin{figure*}
  \centering
 \includegraphics[width=\textwidth,clip]{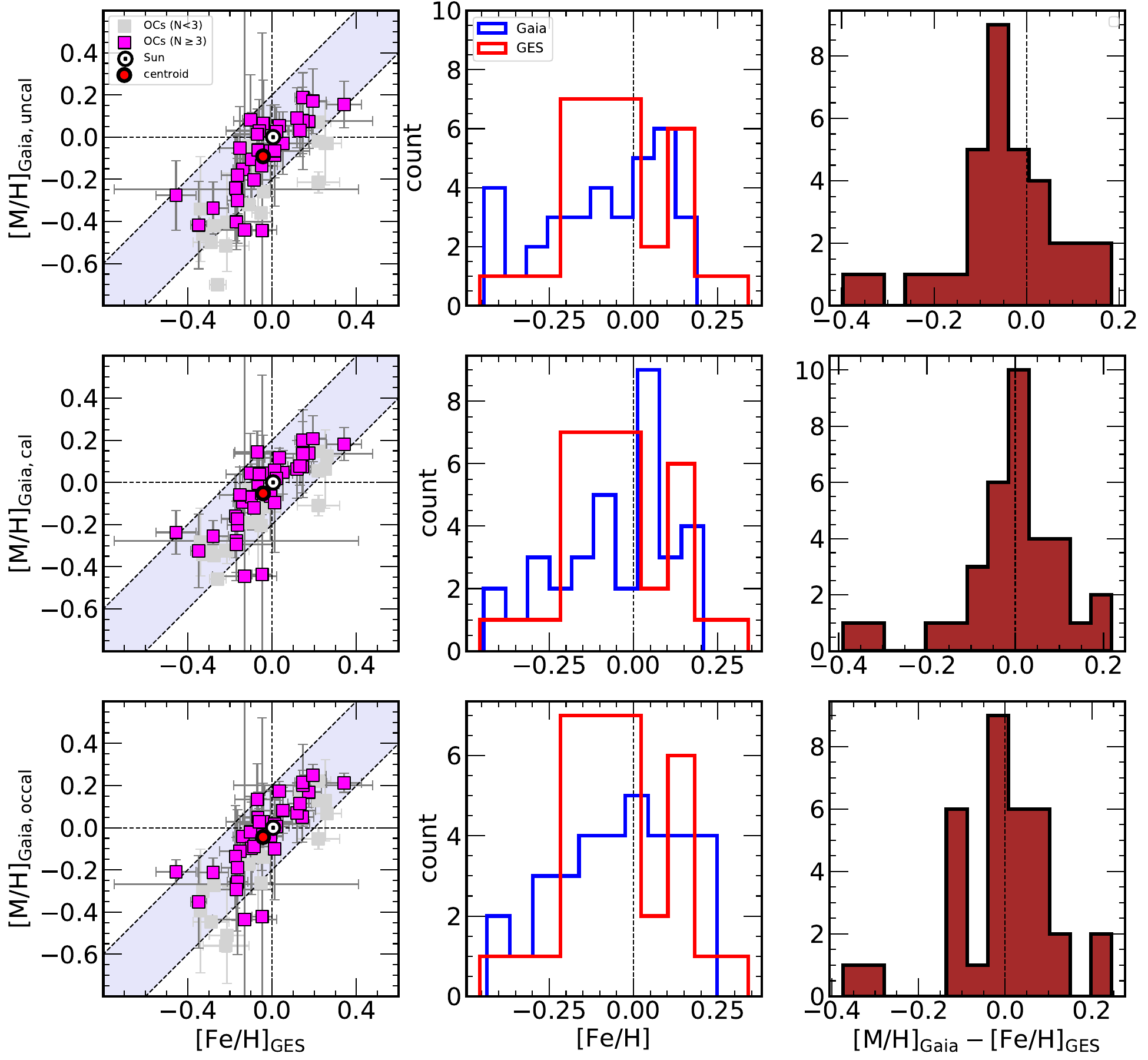}
 \caption{\label{fig:met_clusters_52} \GESsh vs. \Gaia metallicity (uncalibrated in the upper panel, calibrated in the central panel and OC-calibrated in the bottom panel) for \num{52} open clusters in common: \num{34} clusters with \num{3} or more members for both spectroscopic surveys are shown in magenta, while the \num{18} clusters with less than \num{3} members in one of the two surveys are in light grey. Left: \Gaia vs. \GESsh average cluster metallicities; centre: metallicity distributions of the clusters as seen in \GESsh (red) and in \Gaia (blue); right: distribution of the averaged-metallicity differences $\Delta \abratio{Fe}{H} = \abratio{Fe}{H}_{\mathrm{Gaia}} - \abratio{Fe}{H}_{\mathrm{GES}}$.}
 \end{figure*}

In this section, we compare the abundance properties of open clusters in common between \GESsh and \Gaia. To compute the average metallicity and $\abratio{\alpha}{\mathrm{Fe}}$ of clusters observed by \Gaia, we performed the membership analysis starting from the \num{2681} clusters (\num{270487} stars) catalogued in \citet{CG20}. The membership probability was derived using the parallax and proper motion distributions, and then validated using sky-coordinates. We obtained \num{7823} members (\num{1334} OCs) which have \Gaia \gspspec stellar parameters. We selected the stars belonging to the high- and medium-quality samples defined as a combination of \Gaia flags in \citet{gaia2022} and also adopted in \citet{Viscasillas2023A&A...679A.122V}, which reduces the sample to \num{3718} stars (\num{998} OCs). For this test, we select only clusters older than \SI{0.1}{\giga\Year} to avoid problems related to the uncertain cluster membership and to the spectral analysis of young stars (\eg, activity hampering the determination of photospheric abundances). Among the \num{998} OCs, there are \num{52} clusters in common with the \num{62} clusters of \GESlg with age $\ge \SI{0.1}{\giga\Year}$ \citep[see, \eg][]{Magrini2023A&A...669A.119M}. Figure~\ref{fig:met_clusters_52} shows the average metallicities of the sample of \num{52} clusters, computed using for each open cluster \revdeux{the metallicity of} all available member stars in the \GESsh catalogue or in the \Gaia catalogue. In other words, now, we compare the \GESsh and \Gaia average metallicity of given open cluster, computed with a different selection of member stars. In particular, for the \num{52} clusters, we found \num{624} members in \Gaia and \num{679} in \GESlg. We note that the agreement between the \GESsh metallicity scale and the three \gspspec metallicity scales is rather good, and slightly better (smaller bias) when considering the calibrated metallicities. We also examined whether the number of members used to calculate the average metallicity could influence the scatter of the distributions. Clusters with fewer than three members (\num{18} out of \num{52}) are displayed in the background in light gray and are indeed the reason for the scatter around the 1-to-1 line. Among clusters with at least three member stars, there are only three clusters with $\left| \Delta \abratio{Fe}{H} \right| > 0.2$: NGC~6281 and NGC~2516, whose abundance are underestimated by \Gaia compared to \GESlg and NGC~2243, whose abundance is instead overestimated. The above findings are compatible with our remarks in the previous subsections: the computation of \gspspec average abundances gives values well correlated with averages obtained with a higher-resolution survey like \GESlg. \revdeux{In the three clusters with the most significant differences in metallicity, we note: NGC\,2516 has 25 stars in \Gaia and 16 in \GESsh, in which all of the latter are MS stars, as expected for a young cluster; while the former sample contains sources with temperatures from about \num{4000} to \SI{8000}{\kelvin}, and gravities between \num{0.6} and \num{5}. Thus, some of them must be contaminants. Furthermore, since NGC\,2516 is young and close, we do not expect such low abundances as those measured by \Gaia. For NGC\,6821, we have data on three stars from \GESsh and eight stars from \Gaia. The stars observed by \Gaia have $T_{\mathrm{eff}}$ reaching up to about \SI{8000}{\kelvin}, a range in which metallicities are often underestimated, which may explain why \Gaia abundances are lower. Finally, for NGC\,2243 we compare \num{19} stars in \GESsh (both MS and giants) vs. three in \Gaia, which are all giants. The abundances in \GESsh are in better agreement with literature values \citep{francois13}.}

\begin{figure*}
  \centering
  \includegraphics[width=\textwidth,clip]{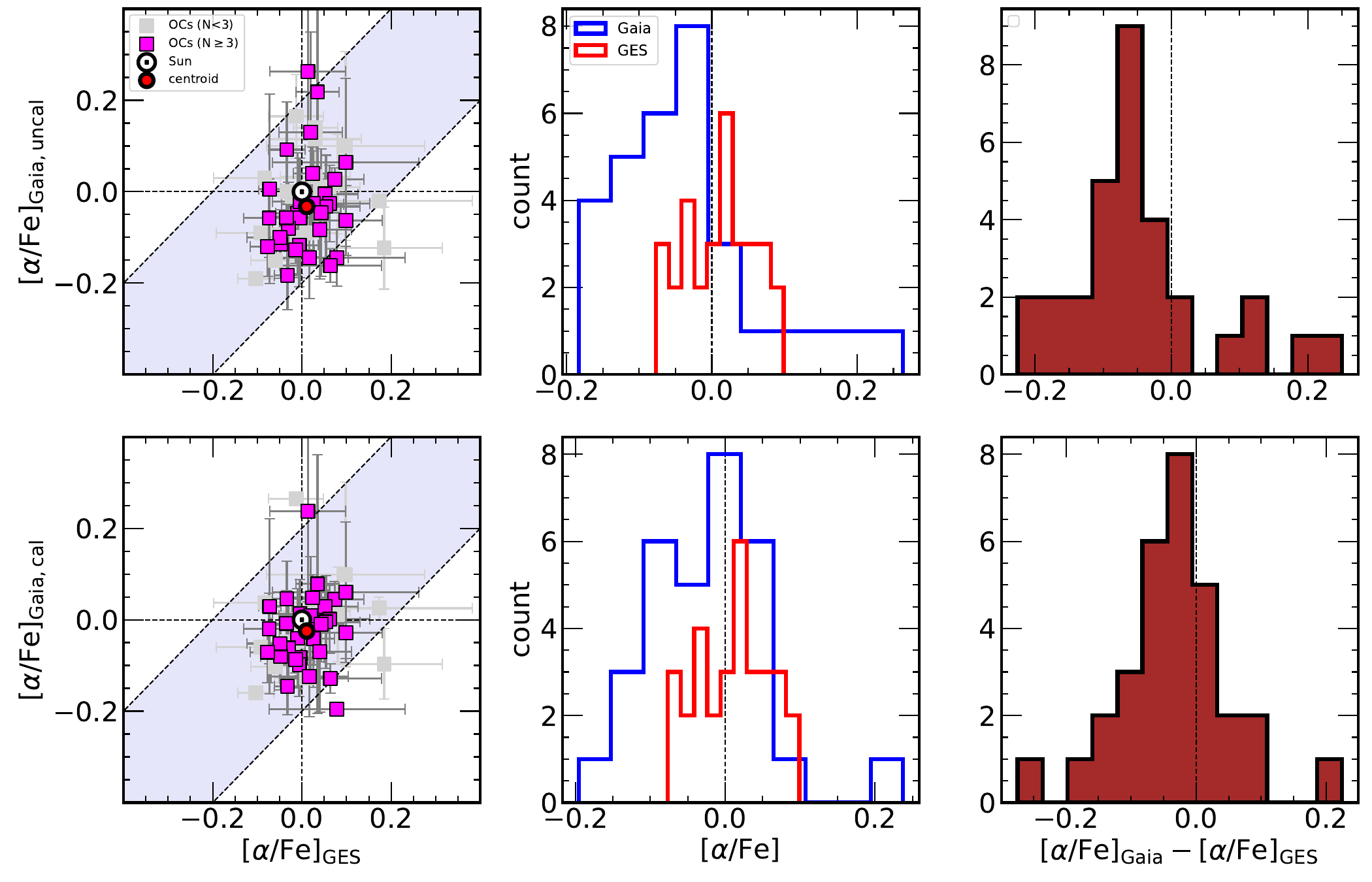}
 \caption{\label{fig:alphafe_clusters_52} \GESsh vs. \Gaia $\abratio{\alpha}{Fe}$ (uncalibrated, calibrated) for \num{52} open clusters in common. Left: average $\abratio{\alpha}{Fe}$ per cluster; centre: histograms of the $\abratio{\alpha}{Fe}$ distributions of both samples; right: histograms of the difference $\Delta \abratio{\alpha}{Fe}$.} 
\end{figure*}

Figure~\ref{fig:alphafe_clusters_52} shows the comparison $\abratio{\alpha}{Fe}$ for the \num{52} open clusters. The values are located in the range $[-0.2, 0.2]$. Given the uncertainties on each mean $\abratio{\alpha}{Fe}$, we note a good agreement between the \GESsh abundance scale and the two \Gaia ones (uncalibrated and calibrated). Nevertheless, the offset is closer to zero when the \GESsh quantities are compared to the \Gaia calibrated ones rather than to the uncalibrated. We identify four clusters in strong disagreement: Tombaugh~2, Blanco~1, NGC~6067 and NGC~6404. \revdeux{To estimate the quality of the \Gaia data, we used the HQ (High Quality) and MQ (Medium Quality) indicators, which are  derived from a combination of \Gaia \gspspec flags and defined in \citet[][see their Appendix B for a complete definition of the ranges of the used \gspspec flags to produce the HQ and MQ samples]{gaia2022}. The MQ sample defined in \citet{gaia2022} contains about $\sim \num{41000000}$ stars with median uncertainty in $\abratio{M}{H}$ of about \SI{0.06}{\dex} and median uncertainty in $\abratio{\alpha}{Fe}$ of about \SI{0.04}{\dex}, while the HQ sample stars ($\sim \num{2200000}$) have with very low parameter uncertainties, in particular a median uncertainty in $\abratio{M}{H} \sim \SI{0.03}{\dex}$ and $\sim \SI{0.015}{\dex}$ in $\abratio{\alpha}{Fe}$.}

For Tombaugh~2, only two member stars with MQ or HQ=1 have an uncalibrated $\abratio{\alpha}{Fe}$ and only one member star has a calibrated $\abratio{\alpha}{Fe}$, while nine member stars have a \GESsh $\abratio{\alpha}{Fe}$. These low statistics is likely the reason of the disagreement. We remind that Tombaugh~2 is one of the most distant known clusters \citep[see, \eg][]{Frinchaboy08}. In Blanco~1, there are six member stars with \Gaia parameters (HQ and/or MQ=1). The average $\abratio{\alpha}{Fe}$ are very similar for the uncalibrated and the calibrated cases, \SI[separate-uncertainty=true]{0.26 \pm 0.14}{\dex} and \SI[separate-uncertainty=true]{0.24 \pm 0.17}{\dex}, respectively, and \SI[separate-uncertainty=true]{0.25 \pm 0.13}{\dex} for $\abratio{Ca}{Fe}$. \GESsh observed \num{16} member stars in Blanco~1, with an average $\abratio{\alpha}{Fe} = \SI[separate-uncertainty=true]{0.015 \pm 0.069}{\dex}$. Therefore, the difference is larger than $3\sigma$. NGC~6067 has eight member stars with \Gaia (HQ and/or MQ=1) $\alpha$ abundance, to be compared to 13 members in \GESsh. The average values of $\abratio{\alpha}{Fe}$ are \SI{-0.14}{\dex} and \SI{-0.19}{\dex} for \Gaia uncalibrated and calibrated, respectively, and \SI{0.08}{\dex} for \GESsh. The agreement of \Gaia $\abratio{\alpha}{Fe}$ with \GESsh $\abratio{Ca}{Fe} = \SI{-0.08}{\dex}$ is better. NGC~6404 has three member stars with a \Gaia (HQ and/or MQ=1) $\alpha$ abundance and four member stars in \GESsh. The average values of $\abratio{\alpha}{Fe}$ are \SI{-0.12}{\dex} (uncalibrated) and \SI{-0.10}{\dex} (calibrated) for \Gaia versus \revun{\SI{0.18}{\dex} for \GESsh}; \revun{the disagreement is not solved (but less drastic) if we compare these \gspspec abundances to \GESsh Ca since \GESsh gives $\abratio{Ca}{Fe} = 0$}. In addition, for three of the four clusters considered (excluding Blanco~1), the observed stars are low-gravity giants. This could lead to a bias introduced by the standard spectroscopic analysis \revdeux{in which usually 1D atmospheric models in Local Thermodynamical equilibrium (LTE) approximation are adopted. These approximations are often not adequate to analyse giant and/or low metallicity stars \citep[see, e.g.][]{casali20b,  Magrini2023A&A...669A.119M}.} The only exception where the discrepancy is not easy to justify is the case of Blanco~1, since both in \GESsh and \Gaia stars with $\log g$ around \num{4} are observed. However, Blanco~1 is a very young cluster, with an age $\sim \SI{0.1}{\giga\Year}$, \ie the lower bound age of our selection. Very young stars are known to present difficulties in their analysis \citet{baratella20, baratella21, spina20}, which may explain the differences in results. As in Figure~\ref{fig:met_clusters_52}, clusters with less than 3 members in both samples (\num{21} out of \num{52}) are in light grey. Their exclusion reduces, as for metallicity, the scatter around the 1-to-1 line.

\begin{figure*}
  \centering
  \includegraphics[width=\textwidth,clip]{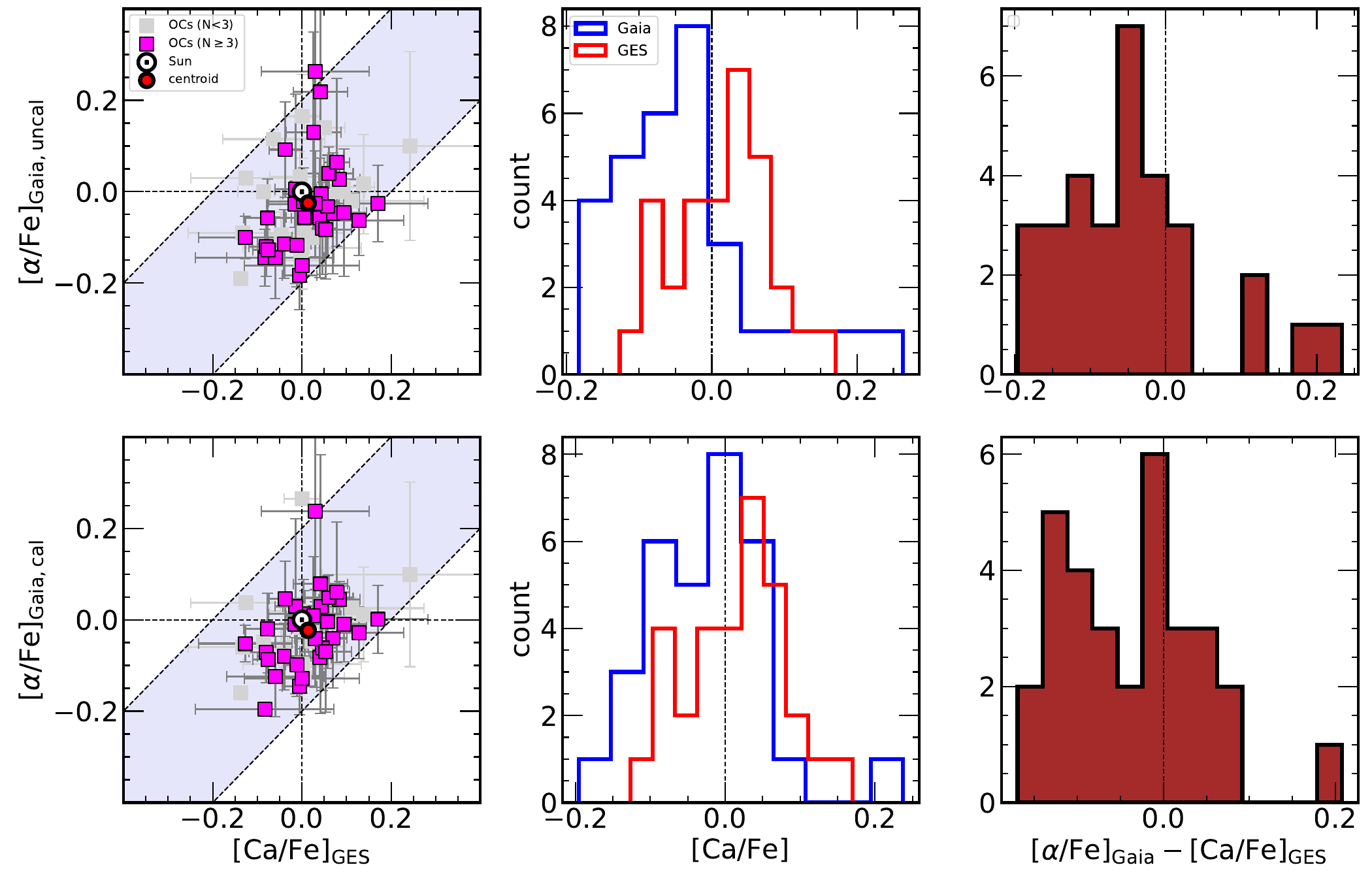}
 \caption{\label{fig:cafe_clusters_52} \GESsh $\abratio{Ca}{Fe}$ vs. \Gaia $\abratio{\alpha}{Fe}$ (uncalibrated, calibrated) for \num{52} open clusters in common. Similar panels as in Fig.~\ref{fig:alphafe_clusters_52}.} 
\end{figure*}

Considering that Ca is the largest contributor to the \Gaia $\abratio{\alpha}{Fe}$ estimate, we now compare the average \Gaia $\abratio{\alpha}{Fe}$ to the average \GESsh $\abratio{Ca}{Fe}$ in Fig.~\ref{fig:cafe_clusters_52}. The overall agreement is also satisfactory. However, Tombaugh~2 and Blanco~1 remain $\alpha$-enhanced in \Gaia compared to \GESsh. The same argument (low gravity stars) as in the previous paragraph can be proposed as the source of the abundance discrepancy.

\section{The properties of the open cluster population from the combined \Gaia-\GESsh datasets}
\label{Sec:science_with_ocs}

\revdeux{The purpose of this section is to illustrate the scientific potential of \Gaia, and the possibility of combining \Gaia RVS results with those of ground-based spectroscopic surveys, with appropriate choices in stellar parameters and abundances.} Given the agreement between the \GESsh and \Gaia abundance ratios, in this section we check what kind of general properties of the open cluster population can be retrieved from the \Gaia abundances, such as the radial metallicity and $\abratio{\alpha}{Fe}$ gradients, the trend of $\abratio{\alpha}{Fe}$ as a function of metallicity, and the age-metallicity relation. {In our analysis, w}e use $\abratio{Ca}{Fe}$ as a proxy for $\abratio{\alpha}{Fe}$ for both \Gaia and \GESsh. We limit our sample to clusters with ages between \num{1} and \SI{3}{\giga\Year} to avoid the problems related to the analysis of the youngest stars and the issue of migration for the oldest clusters as discussed in \citet{Magrini2023A&A...669A.119M}. For the open clusters observed with \Gaia we consider only clusters with at least three member stars.

\revdeux{Open clusters can, indeed, be considered among the best tracers of the chemical properties of the thin-disc stellar populations of our Galaxy, including the spatial distribution of elemental abundances. Since about five decades ago, many works have exploited the use of open clusters to trace radial metallicity and abundance gradients \citep[see e.g.][among many works]{mayor76, janes79, janes88,  friel93, carraro94, friel95, twarog97, friel02, donor20, spina21, Magrini2023A&A...669A.119M, Joshi2024FrASS..1148321J}. The strength of open clusters -- precise ages and distances -- is, indeed, maximised by spectroscopic observations at medium and high spectral resolution. For this reason, many clusters have been observed by large spectroscopic surveys, such as APOGEE, GALAH and \GESsh. An even larger number of open clusters will be spectroscopically observed in the forthcoming years thanks to instruments dedicated to spectroscopic surveys, such as WEAVE \citep[][]{weave} and 4MOST \citep[][]{2019Msngr.175....3D}. A recent review of the state of the art of the radial metallicity gradient obtained with open clusters from the three main current high-resolution spectroscopic surveys (\GESsh, APOGEE, and GALAH) has been presented in \citet{spina22}.}

\begin{figure*}
  \centering
  \includegraphics[width=\textwidth,clip]{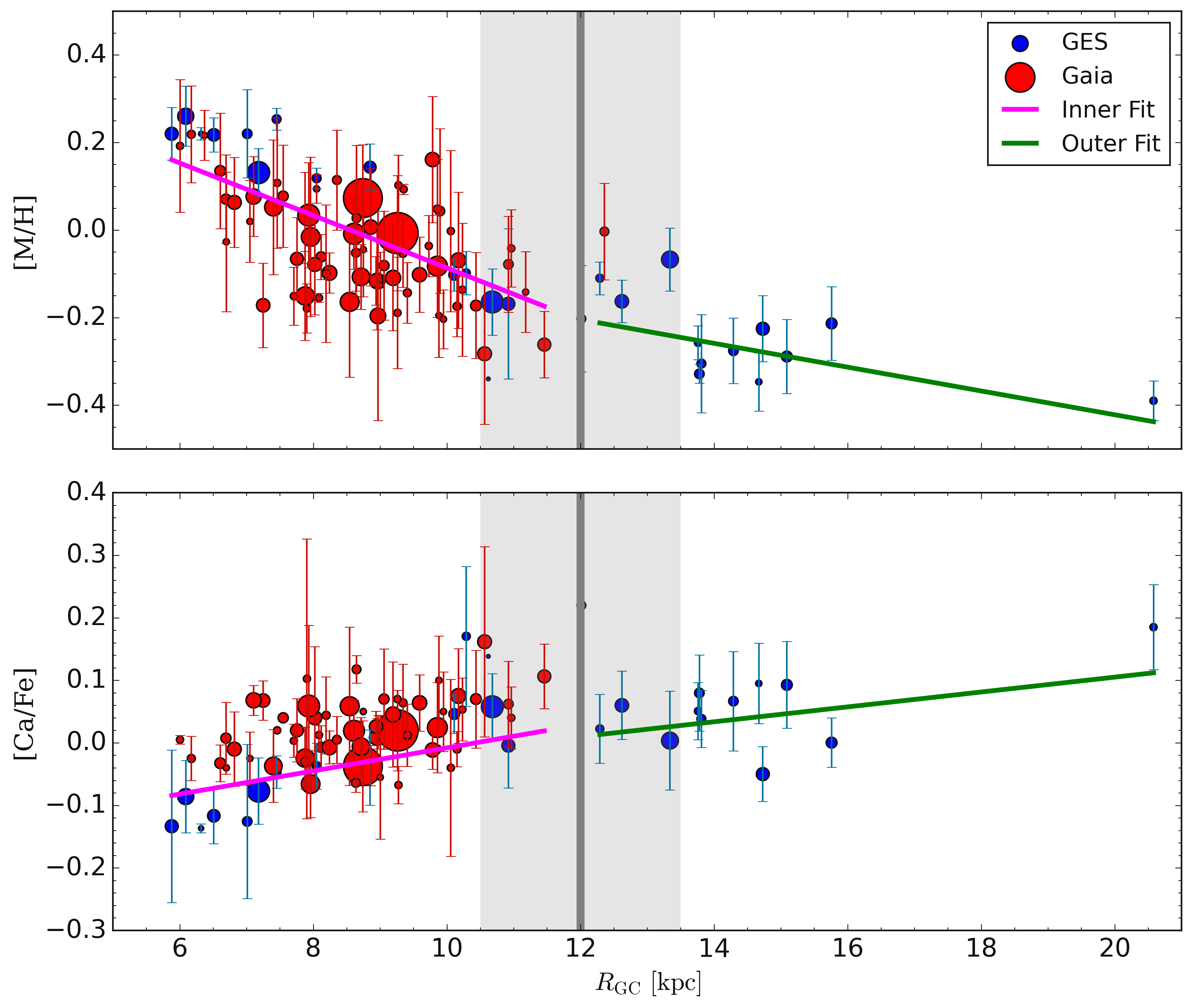}
  \caption{\label{fig:global trends} Metallicity and $\abratio{Ca}{Fe}$ as a function of the Galactocentric radius for the sample of \Gaia and \GESsh open clusters with $\SI{1}{\giga\Year} < \mathrm{age} < \SI{3}{\giga\Year}$. In the upper panel, we show $\abratio{M}{H}$ (calibrated metallicity for \Gaia and $\abratio{Fe}{H}$ for \GESsh) as a function of $R_{\mathrm{GC}}$, while in the bottom panel, we present $\abratio{Ca}{Fe}$ versus $R_{\mathrm{GC}}$. {The weighted linear fits are shown with continuous lines.  The vertical line mark the location of the {\sc elbow}. The size of the symbols is proportional of the number of member stars used to compute the mean values. }}
\end{figure*}

In Figure~\ref{fig:global trends}, we compare the overall trends of the open cluster population observed by \Gaia and by \GESsh. The open clusters observed by \Gaia are located between $\sim 6$ and \SI{12.5}{\kilo\parsec}, while those observed by \GESlg reach further distances. The two samples continuously follow the same decreasing radial \revdeux{metallicity} gradient, and the same increasing trend of $\abratio{Ca}{Fe}$ with $R_{\mathrm{GC}}$. \revdeux{The shape of the metallicity gradient in Fig.~\ref{fig:global trends} shows a bimodal distribution, with a \emph{knee} -- the radius at which there is a change of slope in the gradient from steep to almost flat --  located at around \num{11}-\SI{12}{\kilo\parsec}. The shape of the gradient represents an important observational constraint for defining the timescales of the formation of the Galactic thin disc, the radial variations and efficiency of the star formation rate and of the balance between gas inflow and outflow. Several works confirmed the presence of the gradient break located between \num{10} to \SI{12}{\kilo\parsec} \citep[e.g.][]{bragaglia08, sestito08a, friel10, pancino10, carrera11, yong12, frin13, reddy16, magrini17, casamiquela19, donor20, ZhangLucatello21, netopil22, spina22, myers22, Magrini2023A&A...669A.119M, Carbajo2024arXiv240500110C}. We computed a weighted linear fit to our combined dataset, with a two-slope function. We obtained a slope $\num{-0.060} \pm \SI{0.013}{\dex\per\kilo\parsec}$ in the inner region and $\num{-0.027} \pm \SI{0.011}{\dex\per\kilo\parsec}$ in the outer region, with the separation at about $\num{12} \pm \SI{1.5}{\kilo\parsec}$ (computed as in \citet{Magrini2023A&A...669A.119M} with the \textsc{elbow} method {\footnote{https://www.scikit-yb.org/en/latest/api/cluster/elbow.html}}). The position of the knee identified by our sample is in excellent agreement with that of \citet{spina22}, $R_{\mathrm{knee}} = 12.1 \pm \SI{1.1}{\kilo\parsec}$, as well as the inner slope $\num{-0.064} \pm \SI{0.007}{\dex\per\kilo\parsec}$ and the outer slope $\num{-0.019} \pm \SI{0.008}{\dex\per\kilo\parsec}$.}

\revdeux{In the lower panel, a growth of $\abratio{Ca}{Fe}$ can clearly be seen from the inner part to the outer part of the Milky Way, with an increment of about \SI{0.25}{\dex} in $\abratio{Ca}{Fe}$. This behaviour is also observed in the APOGEE sample \citep[$R_{\mathrm{GC}}$ range \num{7}-\SI{12}{\kilo\parsec},][]{2018AJ....156..142D}, in the OCCASO sample \citep[\num{6}-\SI{11}{\kilo\parsec},][]{casamiquela19} and in the \GESsh sample of \citep[\num{5}-\SI{20}{\kilo\parsec},][]{Magrini2023A&A...669A.119M} for different $\alpha$ elements. The slopes are 0.019$\pm$0.005~dex~kpc$^{-1}$ in the inner part and 0.012$\pm$0.009~dex~kpc$^{-1}$ in the outer one. This, together with the metallicity gradient, is a clear indication of inside-out disc formation \citep{chiappini1997, minchev13, minchev14}, in which inner regions were enriched more rapidly in both iron and $\alpha$ elements, while the outer one had a lower star formation efficiency with a delayed production of elements with longer production time scales, such as iron \citep[see also,][]{1979ApJ...229.1046T,1983MmSAI..54..311G,1986A&A...154..279M}.}

\begin{figure*}
  \centering
  \includegraphics[width=\textwidth,clip]{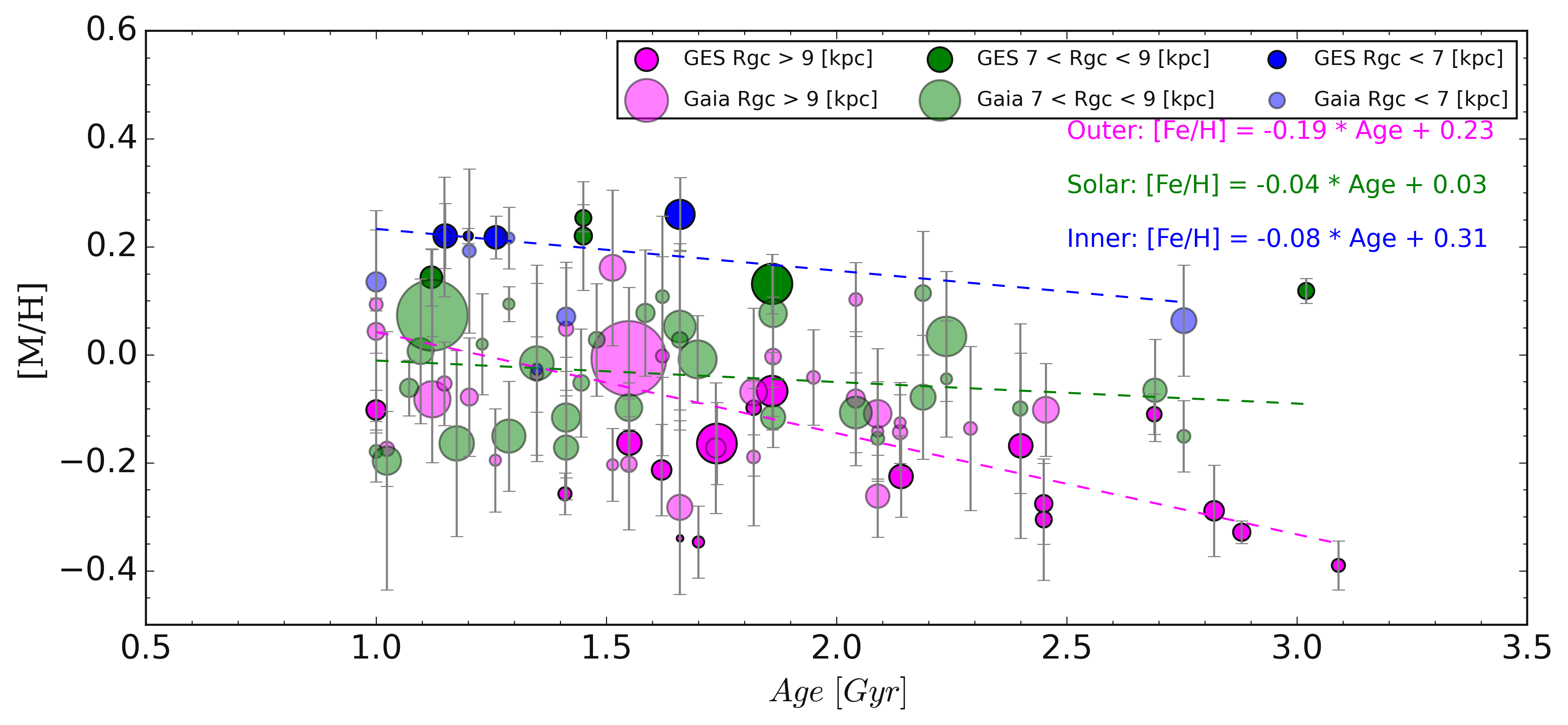}
  \caption{\label{fig:age-met} Age-Metallicity relation for the sample of clusters with ages between \num{1} and \SI{3}{\giga\Year}, divided in three radial bins. The clusters observed by \Gaia are represented with semi-transparent circles, while those observed by \GESsh with solid-filled circles. The linear regressions are shown with dashed lines using the same colour as the corresponding radial bin. {In the plot, we show the coefficients of the three age-metallicity relationships. The size of the symbols is proportional of the number of member stars used to compute the mean values. }}
\end{figure*}

\revdeux{The age-metallicity relation (AMR) in the Galactic disc is crucial to understand how the chemical evolution of the Galaxy proceeded in time. Star clusters provide a useful tool for studying the AMR as they offer a chronological sequence. Previous studies, spanning the last two decades, have explored this relationship using open clusters. While some earlier research found no distinct AMR \citep{friel10, yong12}, others suggested a weak correlation \citep{zhong20}. \citet{Joshi2024FrASS..1148321J} combined a large sample of open cluster parameters, including ages and metallicities, and inferred their AMR. Although there is great scatter in their relationship, they found that for clusters older than \SI{0.25}{\giga\Year}, there is a decreasing trend in metallicity with increasing cluster age. In Figure~\ref{fig:age-met}, we show the age-metallicity relation (AMR) for the sample of open clusters with ages in the range $[\SI{1}{\giga\Year}, \SI{3}{\giga\Year}]$. Considering the clusters all together, we also find a considerable scatter in metallicity at each age. If we divide the clusters into galactocentric distance bins, however, we find much better-defined AMRs. In Figure~\ref{fig:age-met}, our clusters are divided in three radial bins: the inner bin with $R_{\mathrm{GC}} < \SI{7}{\kilo\parsec}$, the central one with $\SI{7}{\kilo\parsec} \leq R_{\mathrm{GC}} \leq \SI{9}{\kilo\parsec}$, and the outer one with $R_{\mathrm{GC}} \geq \SI{9}{\kilo\parsec}$. We note a loose correlation: the youngest clusters tend to have a $\abratio{M}{H}$ higher than the oldest ones. Both the \GESsh and the \Gaia cluster populations support this correlation. Compared to \citet{Joshi2024FrASS..1148321J}, we are considering the end of the tail of the open cluster age  distribution, having only clusters between \num{1}-\SI{3}{\giga\Year}. We confirm their result, but thanks to the radial bin separation, we can explain the origin of the scatter in the AMR.}

\begin{figure*}
  \centering
  \includegraphics[width=\textwidth,clip]{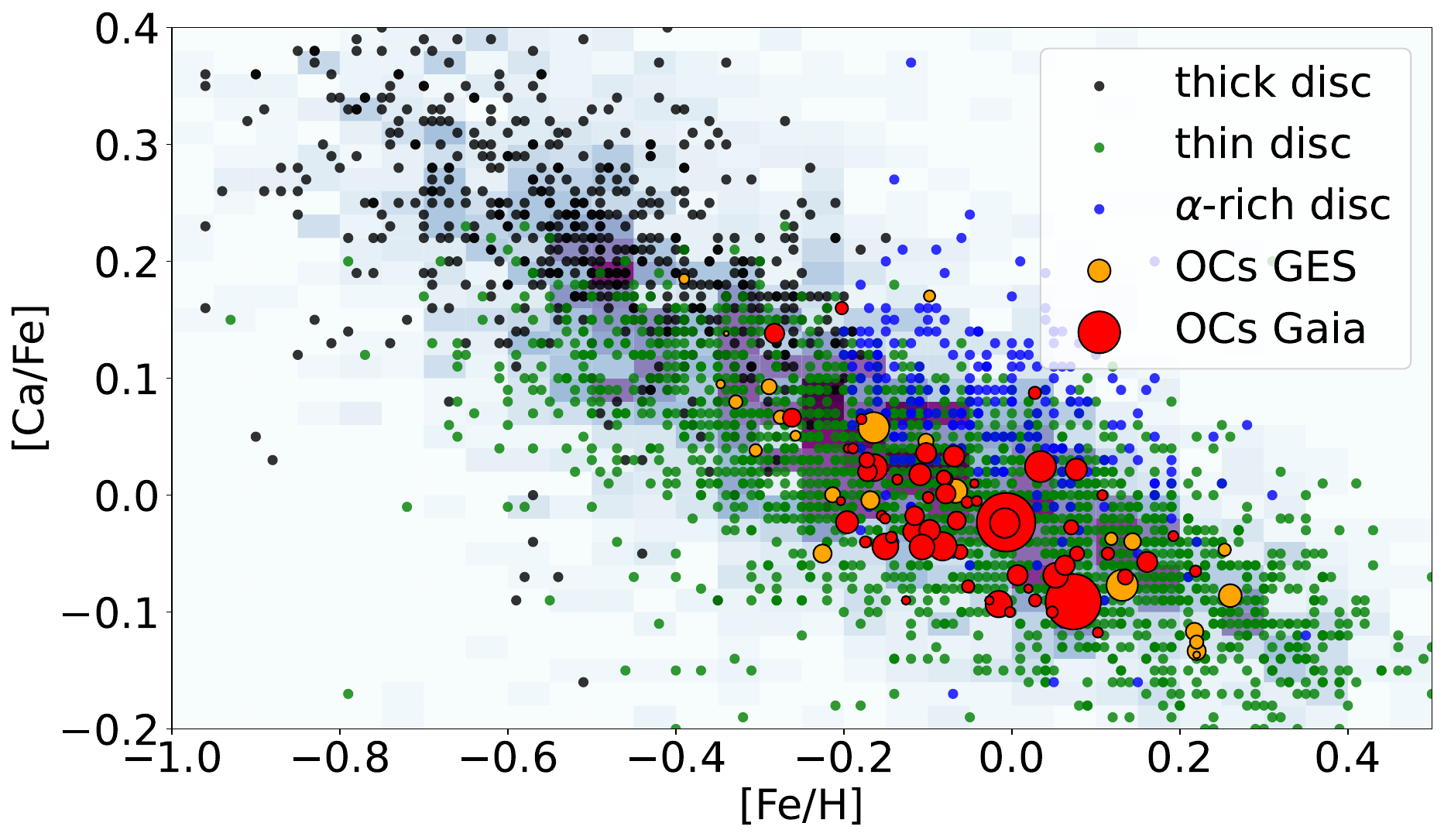}

  \includegraphics[width=\textwidth,clip]{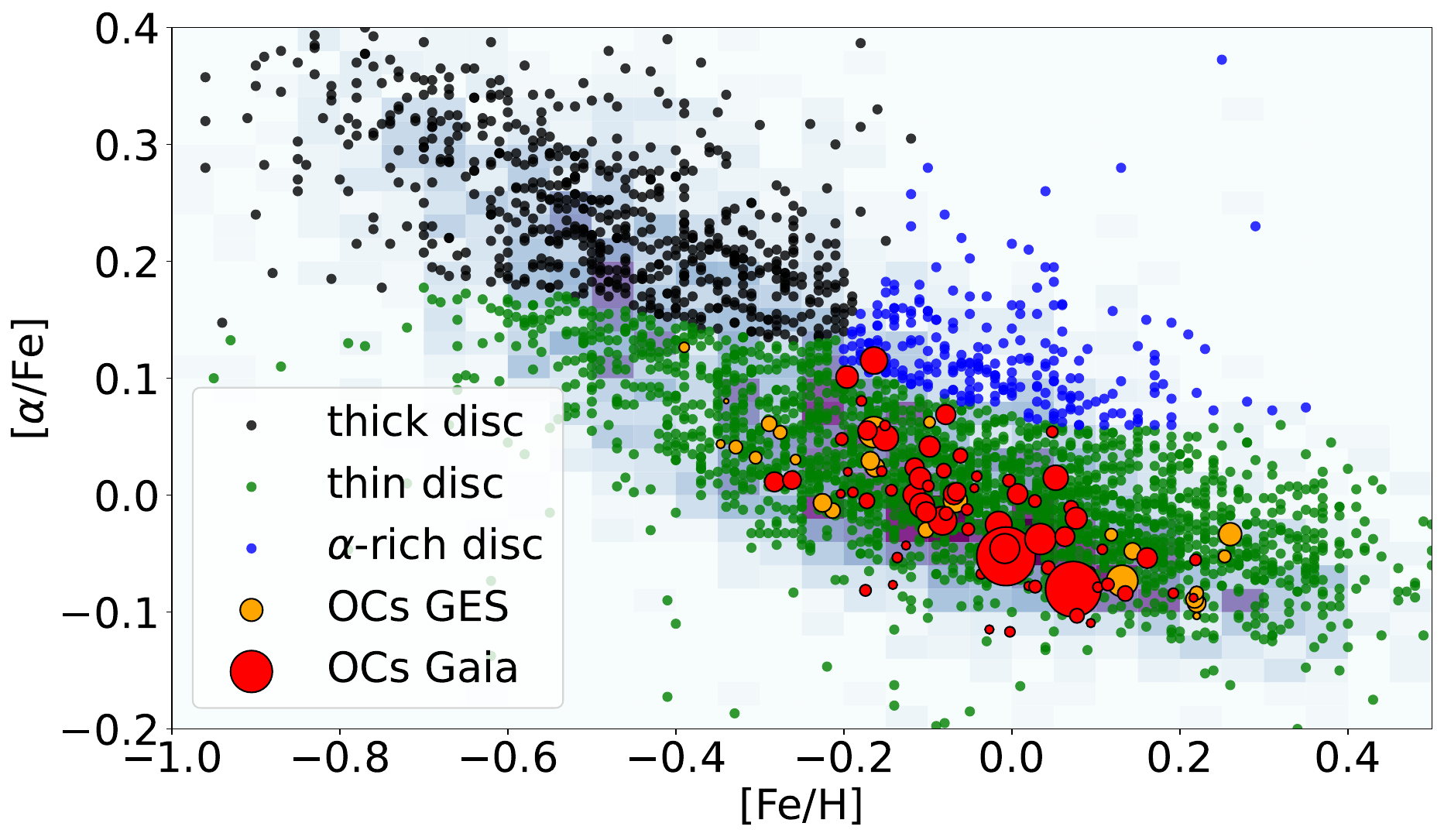}
  \caption{\label{fig:thin_thick} \Gaia and \GESsh clusters in the $\abratio{Ca}{Fe}$ vs. $\abratio{M}{H}$ (upper panel) and in the $\abratio{\alpha}{Fe}$ vs. $\abratio{M}{H}$ (lower panel) planes. In the background (light blue 2D histogram with rectangular cells), we show the abundances of the field stars observed by \GESsh. In black, we show the thick disc population; in green, the thin disc one; in blue, the high-$\alpha$ discs. The open clusters observed by \GESsh are represented by orange circles and those observed by \Gaia by red circles. In both cases the radius of the circles is proportional to the number of members. }
\end{figure*}

In Figure~\ref{fig:thin_thick} we show the \Gaia and \GESsh clusters in the $\abratio{Ca}{Fe}$ vs. $\abratio{M}{H}$ and in the $\abratio{\alpha}{Fe}$ vs. $\abratio{M}{H}$ planes. In the background, we show the abundances of the field stars observed by \GESlg, selected according to these conditions: \texttt{GES\_FLD} is MW, setup is U580, $\snr > 20$, error on $\abratio{Fe}{H} < 0.2$, and errors on the individual abundances $< 0.1$ (about \num{3000} stars). We classified the field stars according to their belonging to the thin, thick and high-$\alpha$ discs using a set of supervised learning techniques such as the Support Vector Machines (SVMs) \citep{Boser92}, already adopted in \citet{viscasillas22a}. We defined a training set based on the sample of \citet{Costa_Silva20}, with $\abratio{\alpha}{Fe}$-$\abratio{Fe}{H}$ derived by \citet{DelgadoMena17}. We included the thin and thick disc populations, as well a high-$\alpha$ metal-rich population (h$\alpha$mr). We obtained an accuracy in the classification of \SI{100}{\percent}. We trained the SVM in the multiclass case with a \textsc{Radial Basis Function} (RBF) and implemented using the \textsc{scikit-learn} package \citep{scikit-learn11}. We calculated the membership probabilities (see Tables \ref{tab:prob_ocs_ges} and \ref{tab:prob_ocs_gaia}) calibrated using Platt scaling extended for multi-class classification \citep{Wu04}; we then transfer the classification probability to the open cluster population. This allows us to place the open clusters of \GESsh and most of \Gaia ones in the thin-disc or h$\alpha$mr populations. The classification is obtained in the $\abratio{\alpha}{Fe}$-$\abratio{Fe}{H}$ plane in which the separation among the populations is more clear, but it can be transferred also to the $\abratio{Ca}{Fe}$-$\abratio{Fe}{H}$ plane. We find that the selected open-cluster population belongs to the thin disc, \revdeux{as expected from their orbital properties \citep[e.g.][]{Wu2009MNRAS.399.2146W}.}

\section{Conclusions}
\label{Sec:conclusion}

This article focuses on assessing the quality and on a scientific usage of the observables measured or inferred from the \Gaia mean spectra recorded with the Radial Velocity Spectrometer of resolving power $R \sim 11500$. The quality assessment is carried out by comparing a number of \Gaia DR3 spectroscopically-derived quantities to their counterparts obtained with the ground-based higher-resolution spectroscopic survey \GESlg. Thus, this study is primarily a new external and independent validation of the \Gaia results. Here are the main results:

\begin{itemize}

\item given the respective uncertainties of the two surveys, we find an excellent agreement between the \Gaia RVS radial velocities and the \GESsh ones. We are not able to explain the discrepancy for only a very small number of objects, about \num{600} out of \num{14692}. This could be due to underestimated uncertainties in one or both surveys, or it can hide a physical origin (jitter, hidden multiplicity);

\item \Gaia DR3 still misses most of the spectroscopic binaries (SB) present in the \GESsh\,--\,\Gaia intersection: only \num{22} out of the \num{2117} \GESsh SB$n$ (preliminary results for the final \GESsh SB census) are flagged as (astrometric or eclipsing or spectroscopic) binaries by \Gaia. We find an empirical efficiency of the DSC classifier compatible with the theoretical efficiency: about \SI{0.2}{\percent} of the unresolved binaries are recovered by this classifier. The \texttt{RUWE} quantity is not a sufficient criterion to spot potential SB$n$;

\item the broadening parameter is loosely correlated with the projected rotational velocity of the stars since the sample under analysis has a bias towards FGK stars that generally have low rotational velocities, and thus we are reaching the instrumental limit in spectral resolution to measure rotational velocities;

\item our comparison sample shows a \revdeux{better agreement between the \GESsh and \gspspec} sets of parameters (effective temperature, surface gravity, metallicity) for objects brighter than $G = \SI{11}{\mag}$ \revdeux{than for objects fainter than $G = \SI{11}{\mag}$}. It tends to indicate that for objects fainter than $G = \SI{11}{\mag}$, the accuracy and/or the precision of the atmospheric parameters of \Gaia \gspspec degrade quickly. This magnitude threshold corresponds to a \snr of the mean RVS spectra of about $70 \pm 20$. The final users should be cautious when they want to work with these parameters for faint objects. The use of the \gspspec \revun{calibrations and} flag systems is mandatory to avoid interpretation mistakes;

\item using asteroseismic-based quantities did not allow us to reconcile unequivocally the different $\log g$ and $\abratio{Fe}{H}$ scales under study. The best agreement is obtained between the calibrated gravity of \Gaia and those from the HR sample of GES, with an offset close to zero for the K2 sample and slightly positive for the CoRoT one. For metallicity, offsets are equally small (but with sign change) in the comparison with  all \Gaia calibrations, both for K2 and CoRoT samples. The scatter tends to increase for the GES MR sample. An effort of the community is still needed to increase the collection of reference stars for which independent techniques are used to estimate their atmospheric parameters that can be used afterwards for cross-survey calibrations;

\item based on our limited sample, the improvement brought by the calibrated \gspspec parameters is important for the surface gravity, while it is not so evident for the abundances;

\item the situation is better for abundance ratios. Indeed, it is empirically known that systematic effects tend to cancel out for abundance ratios. It is therefore probably safer to work with faint objects if one focuses on these quantities. Here again, the \gspspec flags should be used to further clean the selection;

\item in particular, using averaged abundances for open clusters allow us to retrieve well-known properties of this stellar population, providing a scientific check of the \Gaia \gspspec data quality and showing that this spectroscopic survey can be combined to other ground-based spectroscopic surveys to explore the properties of Milky Way stellar populations.
\end{itemize}

\begin{acknowledgements}
\textbf{Institutional:}
The authors thank the referee for their very careful reading and their detailed commentary. It was really of great help to improve this article.
CVV and LS thank the EU programme Erasmus+ Staff Mobility for their support.
CVV and GT acknowledge funding from the Research Council of Lithuania (LMTLT, grant No. P-MIP-23-24).
We acknowledge financial support under the National Recovery and Resilience Plan (NRRP), Mission 4, Component 2, Investment 1.1, Call for tender No. 104 published on 2.2.2022 by the Italian Ministry of University and Research (MUR), funded by the European Union -- NextGenerationEU -- Project `Cosmic POT' Grant Assignment Decree No. 2022X4TM3H by the Italian Ministry of Ministry of University and Research (MUR). We acknowledge support from the INAF Large Grant 2023 ``EPOCH'' (PI: Magrini),  from the INAF Mini Grant 2022 `Checs' (PI Magrini), from the INAF funds to support WEAVE preparation. 

\textbf{Consortium:}
Based on dataproducts from observations made with ESO Telescopes at the La Silla Paranal Observatory under programmes 188.B-3002, 193.B-0936, and 197.B-1074. These data products have been processed by the Cambridge Astronomy Survey Unit (CASU) at the Institute of Astronomy, University of Cambridge, and by the FLAMES/UVES reduction team at INAF/Osservatorio Astrofisico di Arcetri. These data have been obtained from the \Gaia-ESO Survey Data Archive, prepared and hosted by the Wide Field Astronomy Unit, Institute for Astronomy, University of Edinburgh, which is funded by the UK Science and Technology Facilities Council. This work was partly supported by the European Union FP7 programme through ERC grant number 320360 and by the Leverhulme Trust through grant RPG-2012-541. We acknowledge the support from INAF and Ministero dell’Istruzione, dell’Università e della Ricerca (MIUR) in the form of the grant "Premiale VLT 2012".
This work has made use of data from the European Space Agency (ESA) mission Gaia (\url{https://www.cosmos.esa.int/gaia}), processed by the Gaia Data Processing and Analysis Consortium (DPAC, \url{https://www.cosmos.esa.int/web/gaia/dpac/consortium}). Funding for the DPAC has been provided by national institutions, in particular the institutions participating in the Gaia Multilateral Agreement.

\textbf{Databases:}
This research has made use of NASA’s Astrophysics Data System Bibliographic Services.
This research has made a heavy use of the SIMBAD database, the VizieR catalogue access tool and the cross-match service provided and operated at CDS, Strasbourg, France.

\textbf{Softwares:}
This work has made use of python 3.x\footnote{\url{https://www.python.org}} \citep{vanrossum} and of the following python's modules: Astropy\footnote{\url{https://www.astropy.org}}, a community-developed core Python package and an ecosystem of tools and resources for astronomy \citep{2013A&A...558A..33A,2018AJ....156..123A,2022ApJ...935..167A}; matplotlib\footnote{\url{https://matplotlib.org/}} \citep{Hunter:2007}; numpy\footnote{\url{https://numpy.org/}} \citep{harris2020array}; scipy\footnote{\url{https://scipy.org/}} \citep{2020SciPy...NMeth}; Scikit-learn Machine Learning \citep{scikit-learn11}; StatsModels \citep{seabold2010statsmodels}; Seaborn \citep{Waskom2021}. This work has made use of TopCat \citep{Taylor2005}.
\end{acknowledgements}

\bibliographystyle{aa}
\bibliography{Bibliography}

\begin{appendix}
  \section{Additional material}

\begin{table*}
  \centering
  \caption{\label{Tab:main_used_columns} Main columns of the \GESlg and \Gaia public catalogues used in this study. Columns are: catalogue name, column names, motivation, sections where it is mainly used. We remind the reader that the \gspspec calibrated gravities, metallicities and abundances have to be calculated by applying the published relations to the uncalibrated quantities.}
  \begin{tabular}{llll}
    \toprule
    Catalogue & Columns & Motivation & Sections\\
    \midrule
    GES & SNR & physical quantity & -\\
    \Gaia & phot\_g\_mean\_mag & physical quantity & -\\
    \Gaia & grvs\_mag & physical quantity & -\\
    \Gaia & rvs\_spec\_sig\_to\_noise & physical quantity & -\\

    \midrule
    GES & VRAD, E\_VRAD & physical quantity & radial velocity\\
    \Gaia & radial\_velocity, radial\_velocity\_error & physical quantity & radial velocity\\
    GES & SRP & sample cleaning & radial velocity\\
    GES & SRV & sample cleaning & radial velocity\\
    GES & EML & sample cleaning & radial velocity\\
    GES & REC\_SETUP & analysis tracking & radial velocity\\
    \Gaia & RUWE & sample cleaning & radial velocity\\
    \Gaia & phot\_variable\_flag & sample cleaning & radial velocity\\

    \midrule
    GES & PECULI containining 20010-14- or 20010-13 & SB1 selection & binary\\
    GES & PECULI containining 20020-14- or 20020-13 & SB2 selection & binary\\
    GES & PECULI containining 20030-14- or 20030-13 & SB3 selection & binary\\
    GES & PECULI containining 20040-14- or 20040-13 & SB4 selection & binary\\
    \Gaia & non\_single\_star & sample selection & binarity\\
    \Gaia & RUWE & diagnostic & binarity\\
    \Gaia & classprob\_dsc\_combmod\_binarystar & diagnostic & binarity\\
    \Gaia & classprob\_dsc\_specmod\_binarystar & diagnostic & binarity\\
    \Gaia & flags\_msc & diagnostic & binarity\\

    \midrule
    GES & VSINI, E\_VSINI & physical quantity & rotational velocity\\
    \Gaia & vbroad, vbroad\_error & physical quantity & rotational velocity\\

    \midrule
    GES & TEFF, E\_TEFF & physical quantity & temperature\\
    \Gaia & teff\_gspspec, teff\_gspspec\_lower, teff\_gspspec\_upper & physical quantity & temperature\\

    GES & LOGG, E\_LOGG & physical quantity & gravity\\
    \Gaia & logg\_gspspec, logg\_gspspec\_lower, logg\_gspspec\_upper & physical quantity & gravity\\

    GES & FEH, E\_FEH & physical quantity & metallicity\\
    \Gaia & mh\_gspspec, mh\_gspspec\_lower, mh\_gspspec\_upper & physical quantity & metallicity\\

    GES & MG1, SI1, CA1, TI1, E\_MG1, E\_SI1, E\_CA1, E\_TI1 & physical quantity & abundances\\
    \Gaia & alphafe\_gspspec, alphafe\_gspspec\_lower, alphafe\_gspspec\_upper & physical quantity & abundances\\
    \Gaia & cafe\_gspspec, cafe\_gspspec\_lower, cafe\_gspspec\_upper & physical quantity & abundances\\

    \Gaia & flags\_gspspec & sample cleaning & stellar parameters/abundances\\
    \bottomrule
  \end{tabular}
\end{table*}

  \begin{figure}
  \centering
  \includegraphics[width=0.9\columnwidth,clip]{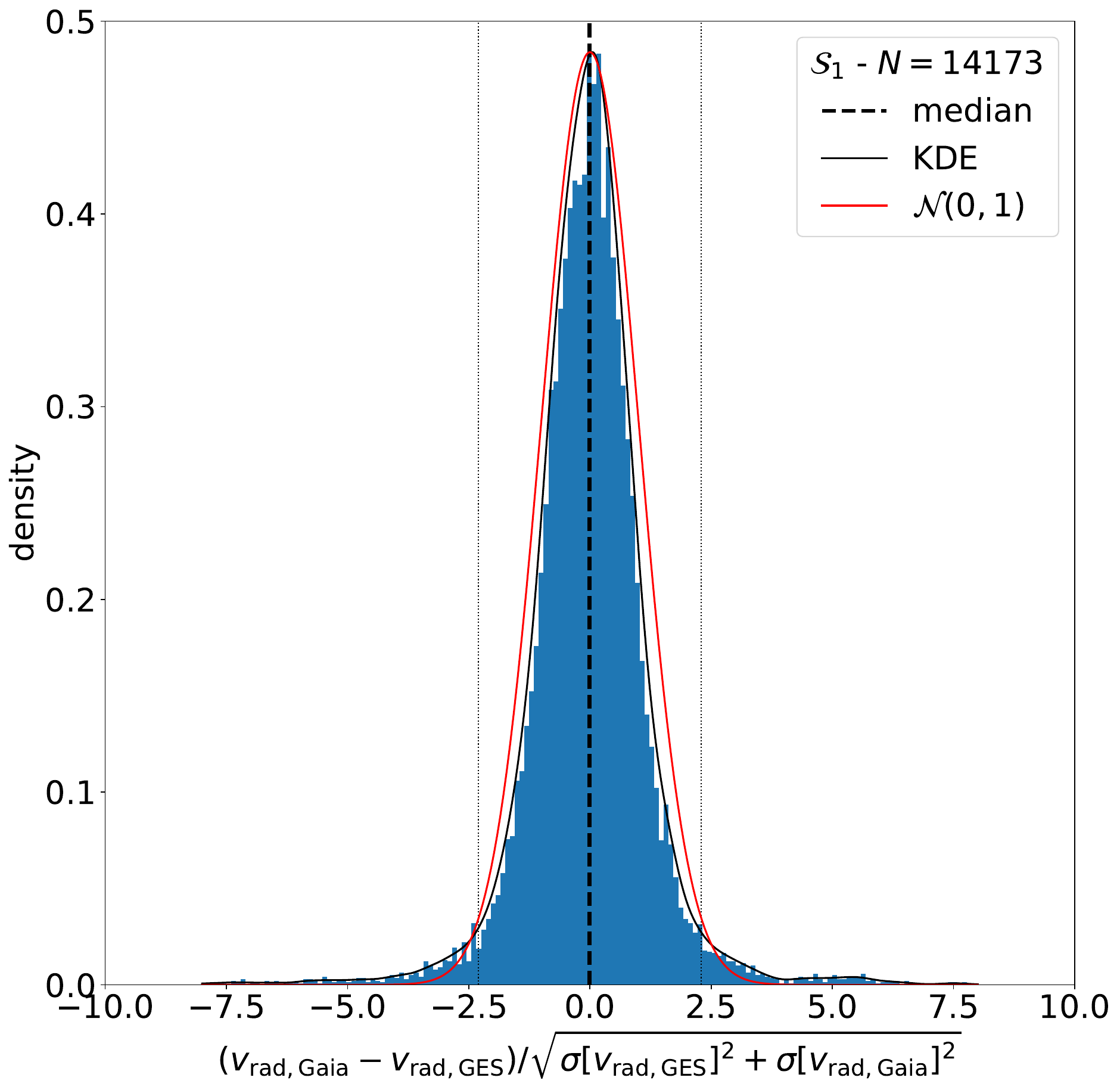}
  \caption{\label{Fig:PDF_difference_normalised_velocities_corr_unc} Probability distribution of the normalised velocity differences. The blue histogram (bin width = \num{0.1}) displays the distribution $\Delta_{\mathrm{norm}} v_{\mathrm{rad}}$ of the difference of the radial velocity differences $v_{\mathrm{rad,Gaia}} - v_{\mathrm{rad,GES}}$ normalised by the propagated errors $\sqrt{\sigma[v_{\mathrm{rad,Gaia}}]^2 + \sigma_{\mathrm{corr}}[v_{\mathrm{rad,GES}}]^2}$ where $\sigma_{\mathrm{corr}}[v_{\mathrm{rad,GES}}]$ is the uncertainty corrected according to \citet{2023A&A...674A..32B}. The black line is the empirical KDE obtained from the sample distribution. The red line is the probability distribution function (PDF) of a the normal law centred in 0 and of unit variance. The dashed vertical black line indicates the mean of the distribution. The black dotted lines have equation $| \Delta_{\mathrm{norm}} v_{\mathrm{rad}} | = 2.3$ and show where the tails of the empirical distribution become heavier than those of the normal law.}
\end{figure}

\begin{figure*}
  \centering
  \includegraphics[width=\textwidth,clip]{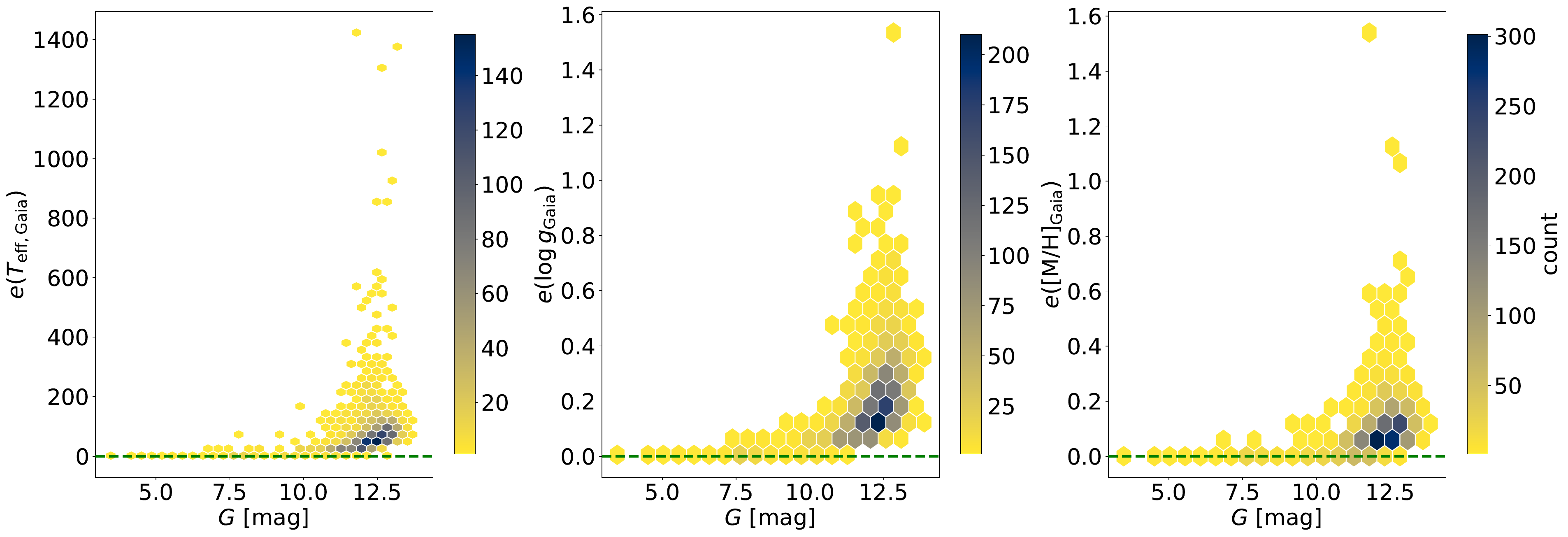}
  \caption{\label{Fig:GSPSpec_error_vs_Grvs_magnitude} Correlation between $e(\mathcal{P}_{\mathrm{Gaia}})$ and $G_{\mathrm{RVS}}$ where $\mathcal{P}$ if either $T_{\mathrm{eff}}$, $\log g$ or $\abratio{M}{H}$ (from left to right).}
\end{figure*}

\begin{figure}
  \centering
  \includegraphics[width=\columnwidth,clip]{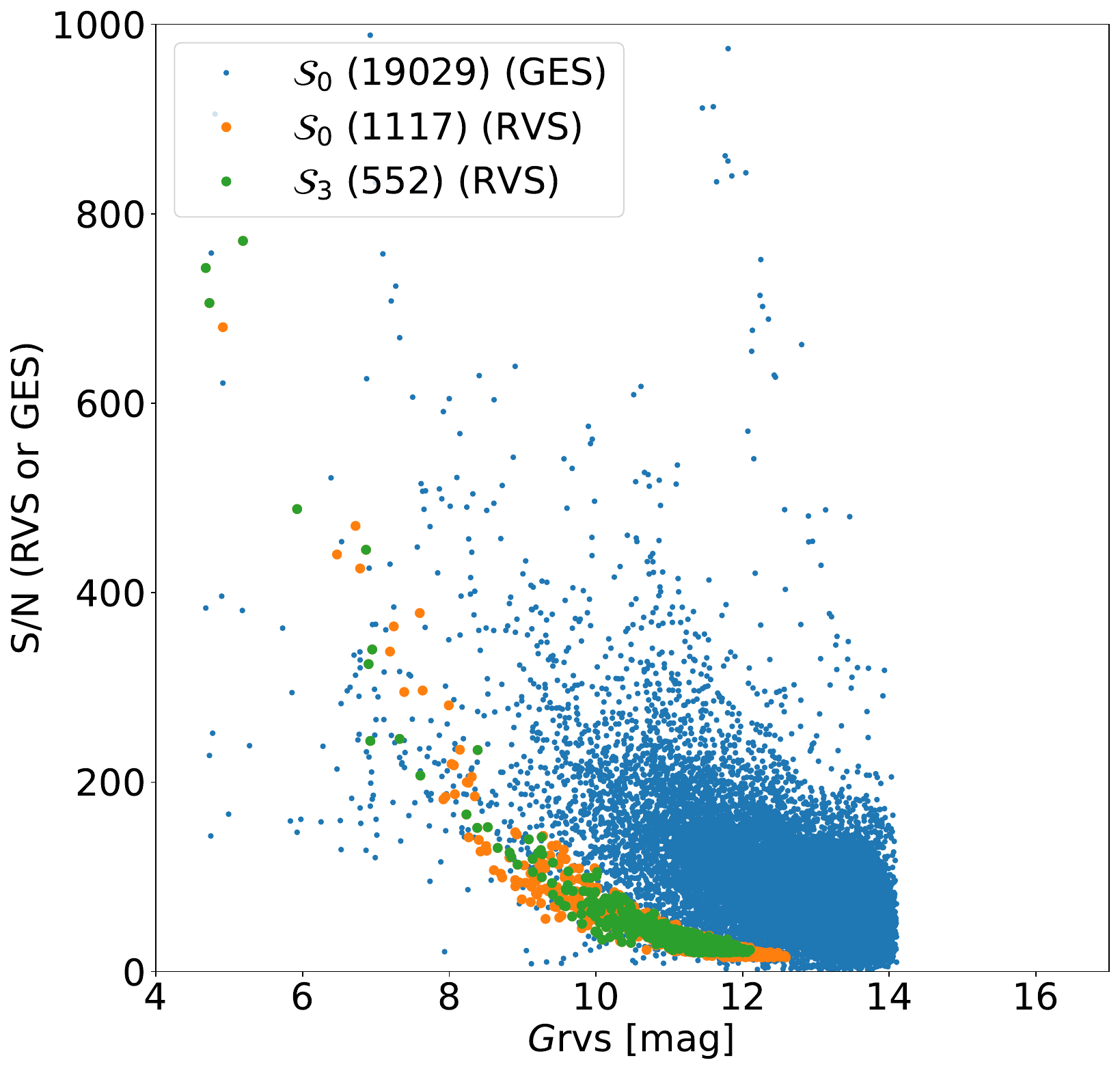}
  \caption{\label{Fig:Grvs_magnitude_vs_snr} \revdeux{\GESsh \snr and RVS \snr as a function of $G$ for respectively \num{19029} (blue) and \num{1117} (orange) objects of $\mathcal{S}_{0}$ with a valid \GESsh (resp., RVS) \snr and RVS \snr as a function of $G$ for the \num{552} (green) objects of $\mathcal{S}_{3}$ with a valid RVS \snr.}}
\end{figure}

\begin{table}
\caption{SVM classification of the GES open clusters (aged between 1 and 3 Gy) according to the given training data and predicted probabilities of belonging to each of of the disk components (thin: P0, thick: P1, and h$\alpha$: P2 )}
\label{tab:prob_ocs_ges}
\centering
\begin{tabular}{l c c c c}
\hline\hline
 GES\_FLD & Pop & P0 & P1 & P2\\
\hline
  Br21 & 0 & 1.0 & 0.0 & 0.0\\
  Br22 & 0 & 0.999 & 0.0 & 0.001\\
  Br25 & 0 & 1.0 & 0.0 & 0.0\\
  Br31 & 0 & 0.999 & 0.0 & 0.001\\
  Br44 & 0 & 0.985 & 0.014 & 0.001\\
  Br73 & 0 & 1.0 & 0.0 & 0.0\\
  Br75 & 0 & 0.999 & 0.0 & 0.001\\
  Br81 & 0 & 0.984 & 0.014 & 0.002\\
  Col110 & 0 & 0.981 & 0.003 & 0.017\\
  Cz24 & 0 & 1.0 & 0.0 & 0.0\\
  Cz30 & 0 & 1.0 & 0.0 & 0.0\\
  NGC2141 & 0 & 0.999 & 0.001 & 0.0\\
  NGC2158 & 0 & 1.0 & 0.0 & 0.0\\
  NGC2355 & 0 & 1.0 & 0.0 & 0.0\\
  NGC2420 & 0 & 0.997 & 0.001 & 0.002\\
  NGC2425 & 0 & 1.0 & 0.0 & 0.0\\
  NGC2477 & 0 & 0.992 & 0.008 & 0.0\\
  NGC2506 & 0 & 0.994 & 0.0 & 0.006\\
  NGC4337 & 0 & 0.986 & 0.013 & 0.001\\
  NGC6005 & 0 & 0.985 & 0.014 & 0.001\\
  NGC6583 & 0 & 0.983 & 0.016 & 0.002\\
  Rup134 & 0 & 0.987 & 0.012 & 0.001\\
  Tom2 & 0 & 1.0 & 0.0 & 0.0\\
  Trumpler20 & 0 & 0.991 & 0.008 & 0.0\\
\hline
\end{tabular}
\end{table}

\begin{table}
\caption{SVM classification of 67 Gaia open clusters (aged between 1 and 3 Gy and with N$\ge$3) according to the given training data and predicted probabilities of belonging to each of the disk components (thin: P0, thick: P1, and h$\alpha$: P2 )}
\label{tab:prob_ocs_gaia}
\centering
\scalebox{0.87}{
\begin{tabular}{l c c c c}
\hline\hline
 cluster & Pop & P0 & P1 & P2\\
\hline

  Alessi\_1 & 0 & 0.999 & 0.001 & 0.0\\
  Berkeley\_68 & 0 & 1.0 & 0.0 & 0.0\\
  Collinder\_110 & 0 & 0.999 & 0.001 & 0.0\\
  Czernik\_12 & 0 & 0.999 & 0.001 & 0.0\\
  ESO\_518\_03 & 0 & 0.995 & 0.004 & 0.0\\
  FSR\_0278 & 0 & 0.992 & 0.008 & 0.0\\
  FSR\_0496 & 0 & 0.999 & 0.001 & 0.0\\
  FSR\_0866 & 0 & 1.0 & 0.0 & 0.0\\
  FSR\_1378 & 0 & 0.999 & 0.001 & 0.0\\
  Gulliver\_13 & 0 & 0.994 & 0.006 & 0.0\\
  Haffner\_22 & 0 & 1.0 & 0.0 & 0.0\\
  IC\_4651 & 0 & 0.996 & 0.004 & 0.0\\
  IC\_4756 & 0 & 0.998 & 0.001 & 0.001\\
  King\_23 & 0 & 0.998 & 0.002 & 0.0\\
  King\_5 & 0 & 1.0 & 0.0 & 0.0\\
  LP\_145 & 0 & 0.991 & 0.009 & 0.0\\
  LP\_2198 & 0 & 0.993 & 0.007 & 0.0\\
  LP\_5 & 0 & 0.999 & 0.001 & 0.0\\
  LP\_930 & 0 & 1.0 & 0.0 & 0.0\\
  NGC\_1245 & 0 & 0.999 & 0.001 & 0.0\\
  NGC\_1817 & 0 & 1.0 & 0.0 & 0.0\\
  NGC\_2112 & 0 & 0.999 & 0.001 & 0.0\\
  NGC\_2141 & 0 & 0.998 & 0.002 & 0.0\\
  NGC\_2158 & 0 & 0.999 & 0.001 & 0.001\\
  NGC\_2204 & 0 & 1.0 & 0.0 & 0.0\\
  NGC\_2354 & 0 & 1.0 & 0.0 & 0.0\\
  NGC\_2355 & 0 & 0.997 & 0.003 & 0.0\\
  NGC\_2360 & 0 & 0.722 & 0.06 & 0.218\\
  NGC\_2420 & 0 & 1.0 & 0.0 & 0.0\\
  NGC\_2423 & 0 & 0.998 & 0.002 & 0.0\\
  NGC\_2477 & 0 & 0.995 & 0.005 & 0.0\\
  NGC\_2506 & 0 & 1.0 & 0.0 & 0.0\\
  NGC\_2509 & 0 & 0.991 & 0.009 & 0.0\\
  NGC\_2627 & 0 & 0.999 & 0.001 & 0.0\\
  NGC\_3680 & 0 & 0.926 & 0.004 & 0.07\\
  NGC\_6208 & 0 & 0.997 & 0.001 & 0.002\\
  NGC\_6811 & 0 & 0.998 & 0.002 & 0.0\\
  NGC\_6819 & 0 & 0.997 & 0.003 & 0.0\\
  NGC\_6939 & 0 & 0.998 & 0.002 & 0.0\\
  NGC\_6940 & 0 & 0.999 & 0.001 & 0.0\\
  NGC\_6991 & 0 & 0.998 & 0.001 & 0.001\\
  NGC\_7044 & 0 & 0.997 & 0.003 & 0.0\\
  NGC\_752 & 2 & 0.172 & 0.09 & 0.738\\
  NGC\_7762 & 0 & 1.0 & 0.0 & 0.0\\
  NGC\_7789 & 0 & 0.998 & 0.002 & 0.0\\
  Ruprecht\_171 & 0 & 0.996 & 0.004 & 0.0\\
  Ruprecht\_68 & 0 & 1.0 & 0.0 & 0.0\\
  Skiff\_J0058+68.4 & 0 & 0.859 & 0.005 & 0.136\\
  Skiff\_J1942+38.6 & 0 & 0.997 & 0.003 & 0.0\\
  Tombaugh\_1 & 0 & 1.0 & 0.0 & 0.0\\
  Trumpler\_20 & 0 & 0.995 & 0.005 & 0.0\\
  Trumpler\_32 & 0 & 0.999 & 0.001 & 0.0\\
  UBC\_1061 & 0 & 0.997 & 0.003 & 0.0\\
  UBC\_141 & 0 & 1.0 & 0.0 & 0.0\\
  UBC\_199 & 0 & 0.999 & 0.001 & 0.0\\
  UBC\_255 & 0 & 0.999 & 0.001 & 0.0\\
  UBC\_284 & 0 & 0.994 & 0.006 & 0.0\\
  UBC\_307 & 0 & 0.988 & 0.011 & 0.001\\
  UBC\_310 & 0 & 0.987 & 0.012 & 0.001\\
  UBC\_374 & 0 & 0.992 & 0.008 & 0.0\\
  UBC\_472 & 0 & 1.0 & 0.0 & 0.0\\
  UBC\_57 & 0 & 1.0 & 0.0 & 0.0\\
  UBC\_577 & 0 & 0.994 & 0.002 & 0.004\\
  UBC\_614 & 0 & 0.998 & 0.002 & 0.0\\
  UFMG\_2 & 0 & 0.985 & 0.014 & 0.001\\
  UPK\_27 & 0 & 0.994 & 0.006 & 0.0\\
  UPK\_84 & 0 & 0.956 & 0.008 & 0.036\\

\hline
\end{tabular}}
\end{table}

\end{appendix}

\end{document}